# Energy Mapping of Existing Building Stock in Cambridge using Energy Performance Certificates and Thermal Infrared Imagery

Yinglong He[a, b, f, 1], Jiayu Pan[b], Ramit Debnath[c,d,e], Ronita Bardhan[b, 1], Luke Cullen[a], Marco Gomez Jenkins[a], Erik Mackie[d], George Hawker[a], Ian Parry[a]

[a] Institute of Astronomy, University of Cambridge, Cambridge, CB3 0HA, UK
[b] Department of Architecture, University of Cambridge, Cambridge, CB2 1PX, UK
[c] EPRG, Cambridge Judge Business School, University of Cambridge, Cambridge, CB2 1AG, UK
[d] Cambridge Zero, Department of Computer Science and Technology and Department of Architecture, Cambridge, CB3 OFD, UK
[e] Division of Humanities and Social Science, California Institute of Technology, Pasadena, 91125, US
[f] School of Mechanical Engineering Sciences, University of Surrey, Surrey, GU2 7XH, UK

## Abstract

Energy performance certificate (EPC) and thermal infrared (TIR) images both play a key role in the energy performance mapping of the urban building stock. In this paper, we developed parametric building archetypes using an EPC database and conducted temperature clustering on TIR images acquired through drones and satellite datasets. We evaluated 1725 EPCs of existing building stock in Cambridge, UK to generate energy consumption profiles. Drone-based TIR images of individual buildings in two Cambridge University colleges were processed using a machine learning pipeline for thermal anomaly detection and investigated the influence of two specific factors that affect the reliability of TIR for energy management applications: ground sample distance (GSD) and angle of view (AOV). The EPC-level results suggest that the construction year of the buildings influences their energy consumption. For example, modern buildings were over 30% more energy-efficient than older ones. Parallelly, older buildings were found to show almost double the energy savings potential through retrofitting than newly constructed buildings. TIR imaging results showed that thermal anomalies can only be properly identified in images with a GSD of 1m/pixel or less. A GSD of 1-6m/pixel can detect hot areas of building surfaces. We found that a GSD > 6m/pixel cannot characterise individual buildings but does help identify urban heat island effects. Additional sensitivity analysis showed that building thermal anomaly detection is more sensitive to AOV than to GSD. Our study informs newer approaches to building energy diagnostics using thermography and supports decision-making for retrofitting at a large scale.

[1] Corresponding authors (Email: yinglong.he@surrey.ac.uk, rb867@cam.ac.uk)

# Impact Statement

Urban building stock energy performance mapping is essential for designing strategies to reduce energy consumption and emissions. However, the sheer scale of urban areas makes estimating the energy performance of every building extremely resource-intensive. New types of ensemble datasets, such as energy performance certificates and thermal infrared imaging via drones and satellite data, can aid in rapid performance evaluation. Here, we integrate these datasets to generate parametric building archetypes for a historical city and assess its energy consumption profiles. Meanwhile, we present a novel method for detecting thermal anomalies using drone-based infrared imaging, which can aid in identifying urban heat island effects. This study informs new approaches to building energy diagnostics utilising thermography and supports large-scale retrofitting decisions.

# 1. Introduction

The built environment is responsible for 25% of the UK's total greenhouse gas emissions (UK Green Building Council, 2021). In 2020, the residential sector emitted 66.3Mt$CO_2$, accounting for around 16% of all UK carbon dioxide emissions, with emissions from the public sector contributing a further 7.4Mt$CO_2$ (2%) (BEIS, 2022). Decarbonising the built environment is therefore a crucial component of the UK's journey towards achieving net-zero emissions by 2050. Given that the majority of these emissions arise from the use of gas for heating and the UK has an ageing housing stock, it is crucial to transform the existing inefficient buildings into more efficient and sustainable forms (i.e., building energy retrofit). Energy retrofit has the capacity to increase system integration and improve resource use at an urban level, and therefore, addresses poor thermal efficiency (linked to climate impacts and fuel poverty), as well as overheating, poor indoor air quality and moisture.

Energy retrofit analyses at an urban scale require building archetypes. With 29 million existing UK homes that need to be retrofitted (Holmes et al., 2019), it is critical to analyse the energy performance of buildings utilising the scant data available from the existing building stock. This analysis can be performed on the basis of certain types of building energy modelling, the outcomes of which can be used to determine the best energy retrofitting strategies (Reinhart and Davila, 2016). Energy modelling at an urban scale usually needs building stock data such as geometry parameters and building physical properties. However, due to a paucity of data and concerns about user privacy, collecting building stock data is challenging (Reinhart and Davila, 2016). To close this gap, it is usual practice to categorise the building stock using a number of building types that represent the related technical, operational, and geometrical attributes of a large collection of buildings (Mata et al., 2014). Building archetypes, in particular, classify the building stock using geometric and non-geometric attributes. All buildings with similar parameters are grouped together and called archetypes (Galante and Torri, 2012; Famuyibo et al., 2012). These archetypes have been created using data from national surveys, which gives a good picture of the nation's building stock (Ali et al., 2019).

In addition to building archetypes, which can facilitate energy modelling and identify priorities for building energy retrofits, reliable approaches for thermal efficiency monitoring and thermal defects detection are important for verifying the retrofit design scenarios and for underpinning the progress towards climate commitments. In building diagnostics, thermal infrared (TIR) imaging has been widely used to detect thermal anomalies (Baldinelli et al., 2018). This is a non-destructive testing approach that measures the infrared radiation emitted from the surface of an object to quantify and visualise the distribution of the surface temperature. The IR camera produces a sequence of two-dimensional and readable IR images (thermographs), where specific colours and tones identify different temperatures (Balaras and Argiriou, 2002). Through the TIR temperature patterns, we can visually identify abnormal temperature distributions, i.e., thermal anomalies that are related to structural features, building materials, and energy problems (Moropoulou et al., 2013; Maldague, 2001). Building thermal anomalies, e.g., thermal bridge, infiltration, and moisture damage within building envelopes, lead to increased heat loss and lower thermal efficiency (Kylili et al., 2014). For example, thermal bridges that occur at the junction of the envelopes are considered to cause up to 30% of the heating energy demand (Erhorn-Kluttig and Erhorn, 2009). Therefore, TIR imaging can help identify potential thermal anomalies and monitor the thermal efficiency of building envelopes (Park et al., 2021).

There are various TIR approaches to conduct building thermal diagnostics and map energy performance of buildings. The hand-held (or walk-through) TIR survey is the most common method, in which a thermographer walks around a building and visually checks the temperature pattern on the envelope with an infrared camera (Fox et al., 2016). Aerial-derived TIR survey utilises an IR camera fixed to a drone, an aeroplane, or a helicopter. It allows users to detect building thermal anomalies at a community scale, especially moisture or energy loss problems on roof surfaces (Lucchi, 2018). Space-derived TIR survey uses earth observation data collected via IR satellite. It has been used to identify the thermal efficiency of the envelopes of large buildings (Smargiassi et al., 2008). The Landsat images at 30x30m resolution have been applied to derive the land surface temperature and predict the energy performance level (Sun and Bardhan, 2023). Furthermore, recent advances in high-resolution IR space telescopes have introduced the capacity to identify the thermal efficiency of individual households and other buildings globally (Ben et al., 2021). With the application of TIR survey from drone, the potential of applying space-derived large-scale high-resolution thermal imagery in detecting building energy performance could be explored.

The goals of this paper are two-fold. Firstly, it aims to develop and analyse building archetypes focusing on the Energy Performance Certificate (EPC) and then convey valuable insights into the energy performance of the existing building stock in Cambridge. Secondly, it attempts to explore the potential of high-resolution thermal infrared (TIR) space telescopes in detecting thermal anomalies of individual buildings, using drone-emulated TIR images with different resolutions and machine learning (ML) image analysis techniques. The approach presented in this paper can have several applications in future data-driven and digital building diagnostics. For example, the development of such TIR image-based analytical pipeline can provide a low-cost non-destructive solution to understand the building characteristics, thermal efficiency, and energy performance for existing building stocks, thus identify the corresponding retrofitting strategies and maximise the energy saving potential to achieve net-zero targets.

The paper is structured as follows: Section 2 gives an overview of the literature on building archetypes development and TIR imaging approaches for thermal diagnostics of buildings.

Section 3 describes the methodology, including data acquisition, the procedure for building archetypes development, and temperature clustering to detect thermal anomalies. Section 4 presents and discusses the results, followed by Section 5 which concludes this study.

# 2. Literature Review

## 2.1. Building Archetypes Development for Energy Assessment

The concept of building archetypes is most frequently applied to energy modelling at the urban scale. For the development of archetypes, there are three different widely accepted methods: synthetical average, real example and real average (Sousa Monteiro et al., 2015). In the synthetical average building approach, the archetypes are chosen using information about the most commonly used materials and systems. Several projects are being established at the European and global levels to describe the national-scale building stocks, for example, TABULA (Loga et al., 2012), BPIE (European Union, 2018) and DOE (US Department of Energy, 2018). In the real example building approach, the building type is selected based on experience, namely, it is chosen according to the expertise of panel experts and other information sources within a real-world climatic situation. The other information sources typically take into account the most widely used materials, construction period, and specific sizes. In the real average building approach, the choice of building type is made through the statistical analysis of data from a large building sample (Famuyibo et al., 2012). Also, image-based approaches, such as street-view images and satellite images of buildings, are increasingly applied in the energy assessment and prediction as parts of model training information (Sun et al., 2022; Sun and Bardhan, 2023, Zhang, C. et al., 2023, Zheng, H. et al., 2020).

## 2.2. Thermal Infrared (TIR) Imaging for Buildings

In the literature, depending on the data acquisition techniques, TIR imaging approaches to monitor building thermal and energy efficiency can be generally classified into three categories: hand-held, aerial-derived, and space-derived.

### *1) Hand-Held TIR Survey*

In the hand-held (or walk-through) TIR survey, a thermographer walks around a building and conducts systematic scans of the building surfaces (Thumann and Younger, 2008). To identify thermal anomalies, a series of requirements are specified in standard procedures including (Lucchi, 2018; RESNET, 2012; IOS, 1983; Pearson, 2011): (i) air temperature difference of at least 10°C between the internal and external spaces; (ii) wind speeds less than 5m/s; (iii) surfaces not exposed to direct sunlight both during the testing phase and in the 8 or 12 hours prior to the survey; and (iv) overcast situations to prevent the reflection of a clear sky. In addition, practical instructions suggest including the heat flux metre (HFM) measurements for assessing the thermal efficiency of buildings (Hart, 1991). The time-lapse survey uses recorded movies or time-lapse images to inspect changes in surface temperature patterns over time (Fox et al., 2016; Fox et al., 2015).

### *2) Aerial-derived TIR Survey*

Unmanned Aerial Vehicles (UAVs) such as drones enable professionals to perform building thermography rapidly and accurately while reducing operational costs and minimising safety risks (Rakha and Gorodetsky, 2018). For example, previous studies use UAVs with IR cameras to analyse solar radiation on roofs in order to determine the optimal placement of solar panels (López-Fernández et al., 2015; Schuffert et al., 2015). The window thermal heat losses in buildings have been examined and analysed using data collected through UAVs (Martinez-De Dios and Ollero, 2006). A novel workflow for accurate building envelope defect characterization has been proposed using drone aerial time-lapse IR data (Rakha et al., 2022). In addition, aerial thermography has been demonstrated to be well placed for detecting moisture over flat roof surfaces (Stockton, 2013).

### *3) Space-derived TIR Survey*

TIR earth observation satellites in low earth orbit can monitor building energy output, making them an effective tool for ensuring that governments, businesses, and even individuals are on track to reach internationally agreed carbon emission reduction targets (Mo et al., 2018). In the literature, IR satellite images across the city have been used together with individual building characteristics to predict indoor temperature (Smargiassi et al., 2008) and energy performance (Sun and Bardhan, 2023). A statistical approach is proposed to exploit both spatially explicit satellite TIR data and time-varying meteorological data for estimating surface temperature, which can be used to further assess indoor exposure to heat by taking into account building characteristics (Kestens et al., 2011). Moreover, recent advances in high-resolution TIR space telescopes can provide a ground sample distance (GSD) of less than 7 m/pixel with a daily revisit rate, introducing the capacity to identify the thermal efficiency and anomalies of individual buildings, even for residential ones (Ben et al., 2021).

# 3. Methodologies

## 3.1. Data Acquisition

### *1) Building Geometry and Non-geometry Data*

Several data inputs, including building geometry and non-geometry information, are required for the building archetypes (Ali et al., 2018; Wate and Coors, 2015).

Geometry input data consist of shapes, heights, building envelopes, the number of floors, walls, etc. Typically, geometric building data is collected from building stock data. The building footprint is gathered through the GIS data model. The most widely accepted standard format model is the geospatial vector data format (i.e., shapefile), which includes geometry data like polygons, lines, and points.

Non-geometric building properties are also necessary for creating building archetypes, including energy performance certificate (EPC), property type, built form, usage pattern, HVAC systems, etc. The accessibility of the above information is the major challenge. Non-geometric building data is usually collected via national census databases or statistical surveys.

Table 1 gives the sources and attributes of the different building datasets, which are employed in the development of the building archetype in this work.

**Table 1.** Data sources for building archetype development

| | **Provider - Source** | **Main attributes** |
|---|---|---|
| *Geometric Data* | | |
| 1 | Verisk - UKBuildings [1] | • Property area ($m^2$)<br>• Building area ($m^2$)<br>• Height (m)<br>• Age (e.g., historic)<br>• Use (e.g., residential)<br>• etc. |
| 2 | Digimap - Boundary-Line [2] | • District (i.e., ward) |
| 3 | UK Ministry of Housing, Communities and Local Government - Lower Super Output Area (LSOA) IMD2019 (WGS84) [3] | • LSOA code<br>• LSOA name<br>• Shape area ($m^2$)<br>• Shape length (m)<br>• etc. |
| 4 | Office for National Statistics - National Statistics Postcode Lookup (NSPL) [4] | • Postcode<br>• Longitudinal<br>• Lateral<br>• etc. |
| 5 | USGS/NASA - Landsat 8 [5] | • Land surface temperature (LST, °C)<br>• Normalised difference vegetation index (NDVI)<br>• etc. |
| *Non-Geometric Data* | | |
| 6 | Consumer Data Research Centre (CDRC) - Dwelling Age Group Counts (LSOA) [6] | • LSOA code<br>• Age group counts |
| 7 | Department for Levelling Up, Housing & Communities - Energy Performance of Buildings Data [7] | • Postcode<br>• Property type (e.g., flat)<br>• Built form (e.g., detached)<br>• Energy performance certificate (EPC, current/potential)<br>• Energy efficiency rating (current/potential)<br>• $CO_2$ emissions (current/potential)<br>• Solar water heating flag<br>• Floor height<br>• Glazed area/type<br>• Lightning/heating/hot water costs (current/potential)<br>• Transaction type (e.g., rental)<br>• Number of habitable/heated rooms<br>• etc. |

*Notes*:

[1] https://www.verisk.com/en-gb/3d-visual-intelligence/products/ukbuildings/

[2] https://digimap.edina.ac.uk/help/our-maps-and-data/os_products/#boundary-and-location-data

[3] https://data-communities.opendata.arcgis.com/datasets/lower-super-output-area-lsoa-imd2019-wgs84

[4] https://geoportal.statistics.gov.uk/datasets/ons::national-statistics-postcode-lookup-february-2022

[5] https://developers.google.com/earth-engine/datasets/catalog/landsat-8

[6] https://data.cdrc.ac.uk/dataset/dwelling-ages-and-prices/resource/dwelling-age-group-counts-lsoa

[7] https://epc.opendatacommunities.org/

### *2) Drone Thermal Mapping of Building Rooftops*

Images are captured from the drone flying at 120m elevation. The sensor properties include a field of view of 69°x56° and format of 640x512 pixels, resulting in a resolution of 0.26m which refers to the diameter of each individual pixel in the TIR image. Images were at nighttime to avoid the effects of shadows and the sun's heat on reflective surfaces. The valid temperature range for the camera is -25°C to 135°C at 20°C ambient temperature. Future satellite imagery, due to the altitude of orbit, will be of significantly lower resolution but cover a larger area with each image. To emulate this, first, we stitch together a large number of drone images to obtain a larger area than a single image, and second, we downsample the resulting image to various potential satellite resolutions.

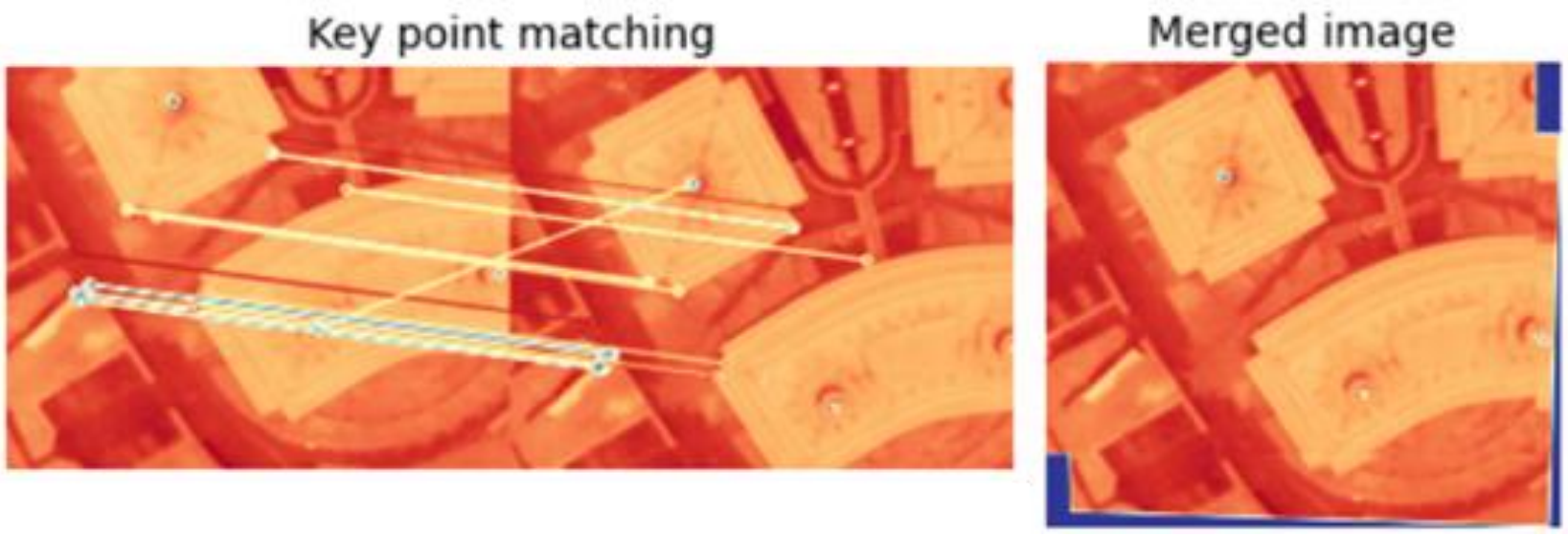

**Figure 1.** Key point matches between images identified by the RANSAC algorithm: (left) raw images and (right) the resulting stitched image.

Stitching drone images presents several challenges including image distortion from the fisheye lens and uneven scaling across images. Pre-processing across all images uses the equations defined in the literature (Tattersall, 2019) to convert Forward Looking InfraRed (FLIR) measurements to surface temperature values with scaling adjusted to align adjacent image distributions. Discarding image contours to avoid the fisheye effects (López et al., 2021), the images are then individually input to the Random Sample Consensus (RANSAC) algorithm (Derpanis, 2010), as shown in Figure 1.

### *3) Regional Land Surface Temperature (LST) from Landsat Satellite Images*

Figure 2 illustrates the diagram for deriving Land Surface Temperature (LST) from four satellite databases including Landsat 8 Surface Reflectance (SR), ASTER Global Emissivity Dataset (GED), Landsat 8 Top of Atmosphere (TOA) Reflectance, and NCEP Total Column Water

Vapour (TCWV), which have spatial resolutions of 30m, 100m, 30m, and 278,300m, respectively.

In this diagram, as indicated by the left-most block, we first provide the aforementioned satellite databases, the date range, and the region of interest (ROI) for subsequent processes. Covering the user-selected dates and regions, the program then loads images from Landsat 8 SR and TOA databases. A cloud mask is applied to both SR and TOA images using their quality information bands. For each TOA image, the two closest TCWV NCEP analysis times are selected and interpolated to the Landsat observation time. The SR data are utilised to estimate Normalised Difference Vegetation Index (NDVI), which quantifies the health and density of vegetation using near-infrared ("NIR", reflected by vegetation) and red light ("Red", absorbed by vegetation). NDVI is then converted to Fractional Vegetation Cover (FVC) values (Carlson and Ripley, 1997). These FVC values are then used together with previously computed ASTER emissivity values for the bare ground to obtain the corresponding Landsat emissivity (Ermida et al., 2020). Finally, the Statistical Mono-Window (SMW) algorithm is applied to the TIR bands of the TOA images (Freitas et al., 2013). The coefficients of the SMW algorithm are mapped onto the Landsat image based on the NCEP TWVC data (Malakar et al., 2018). Figure 3 shows a resulting LST image of Cambridge city.

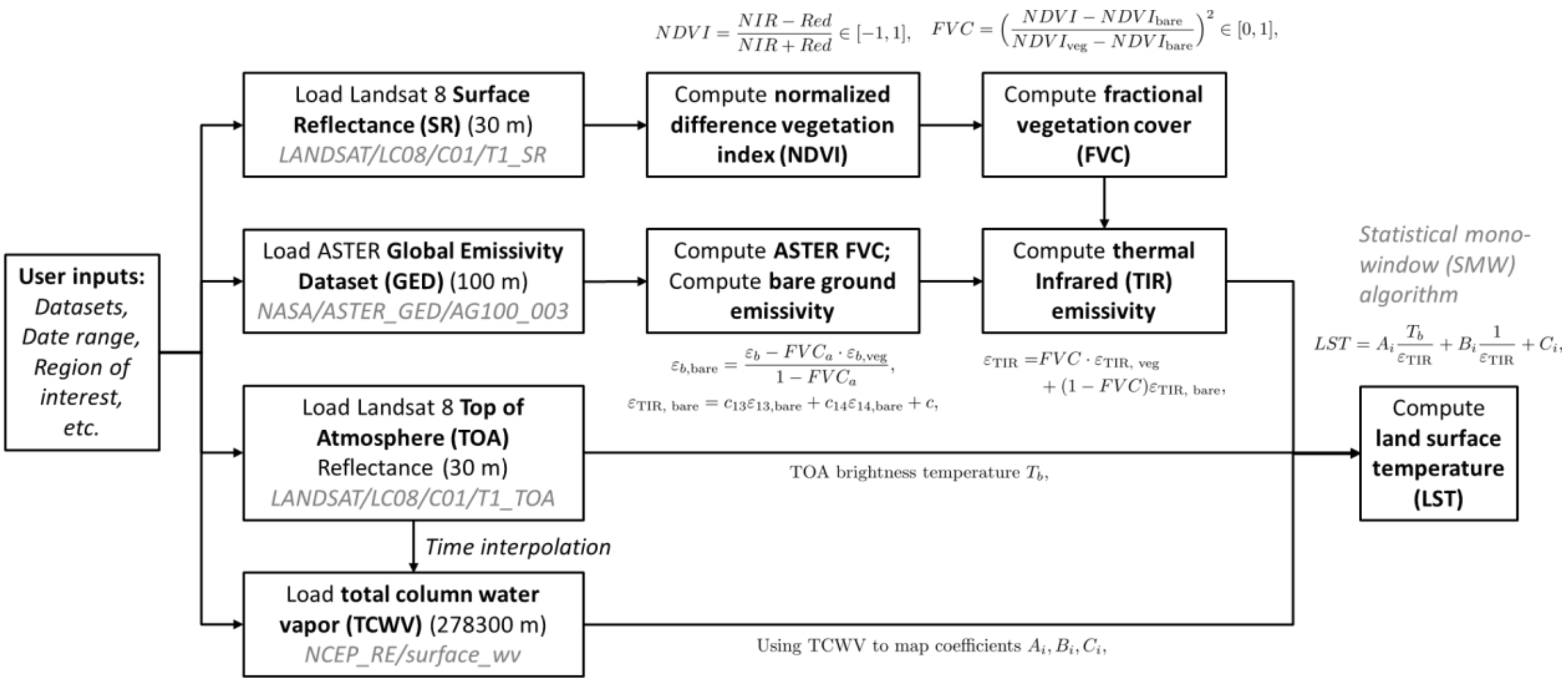

**Figure 2.** Diagram for estimating Land Surface Temperature (LST) from Landsat satellite imagery datasets

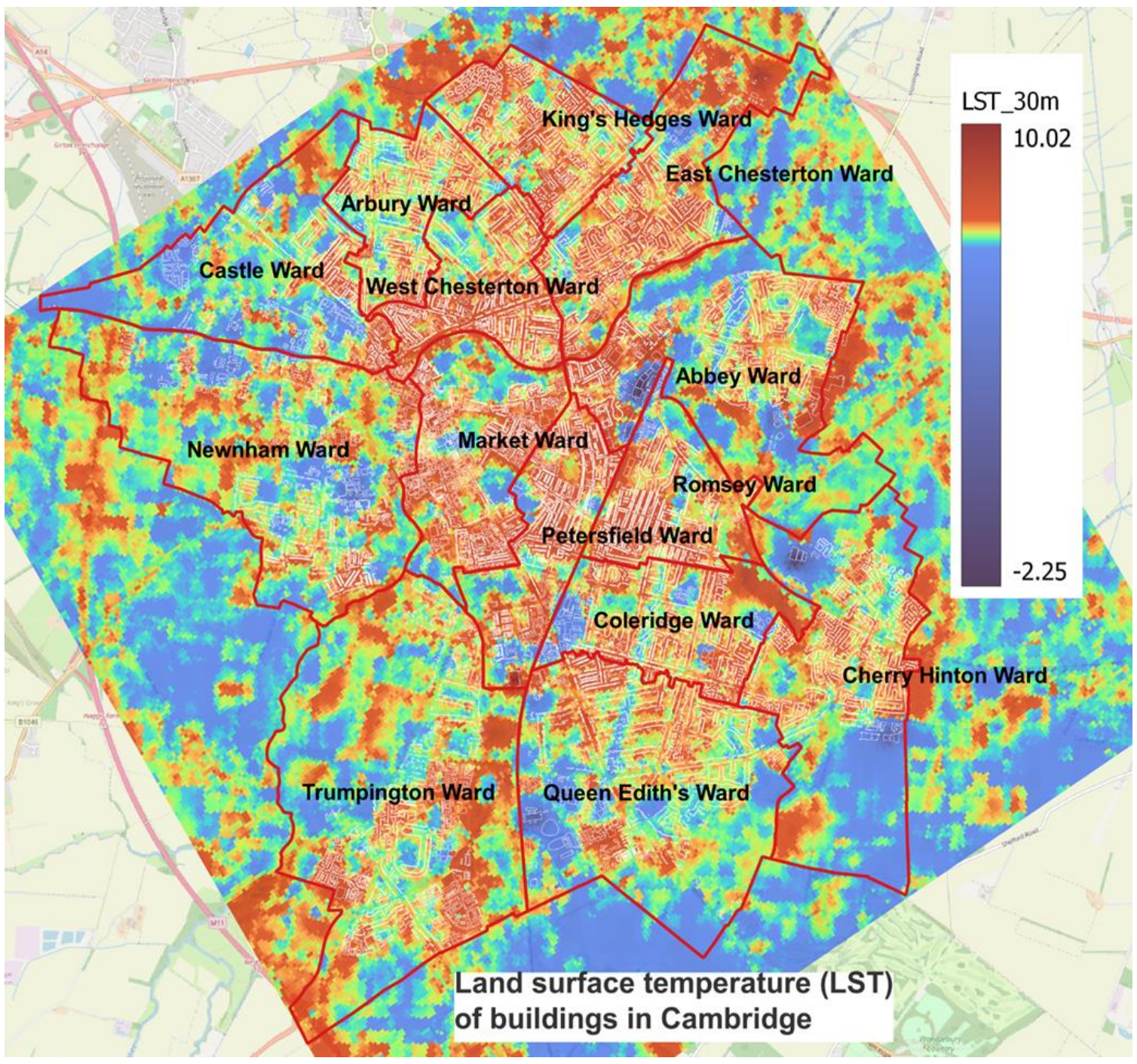

**Figure 3.** Satellite-derived Land Surface Temperature (LST) of Cambridge

## 3.2. Building Archetypes Development

As illustrated in Figure 4, which is adapted from (Ali et al., 2018), the sequential phases of data prepossessing, feature selection, outlier detection, and aggregation are included in the building archetypes development process. Data preprocessing is a data mining method that entails converting raw or real-world data into a usable format. Data cleaning, data integration, data transformation, data reduction, and data discretization are the most commonly adopted steps in preprocessing. Feature selection is the practice of choosing a portion of the most important variables or properties. To achieve accurate results, the feature selection method eliminates irrelevant and redundant attributes. Outlier detection refers to the process of identifying observations in data that deviate significantly from a given set of data. The most popular outlier detection methods include distance-based, density-based and Local Outlier Factor (LOF). The aggregation procedure aims to group the data and then perform arithmetic or geometric mathematical operations on it. The resulting aggregated value represents the properties of a single building archetype.

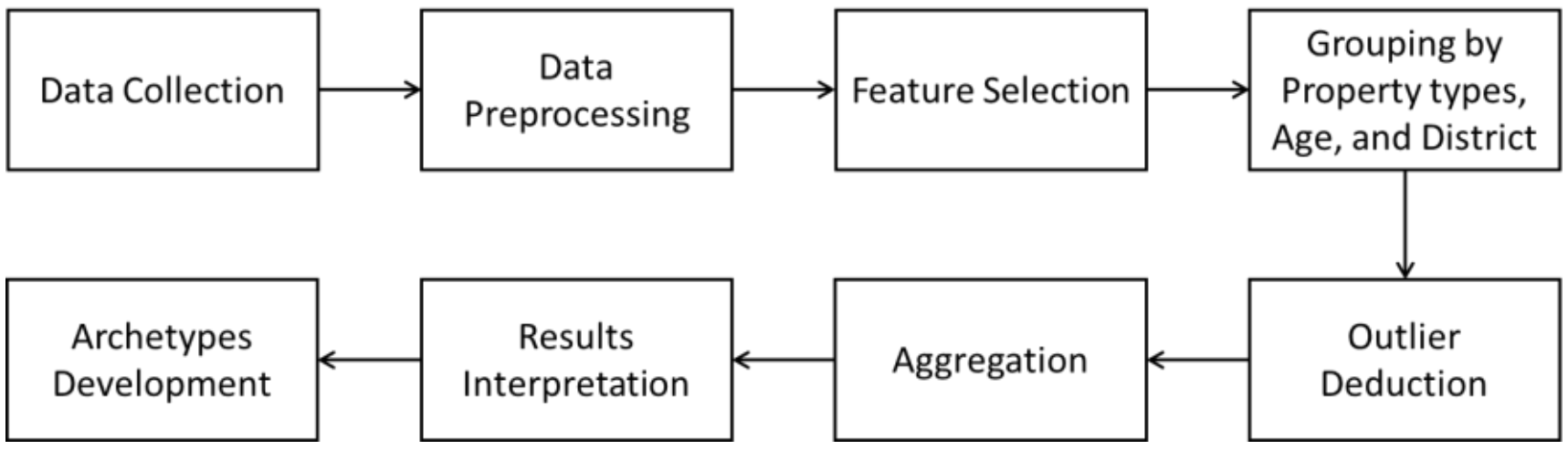

**Figure 4.** Diagram for building archetypes development

For each EPC instance in the developed building archetype, the variables collected are summarized in Table 2.

**Table 2.** Variables in the developed building archetype for EPC analysis

| Variable | Symbol | Unit | Comment |
|---|---|---|---|
| Energy rating | | | Expressed as a letter ranging from A to G, A being the most energy efficient, G the least efficient |
| Current energy consumption | $E_c$ | kWh/m$^2$ | Current annual energy consumption for the property |
| Potential energy consumption | $E_p$ | kWh/m$^2$ | Potential annual energy consumption for the property after specific improvements have been carried out |
| Energy saving potential | $\delta E_p$ | % | Derived from the current and the potential energy consumption, namely, $\delta E_p = (E_c - E_p)/E_c \times 100\%$ |
| Use type | | | Including residential (92.9%) and non-residential (7.1%) |
| Property type | | | Including flat (30.2%) and house (69.8%) |
| Construction year | | | Including <1918 (historic, 21.7%), 1918-1939 (interwar, 14.5%), 1939-1960 (postwar, 26.4%), 1960-1980 (sixties & seventies, 16.1%), >1980 (modern, 21.3%) |
| Lighting cost | | GBP | Annual energy costs for lighting the property |
| Hot water cost | | GBP | Annual energy costs for hot water |
| Heating cost | | GBP | Annual energy costs for heating the property |

## 3.3. Temperature Clustering to Detect Thermal Anomalies

Thermal anomalies on building envelopes often show considerably different temperature distributions than those surrounding normal areas (Park et al., 2021). Therefore, TIR images from infrared thermography can be used to identify thermal anomalies of buildings (Lucchi, 2018). Previous studies in the literature have demonstrated that, if the target domain (e.g., a wall or a roof with the same material) contains both normal and abnormal (i.e., with thermal anomalies) areas, its temperature distribution follows a multi-modal Gaussian (or normal) distribution (Garrido et al., 2018). The Gaussian Mixture Model (GMM) is an unsupervised machine learning (ML) approach to representing a dataset with a weighted average of all individual normal (or Gaussian) distributions called mixture components (Debnath et al., 2022; Garg et al., 2013). It is said that a d-dimensional random variable x follows a k-component GMM if its probability density function can be written as:

$$p(x) = \sum_{i=1}^{k} \omega_i \cdot N_i(x|\mu_i, \Sigma_i), \quad \omega_i \geq 0 \text{ and } \sum_{i=1}^{k} \omega_i = 1 \text{ for } i \in [1, \dots, k], \tag{1}$$

where $N_i(\cdot)$ is the i-th mixture component that is a Gaussian distribution; and $\omega_i$, $\mu_i$, and $\Sigma_i$ are its weight, mean, and variance, respectively. The Expectation – Maximisation (EM) algorithm is commonly utilised to fit GMM, namely, to estimate the values of $\omega_i$, $\mu_i$, and $\Sigma_i$ for ensuring that the GMM has the maximum likelihood.

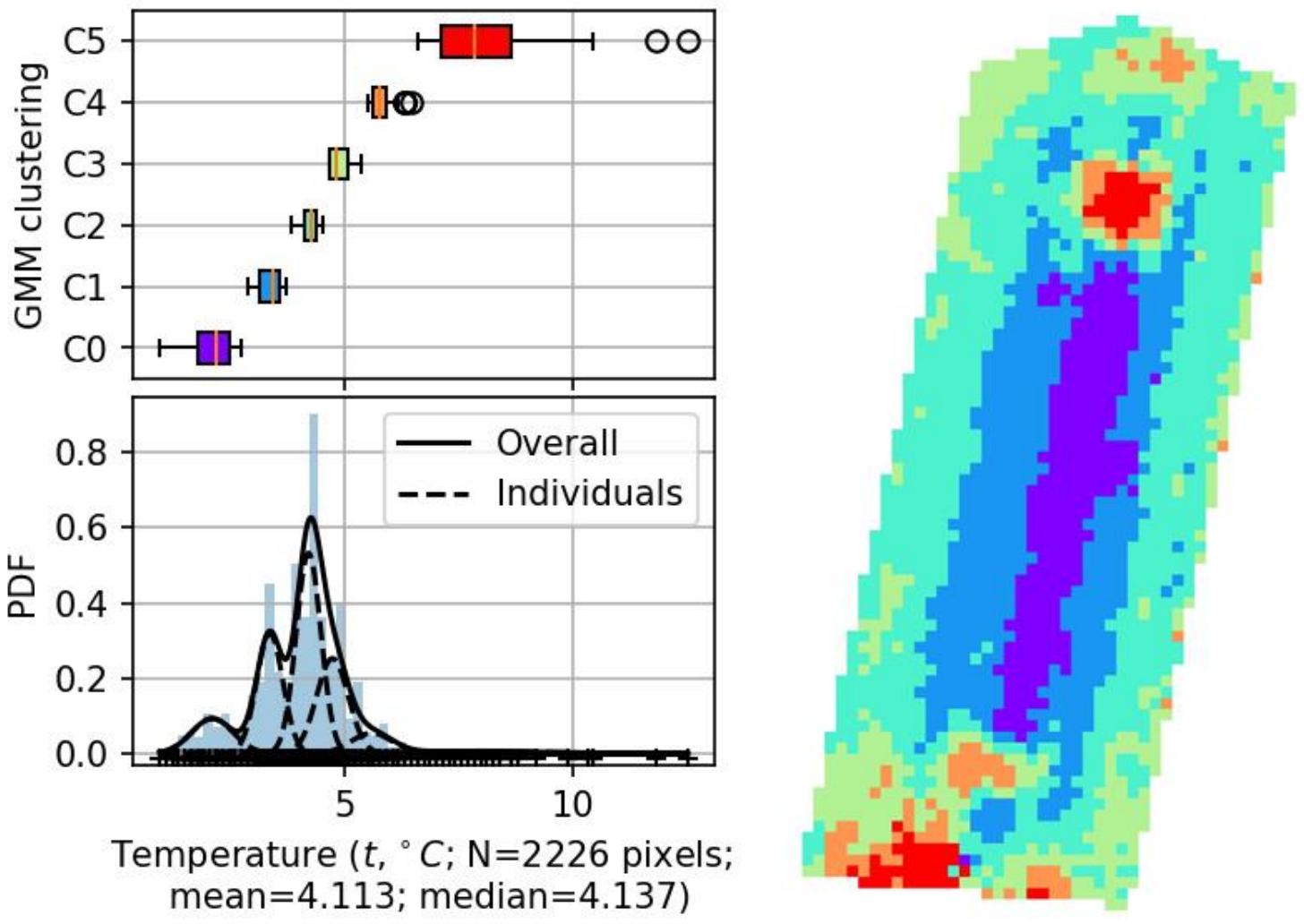

**Figure 5.** Temperature clustering to detect thermal anomalies: (left) Gaussian Mixture Model (GMM) clustering and (right) segmentation result.

The temperature clustering based on GMM is performed as follows. First, clustering is performed by applying GMM to the temperature data of all pixels constituting the input TIR image. As an example, Figure 5 (left bottom) illustrates a histogram of the temperature data, along with the best-fit model for a mixture with six components, the probability density functions (PDFs) of which are represented by dashed lines. Figure 5 (left top) shows the temperature distribution in colour-coded box plots for six mixture components. Figure 5 (right) indicates the spatial (pixel) locations of each distribution (or cluster) in the thermal image. Figure 5 (left top) and (right) share the same colour code. As shown in the figure, temperatures belonging to the same mixture component are regarded as clusters with the same temperature characteristics. The Akaike information criterion (AIC) is applied to optimise the hyperparameter (k), which is the number of mixture components. As the most widely used method to determine the optimal k of GMM, AIC indicates the relative distance between the unknown true likelihood function of the data and the fitted likelihood function of the model (Li et al., 2018). A lower AIC means the model is closer to reality. In this work, we used a range from 2 to 8 for the hyperparameter k, following previous studies (Kim et al., 2021), and the k value with the lowest AIC value is selected for the GMM clustering.

# 4. Results and Discussion

## 4.1 EPCs of Buildings in Cambridge

Energy performance certificate (EPC) databases are a vital resource for mapping the urban building stock and promoting higher overall energy efficiency. This section provides an overview of the energy performance of existing buildings in Cambridge, through an analysis of 1725 EPCs collected by the UK Department for Levelling Up, Housing & Communities. All information that could identify the building, the building unit, the owner, or the technician was removed due to confidentiality concerns.

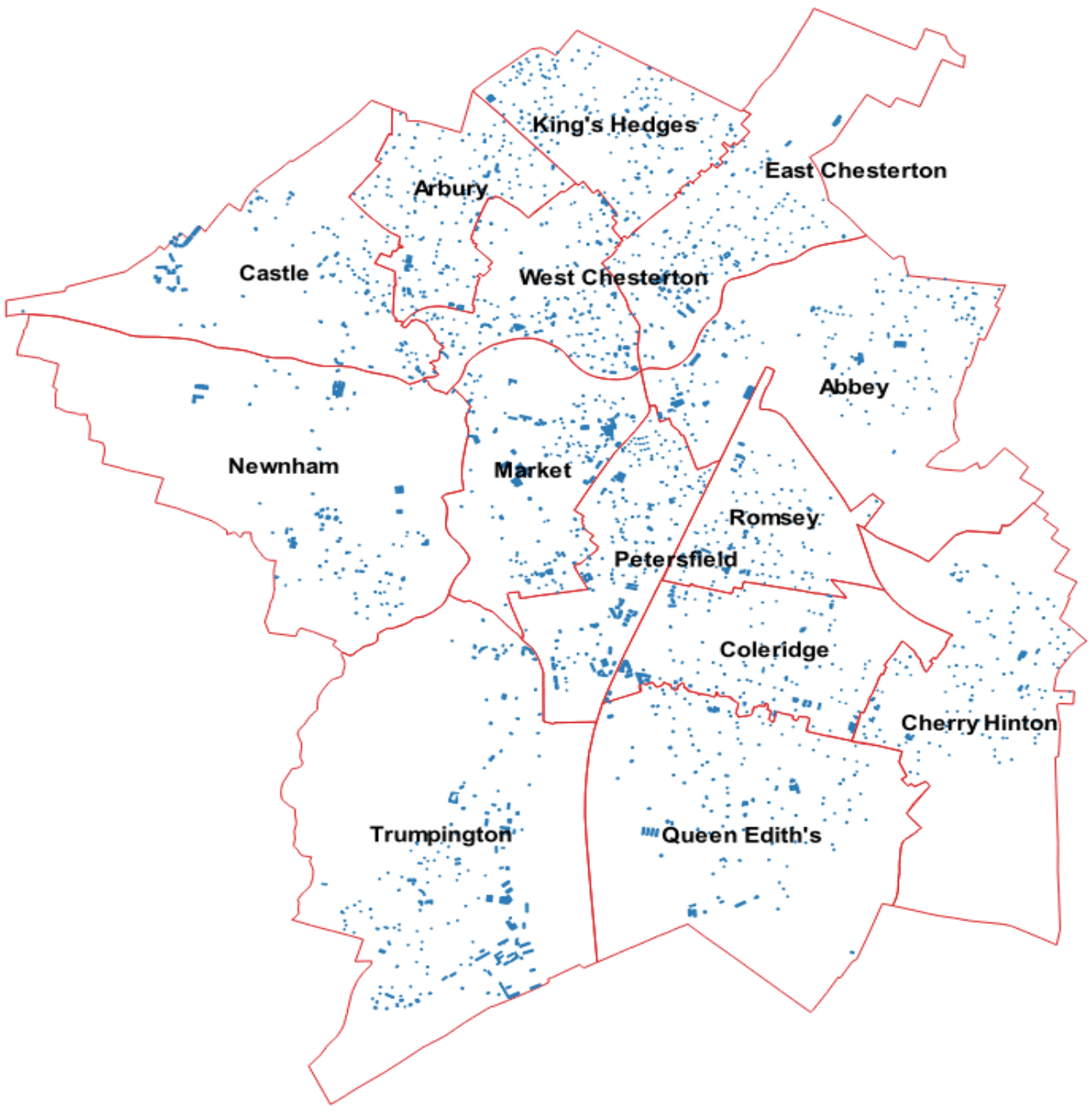

**Figure 6.** Spatial distribution of 1725 individuals in the building archetype for EPCs analysis.

Figure 6 shows the spatial distribution of these 1725 individual buildings across 14 wards in Cambridge. From this map, we can find that these buildings are approximately evenly distributed in the whole city, making the EPCs sample data representative and leading to reliable conclusions.

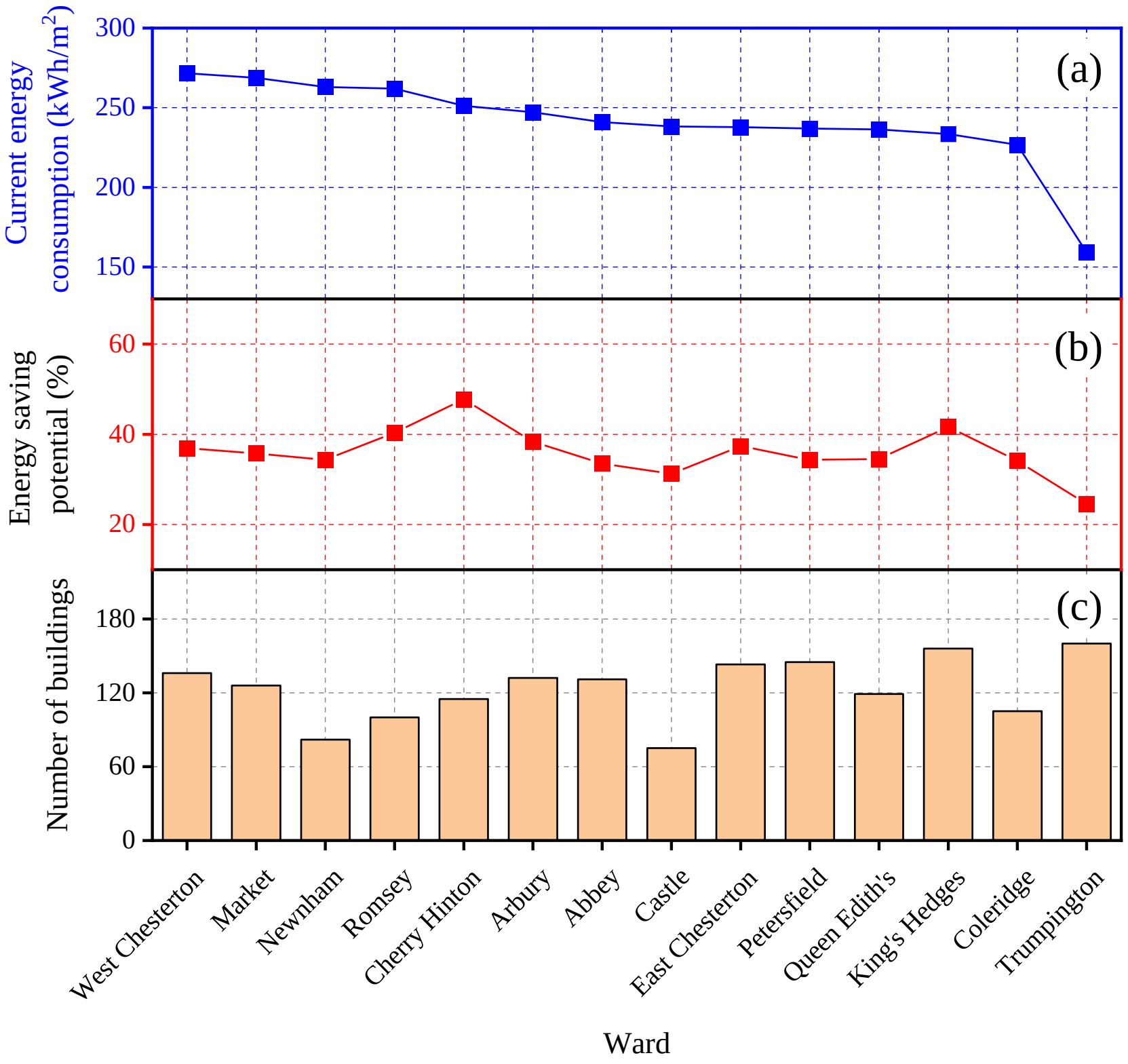

**Figure 7.** Current energy consumption (a), energy saving potential (b), and number of buildings (c) grouped by ward.

In Figure 7, the bar plot at the bottom shows the number of buildings in each ward, in which Trumpington ward has the greatest number of buildings (160); while the Castle ward has the least (75). The data suggest that each ward has enough samples to generate statistically significant findings. The two line plots at the top represent the average current energy consumption ($E_c$, kWh/m$^2$, blue) and the average energy saving potential ($\delta E_p$, %, red) grouped by ward. What stands out in these data is that Trumpington ward hits the minimum values of both line plots, where its $E_c$ and $\delta E_p$ are 159.2kWh/m$^2$ and 24.5%, respectively. A possible explanation for this might be that, in the database, almost 50% of buildings in this ward were built after 1980. Thermal regulations have changed over time, and as a result, the energy performance requirements in recent building code updates have become stricter.

According to the current energy consumption data, we found that buildings in four wards, specifically Market (268.8kWh/m$^2$), Newnham (263.1kWh/m$^2$), Romsey (261.9kWh/m$^2$), and West Chesterton (271.7kWh/m$^2$), exhibit higher average energy consumption compared to buildings in other wards, which have an average energy consumption of 228.84kWh/m$^2$. This discrepancy can be attributed to the construction year of the buildings in these wards, as they generally have a higher proportion of samples built before 1939, namely 71.4%, 58.5%, 65.0%, and 63.2%, respectively, in contrast to the average of 26.2% for other wards.

With regard to energy saving potential, Cherry Hinton ward exhibits the maximum value (47.8%). It may be explained that 85.2% of sample buildings in this ward are houses, which is significantly higher than the average level (69.8%). While houses generally possess more energy saving potential than flats, which will be further elaborated through subsequent data in Table 3.

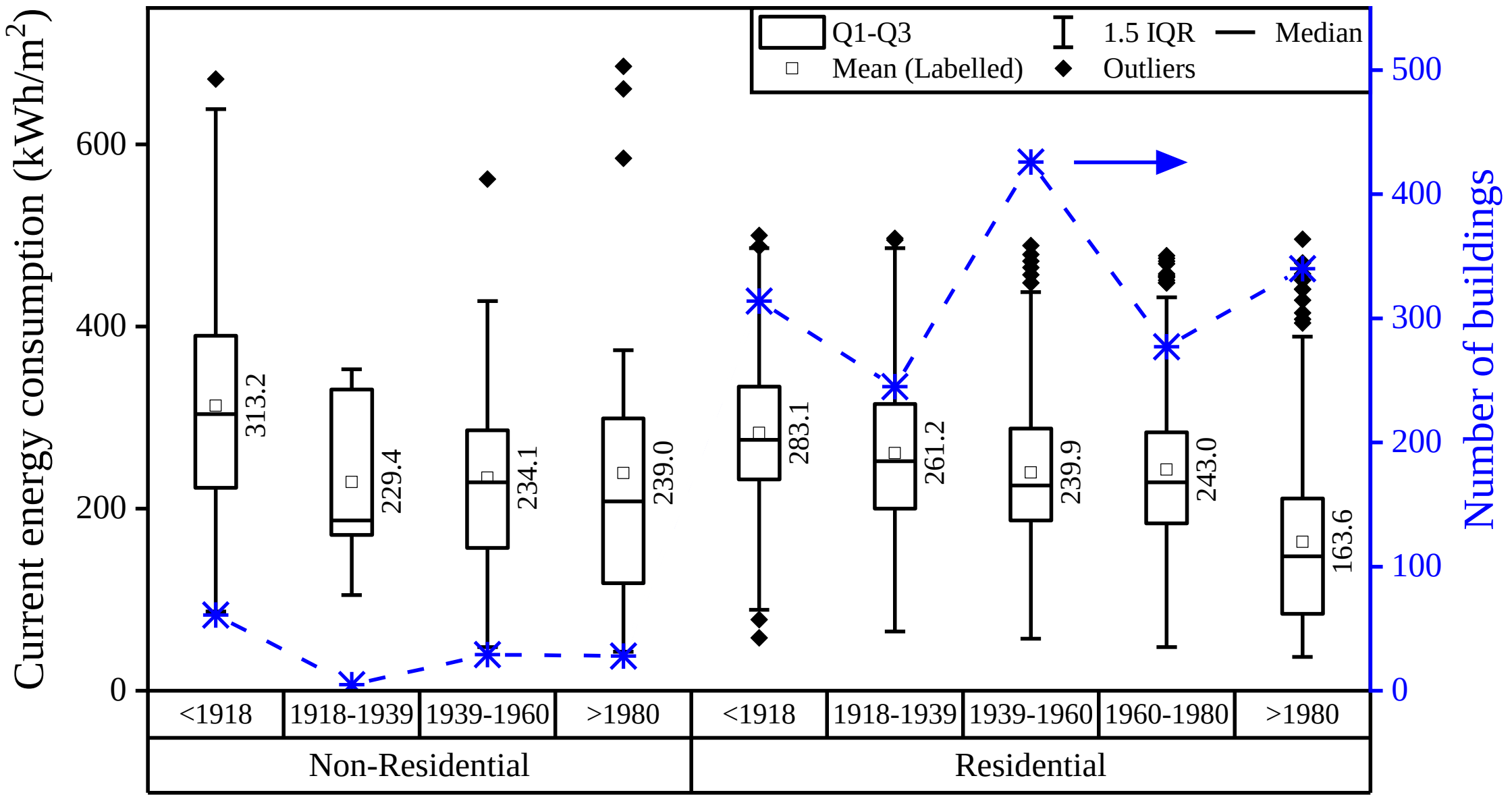

**Figure 8.** Energy consumption grouped by use type and construction year.

The average energy use, separated into use types and construction periods, is presented in Figure 8. In this box-and-whisker diagram, the box indicates the interquartile range (IQR) between the 1st quartile (Q1, 25%) and the 3rd quartile (Q3, 75%), i.e., IQR = Q3 – Q1, which are derived from *n* building samples in the database and *n* is the number of buildings as illustrated by the marked blue curve. The band and the square inside the box are the median (Q2, 50%) and the mean (with a data label next to the box), respectively. While diamonds beyond the whiskers denote outliers. The data of non-residential buildings lack a clear overall trend due to the limited sample size, but the ones built after 1980 (modern) generally consume less energy than those built before 1918 (historic), saving about 23.7% energy on average. In contrast, the residential building category has a much greater number of samples, and there is a clear tendency toward reduced energy consumption levels. Specifically, buildings erected after 1980 are about 42.2%, 37.4%, and 31.8% more energy efficient than those built before 1918, between 1918 and 1939, and between 1939-1960, respectively.

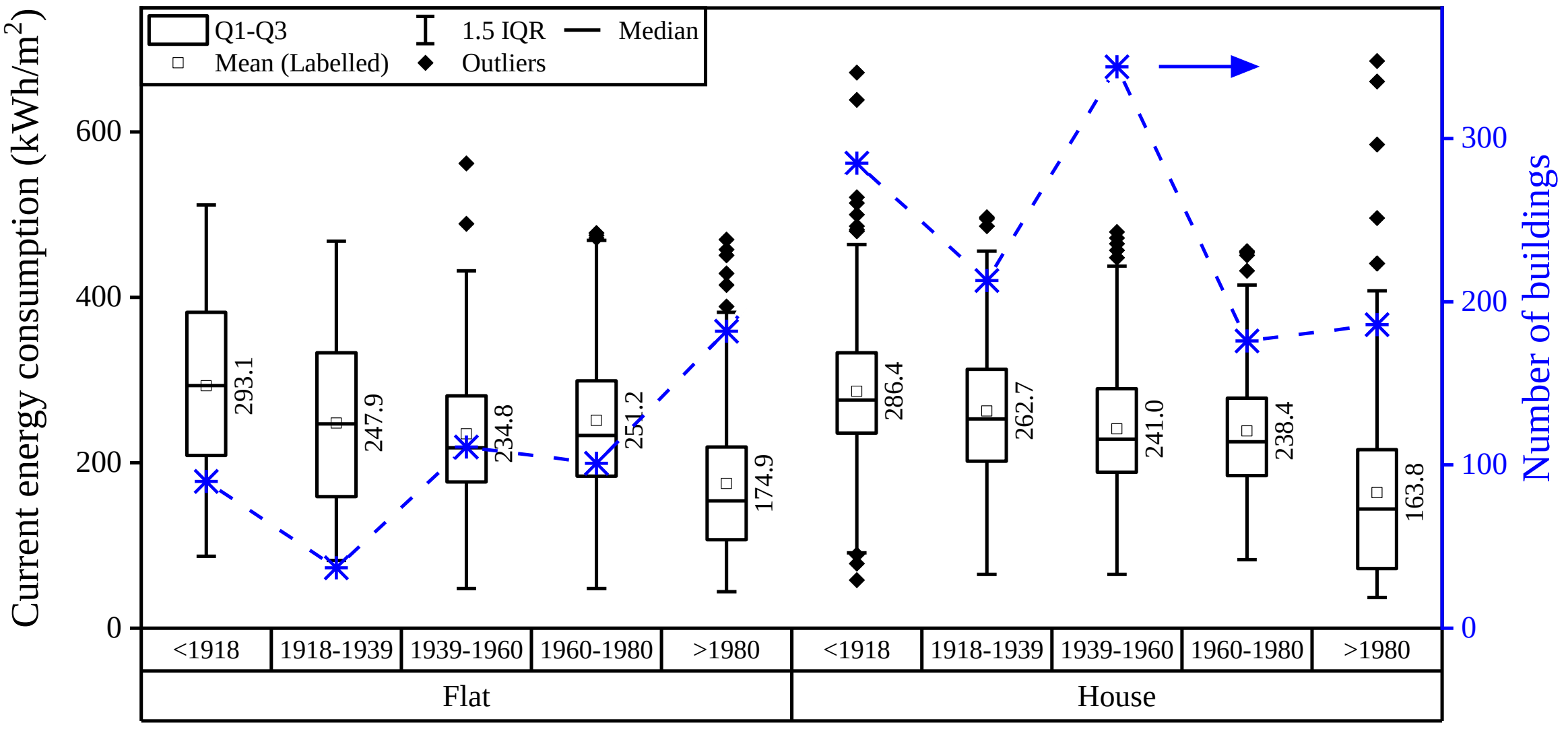

**Figure 9.** Energy consumption grouped by property type and construction year.

Figure 9 gives the energy consumption grouped by property types and construction periods. The results suggest that flats are generally more energy-intensive than houses, for example, regarding those built after 1980, flats and houses averagely consume energy of 174.9 and 163.8kWh/m$^2$, respectively. In addition, both flat and house samples demonstrate that there is a clear relationship between the amount of energy consumed and the year of construction. More specifically, the older the building is, the more energy it consumes.

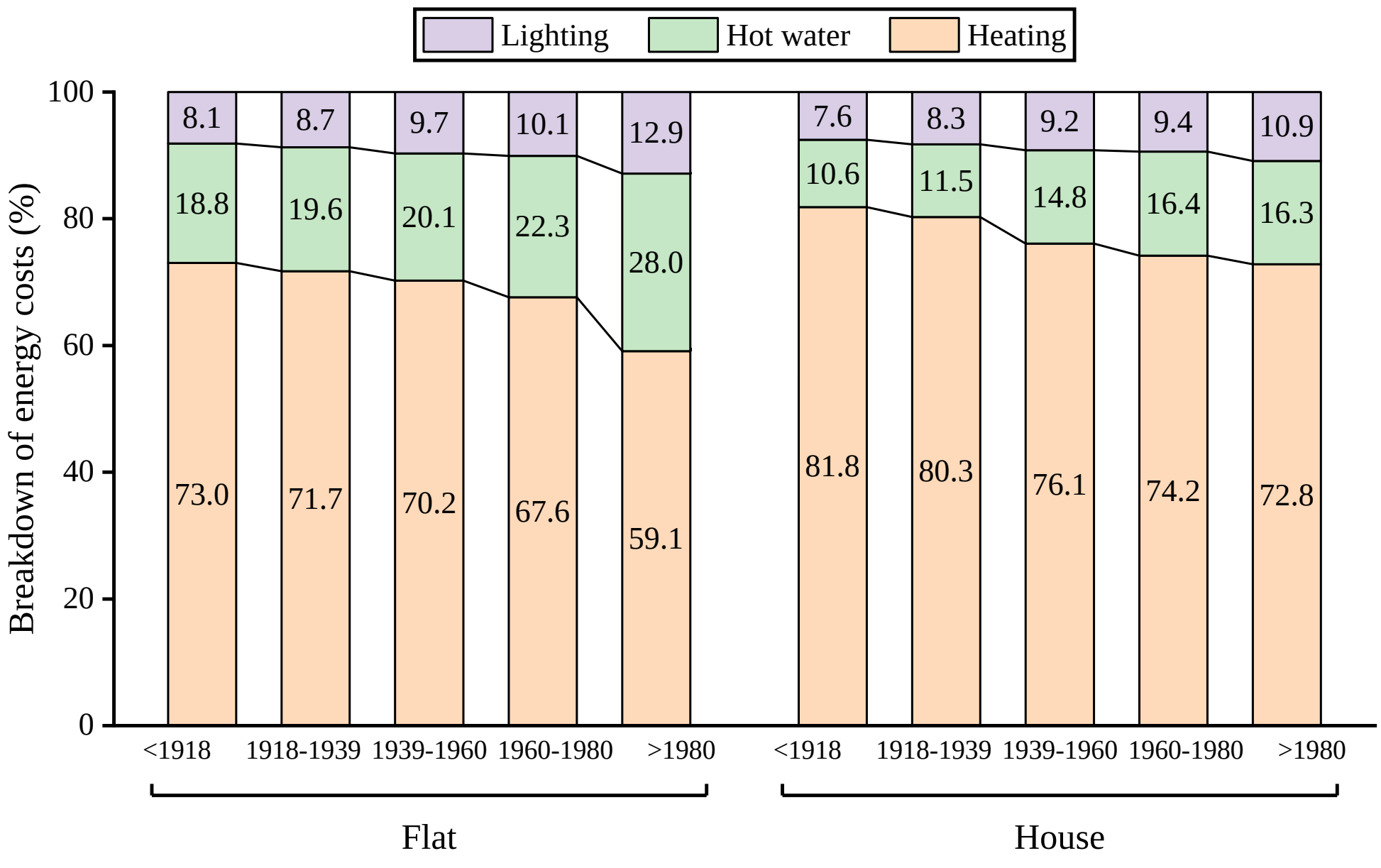

**Figure 10.** Breakdown of energy costs (or end-uses) grouped by property type and construction year.

Figure 10 provides a closer inspection of energy end-uses in both flats and houses built in different periods. The results reveal that, in flats, most of the energy is used for heating (59.1-73.0%). As expected, in houses, energy consumption associated with heating is even higher (72.8-81.8%). Moreover, the heating system in newer buildings uses less energy since thermal regulations have changed regularly over time. About 20% and 15% of the total energy consumed by flats and houses, respectively, are attributed to the hot water system. Lighting accounts for about 10% in both flats and houses.

Figure 11 illustrates the distribution of energy rating categorised by property type and construction year. In older buildings (erected before 1980), most of the houses show a poor energy performance profile. Particularly, D class is the most predominant energy label within house units (47.7-52.3%), followed by C class (11.6-41.5%) and E class (7.4-29.1%). While flat units have better energy performance in general, where the predominant label is C class (29.7-56.8%), followed by D class (26.1-47.8%). When it comes to modern buildings (built after 1980), the energy labels of both flats and houses are dominated by C and B classes (about 80% in total). On the whole, the percentages of different energy labels largely depend on the construction year and the property type.

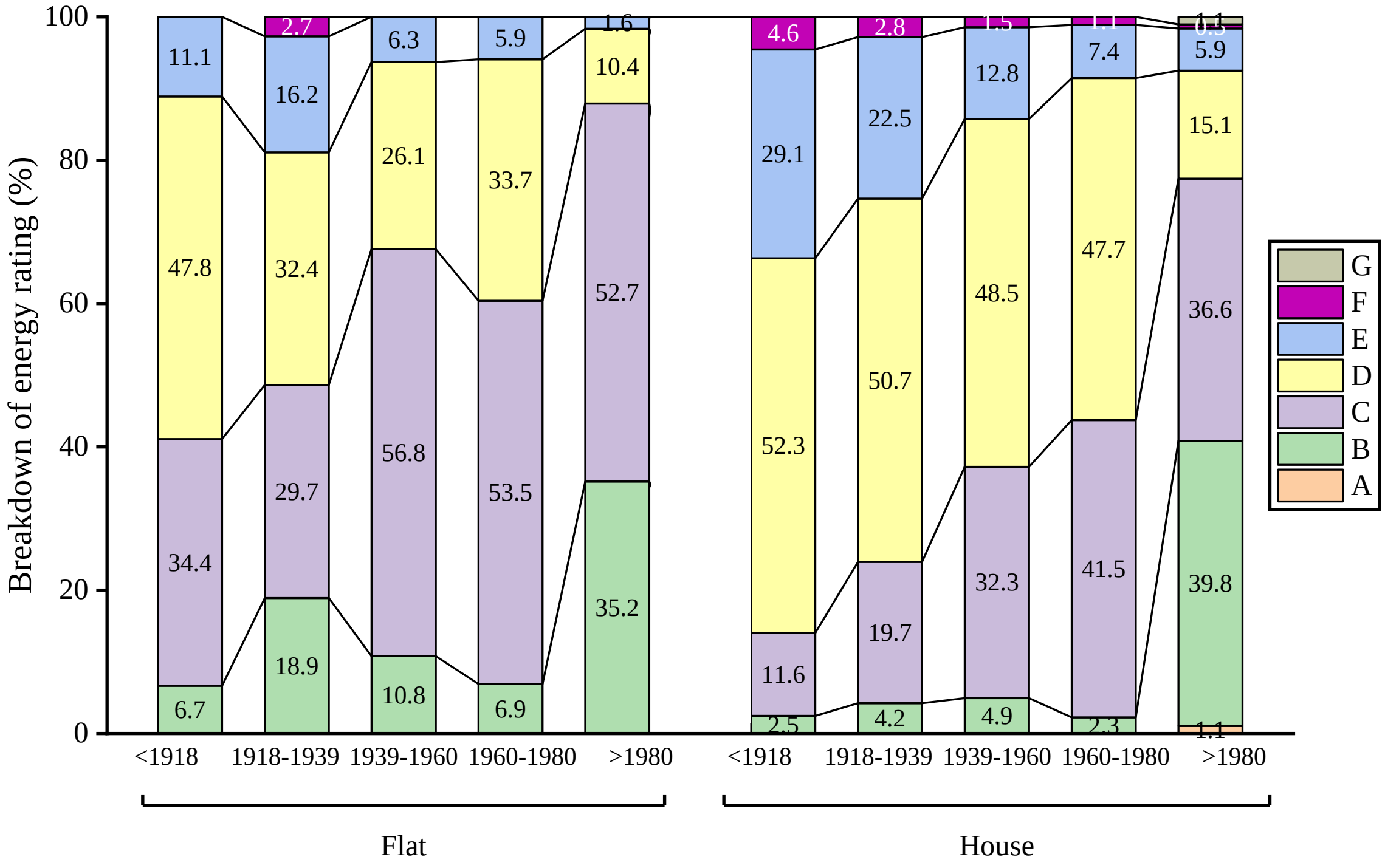

**Figure 11.** Breakdown of energy ratings grouped by property type and construction year.

Table 3 summarises the energy saving potential (mean ± std) of buildings grouped by construction year and use/property type. It is worth noting that the data of non-residential buildings in this table are not statistically significant due to the limited sample size as mentioned above. Despite relatively broad uncertainty ranges, some distinct trends are apparent. The results suggest that, compared with modern buildings (>1980), the older ones have the much greater potential to cut energy use by implementing efficiency measures. For example, in residential buildings, the energy saving potential of older ones (built before 1980) is about 40%, which is double the value of modern ones. In addition, the house units are more likely to have an outsized benefit from energy retrofit interventions. Specifically, through energy improvement measures, the older houses (built before 1980) can save almost half of the energy use, while the older flats can only reduce energy consumption by about 20% on average. Even for the modern houses erected after 1980, the amount of energy consumed may fall by 31.6%.

**Table 3.** Energy saving potential (mean ± std, %) of buildings

| | <1918 | 1918-1939 | 1939-1960 | 1960-1980 | >1980 |
|---|---|---|---|---|---|
| *Grouped by use type* | | | | | |
| Non-residential | 28.1 ± 25.9 | 42.4 ± 28.0 | 25.7 ± 24.8 | - | 18.7 ± 23.5 |
| Residential | 43.2 ± 26.2 | 45.9 ± 24.8 | 38.9 ± 24.4 | 39.3 ± 23.9 | 19.5 ± 22.9 |
| *Grouped by property type* | | | | | |
| Flat | 22.3 ± 20.8 | 21.2 ± 21.2 | 15.9 ± 14.3 | 21.5 ± 16.6 | 7.0 ± 11.3 |
| House | 46.6 ± 25.8 | 50.1 ± 22.8 | 45.2 ± 23.0 | 49.5 ± 21.3 | 31.6 ± 24.9 |

## 4.2 Effects of Resolution (GSD) on TIR Data of Buildings

This section investigates the effects of image resolution on the thermal mapping of buildings. Figure 12 (a) shows the drone TIR image of Wolfson College, with a GSD of 0.26m/pixel. The black-coloured polygons (i.e., WC-1 – 16) encapsulate the built-up area of the footprint of 16 different buildings. It is worth noting that, in this imagery, some buildings' data are partially affected by vegetation covers, e.g., WC-2 and WC-5. These affected areas are therefore excluded when drawing the buildings' polygons.

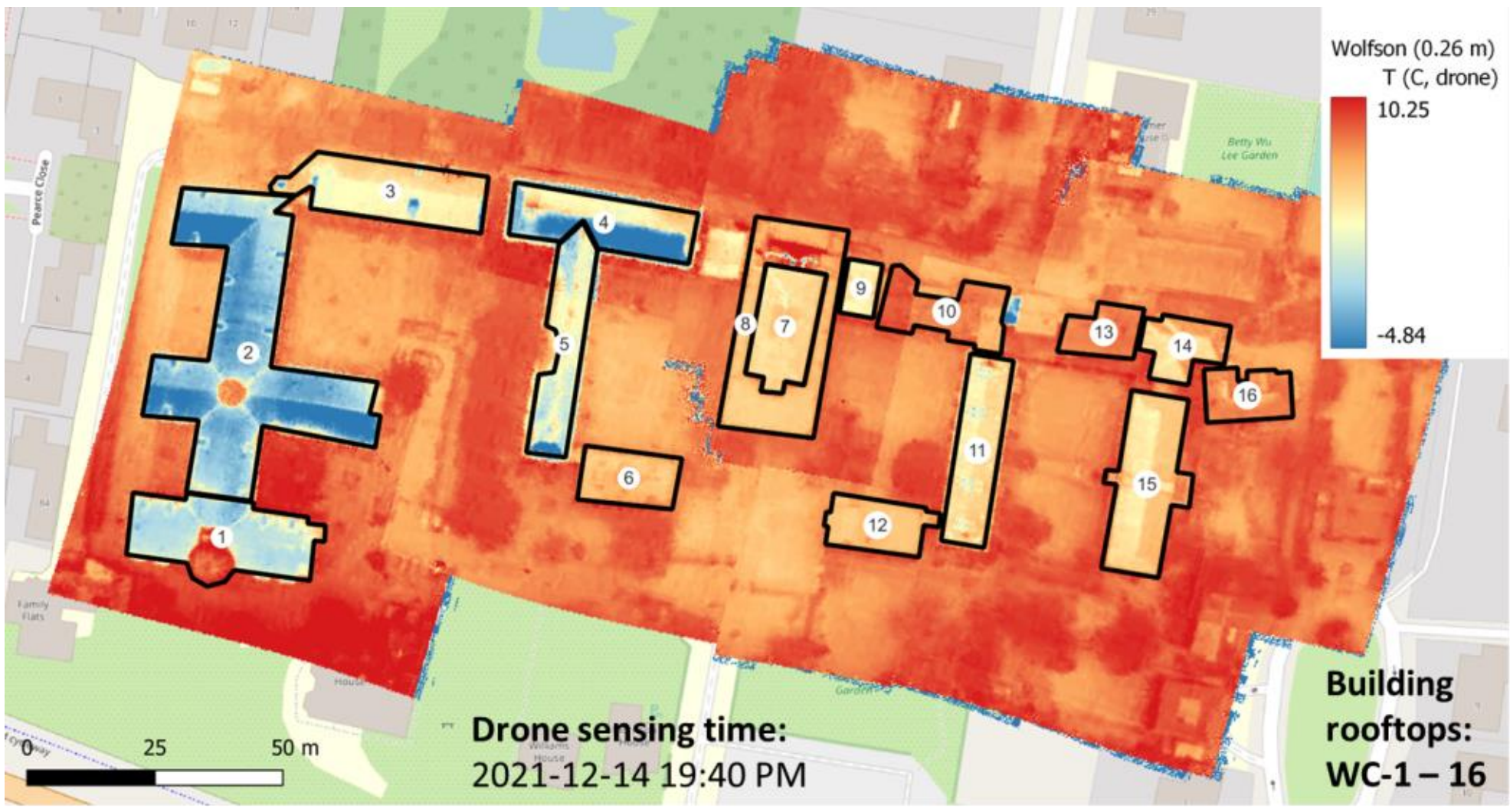

(a)

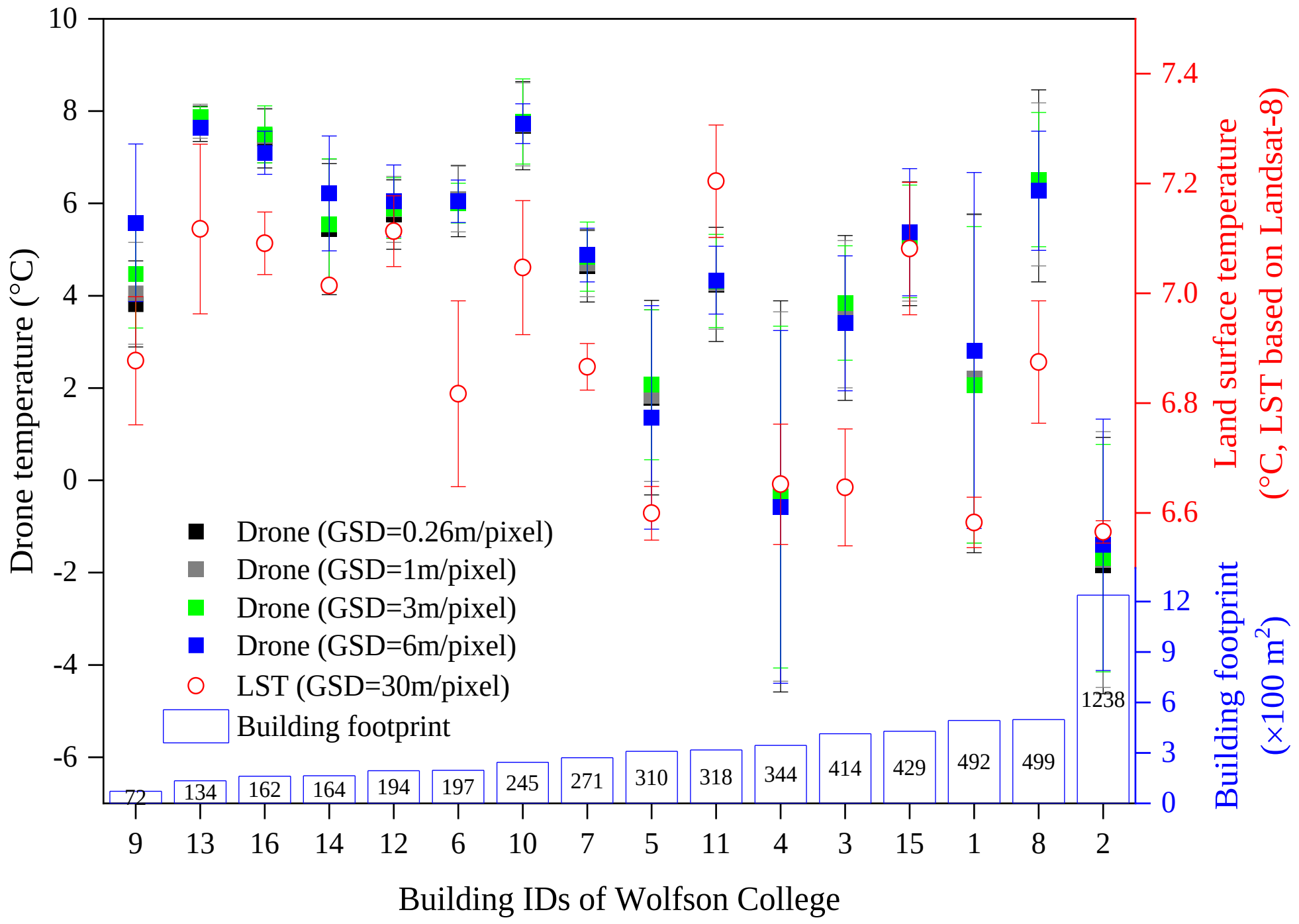

(b)

**Figure 12.** Drone thermal mapping of Wolfson College (a) and corresponding temperature results at different resolutions (b)

Figure 12 (b) compares the temperature and the footprint of 16 Wolfson College buildings. In particular, four solid lines with error bars (in different colours) show these buildings' temperature values (mean ± std) that are derived from the drone TIR imageries at four different resolutions (i.e., GSD = 0.26, 1, 3, and 6m/pixel). Meanwhile, the dashed line with the circle marker gives the land surface temperature (LST) values of the buildings. LST data is calculated based on the Landsat-8 database, and therefore, has a much lower resolution (GSD = 30m/pixel). In addition, at the bottom of this chart, the buildings' footprints are provided, ranging from 70 to 1300m$^2$, to reveal the impact of the building footprint on the analysis of buildings' thermal characteristics at different resolutions. The results suggest that downsampling is more likely to distort smaller buildings' results (e.g., ID = 9 with a footprint of 72m$^2$), which implies that the space telescope with a GSD of 6m/pixel is probably not reliable for monitoring residential buildings' temperature.

**Table 4.** Pearson correlation (ρ) of mean temperature values of Wolfson College buildings

| **Resolutions** | **0.26m/pixel** | **1m/pixel** | **3m/pixel** | **6m/pixel** |
|---|---|---|---|---|
| **30m/pixel** | 0.774 | 0.774 | 0.770 | 0.778 |

In addition, Table 4 calculates the Pearson correlation values between the LST data (GSD = 30m/pixel) and drone-emulated data (GSD = 0.26, 1, 3, and 6m/pixel). Although the temperature values of these two data sources lie in different ranges, see Figure 12 (b), possible reasons include different sensing times, distances, and devices, they show sensible correlations, in which their Pearson values range from 0.770 to 0.778.

Figure 13 (a) shows the TIR mapping data of Peterhouse, with a GSD of 0.26m/pixel. The black-coloured polygons (i.e., PH-1 – 15) encapsulate the built-up area of the footprint of 15 different buildings. Figure 13 (b) compares the building rooftop temperature values, including drone-derived TIR data and satellite-derived LST data. The footprints of these 15 buildings are given at the bottom. The drone TIR results at four different resolutions (GSD = 0.26, 1, 3, and 6m/pixel) further support the findings in Figure 12 (b), namely, downsampling is more likely to distort smaller buildings' results (e.g., ID = 2).

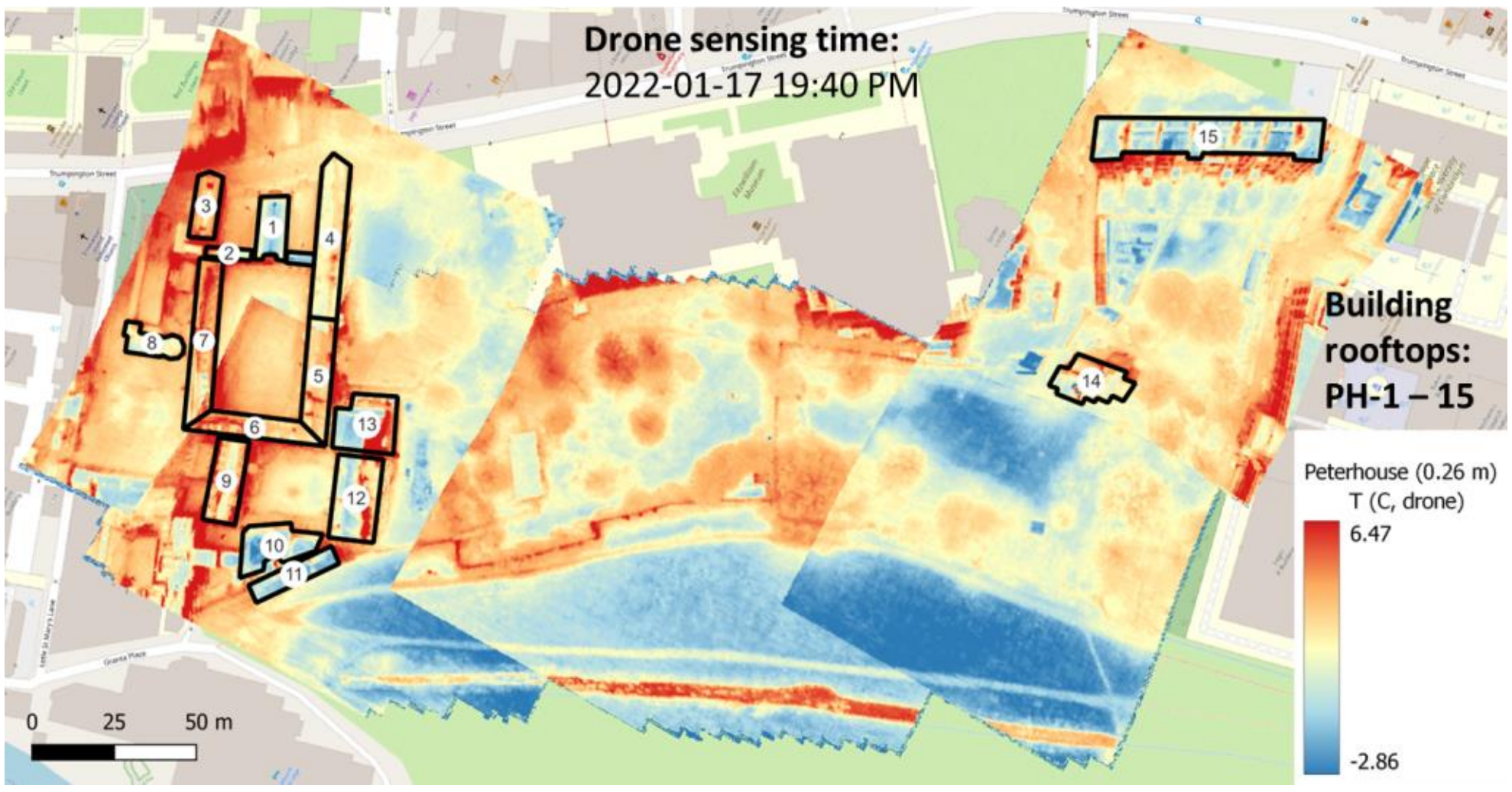

(a)

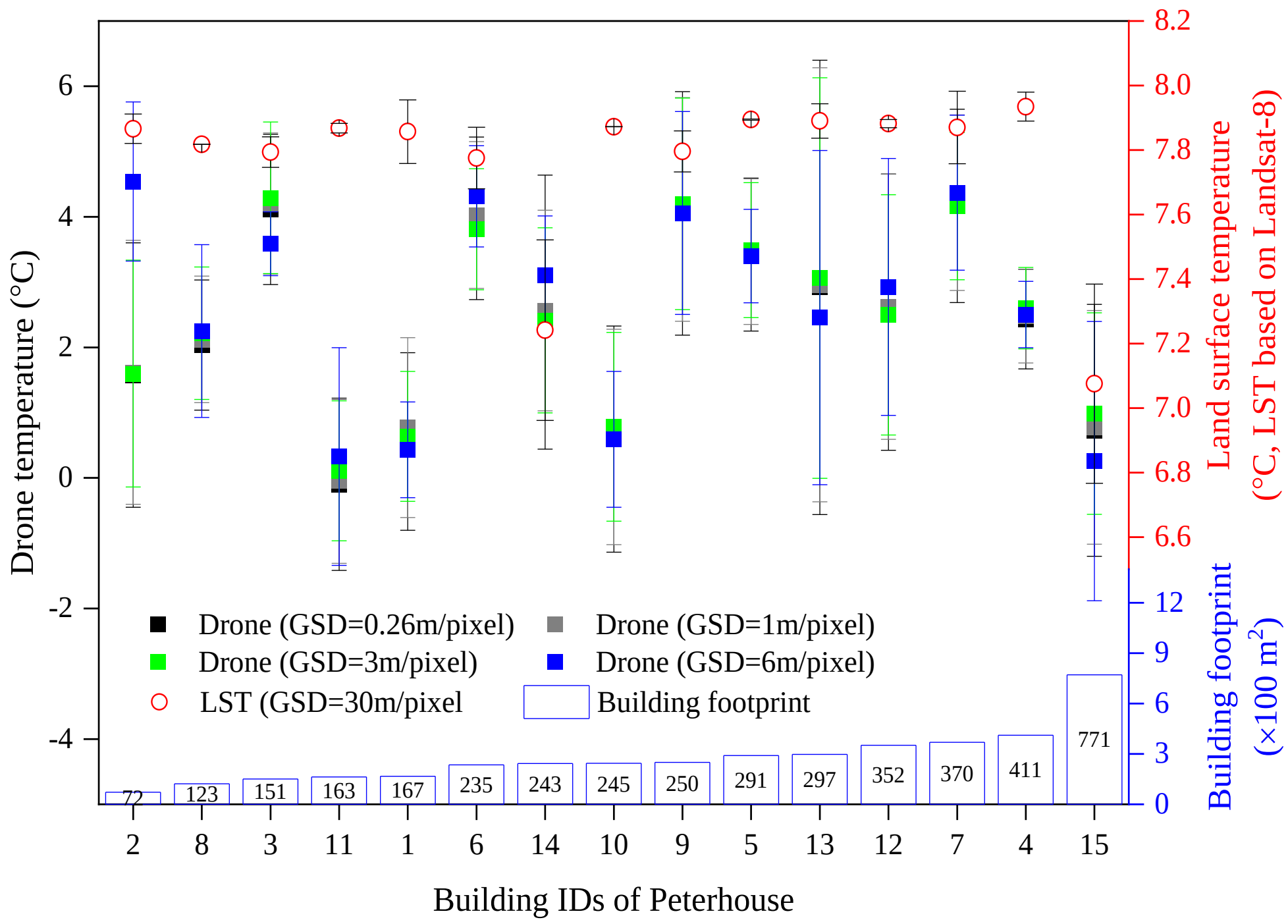

(b)

**Figure 13.** Drone thermal mapping of Peterhouse (a) and corresponding temperature results at different resolutions (b)

Table 5 calculates the Pearson correlation values between the LST data (GSD = 30m/pixel) and drone TIR data (GSD = 0.26, 1, 3, and 6m/pixel). However, the temperature values from these two data sources generally show trivial correlations, lying between 0.17 and 0.24, which may be caused by different sensing times, distances, and devices.

**Table 5.** Pearson correlation (ρ) of mean temperature values of Peterhouse buildings

| **Resolutions** | **0.26m/pixel** | **1m/pixel** | **3m/pixel** | **6m/pixel** |
| --- | --- | --- | --- | --- |
| **30m/pixel** | 0.173 | 0.181 | 0.181 | 0.232 |

## 4.3 Effects of Angle of View (AOV) on TIR Data of Buildings

In order to disclose the impacts of the angle of view (AOV) and the resolution (i.e., GSD) on thermal imagery of building facades, Table 6 compares regional temperature results of the north facade of St. Peter's Terrace of Peterhouse. In this table, from top to bottom, different graphs' AOVs decrease (i.e., extra large, large, medium, and small); while from left to right their GSDs increase (0.26, 1, 3, and 6m/pixel). In addition, the zonal statistics results are provided below each graph, giving statistics (e.g., mean value, count, and standard deviation [std]) on pixels of the thermal image that are within the selected polygons/zones. The results suggest that AOV has significant impacts. In particular, temperature statistical results are more sensitive to AOV than to GSD. Moreover, the use of fractions of pixels for the lowest resolution (GSD=6m/pixel) leads to significant variations in std.

**Table 6.** Thermal imagery at different angles of view (AOVs)

| **AOV** | **Zonal statistics** | **Resolutions (GSD, m/pixel)** | | | |
| --- | --- | --- | --- | --- | --- |
| | | **0.26** | **1** | **3** | **6** |
| Extra large (XL) | | 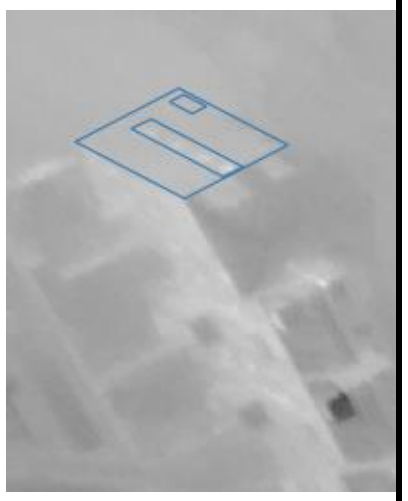 | 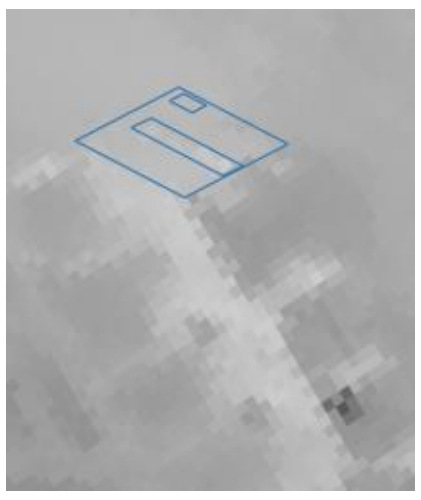 | 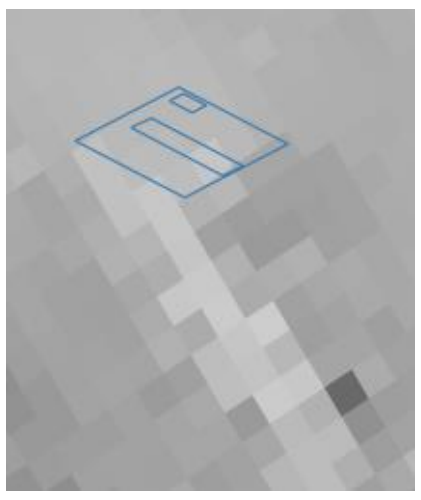 | 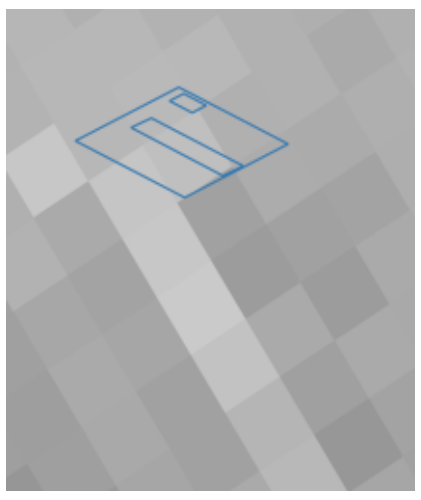 |
| | count | 1167 | 77 | 7 | 2.16 |
| | mean ± std | 9.05 ± 0.48°C | 9.07 ± 0.52°C | 9.17 ± 0.21°C | 8.96 ± 1.45°C |
| Large (L) | | 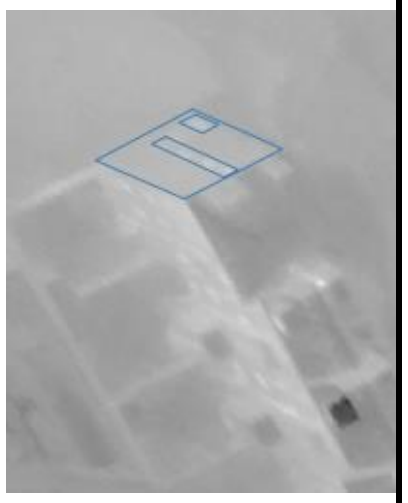 | 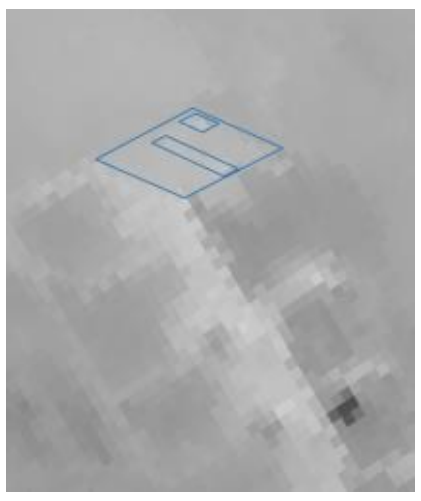 | 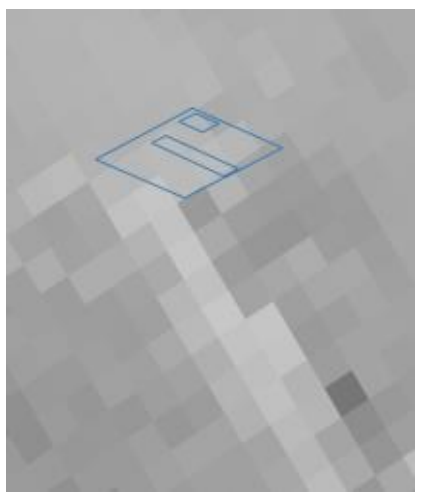 | 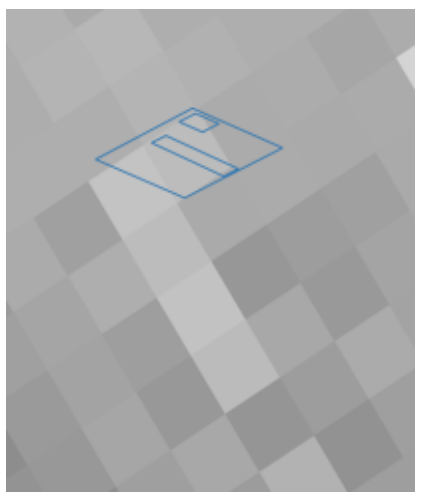 |
| | count | 1029 | 56 | 5 | 1.58 |
| | mean ± std | 8.69 ± 0.47°C | 8.71 ± 0.49°C | 8.91 ± 0.22°C | 8.75 ± 1.31°C |
| Medium (M) | | 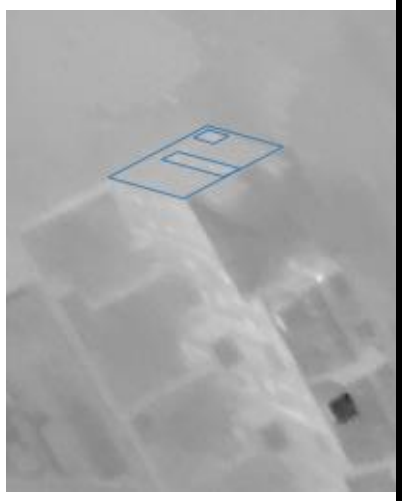 | 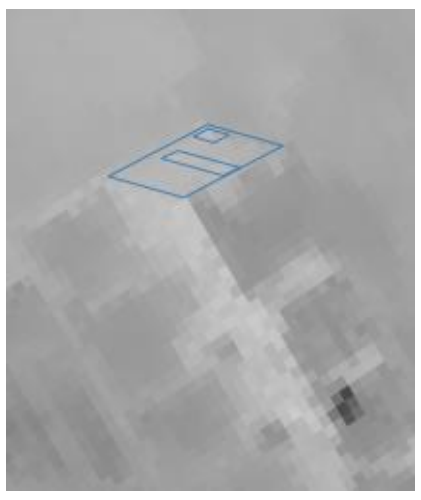 | 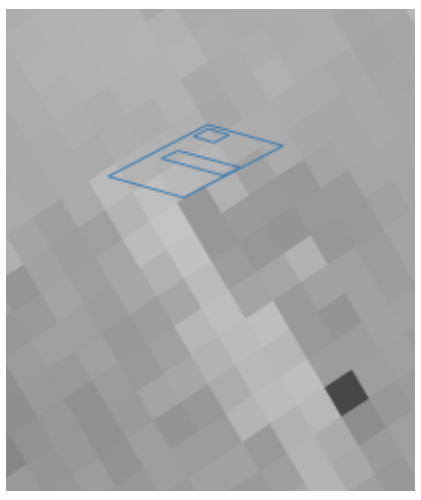 | 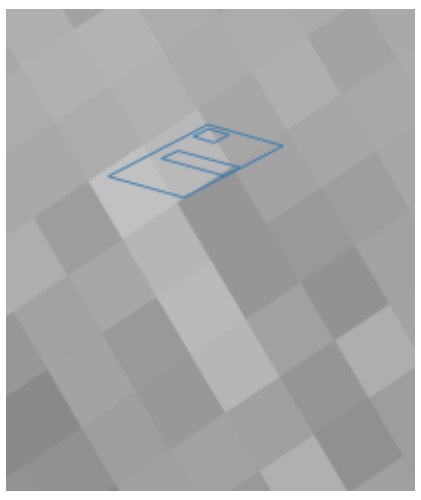 |
| | count | 870 | 42 | 4 | 1.13 |
| | mean ± std | 8.35 ± 0.51°C | 8.22 ± 0.59°C | 8.15 ± 0.48°C | 8.49 ± 2.36°C |
| Small (S) | | 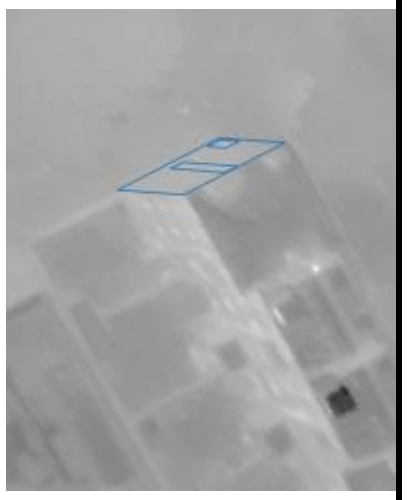 | 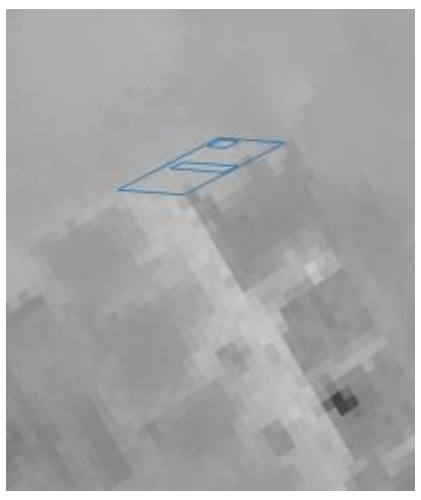 | 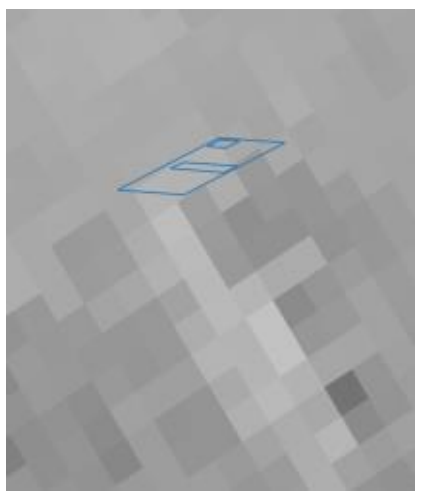 | 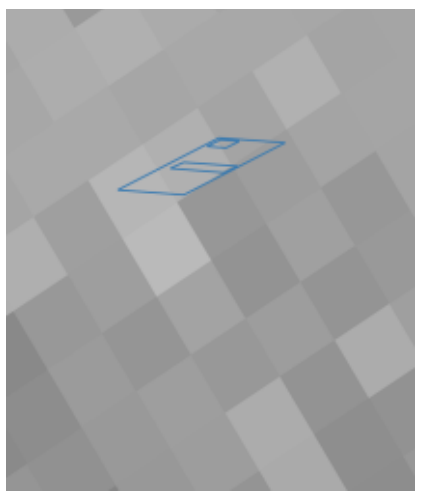 |
| | count | 677 | 29 | 2 | 0.77 |
| | mean ± std | 7.88 ± 0.54°C | 7.78 ± 0.82°C | 8.16 ± 0.06°C | 7.40 ± 1.74°C |

## 4.4 Temperature Clustering of Building Rooftops

Table 7 gives the pixel-based temperature clustering results of Wolfson College (WC) and Peterhouse (PH) at four different resolutions (GSD = 0.26, 1, 3, and 6m/pixel), based on the drone-emulated data and the Gaussian Mixture Model (GMM) clustering method. In addition, more detailed temperature clustering results regarding individual building rooftops are given in Table A1 (in Appendix A). Clusters are displayed in different colours, and the pixels of the same colour are in the same cluster. It is obvious that thermal anomalies can only be properly identified through images with GSD <= 1m/pixel. On the other hand, images with 1m/pixel < GSD <= 6m/pixel can be used to detect hot areas of building surfaces, instead of more detailed thermal features (e.g., thermal anomalies or bridges). For example, in Table A1, the case WC-8 highlights a thermal anomaly (in red) in the region corresponding to the rooftop of the kitchen of the Clubroom in Wolfson College. When GSD = 3 or 6m/pixel, this region is represented by only several or even one pixel(s), making it statistically non-significant for detecting thermal anomalies in practice.

**Table 7.** GMM clustering of rooftops TIR data of Wolfson College and Peterhouse

| | Wolfson College (WC) | | Peterhouse (PH) | |
|---|---|---|---|---|
| GSD | GMM clustering | GMM components | GMM clustering | GMM components |
| 0.26m /pixel | 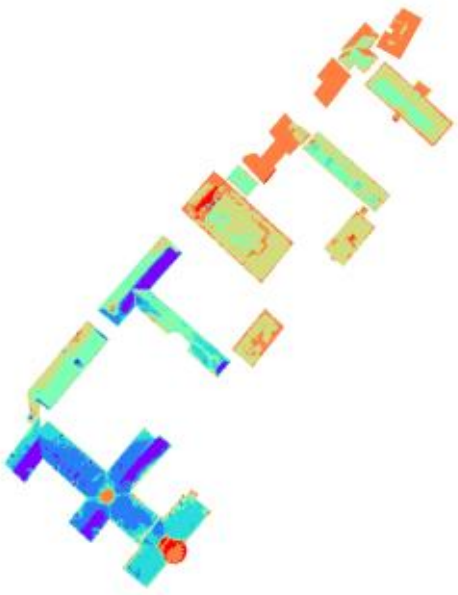 | 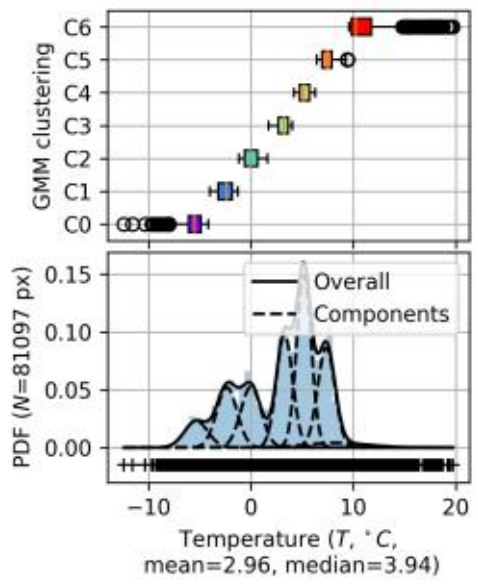  | 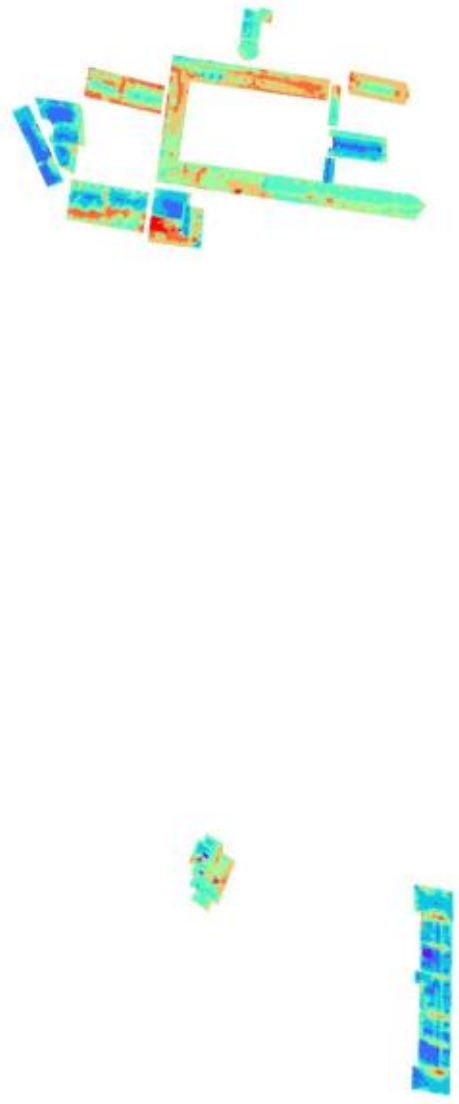 | 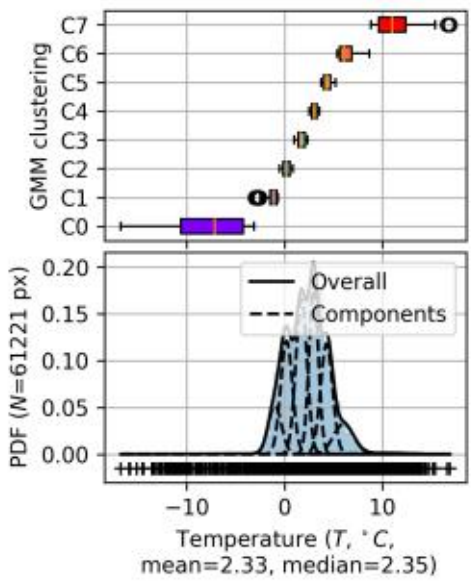  |
| 1m/pi xel | 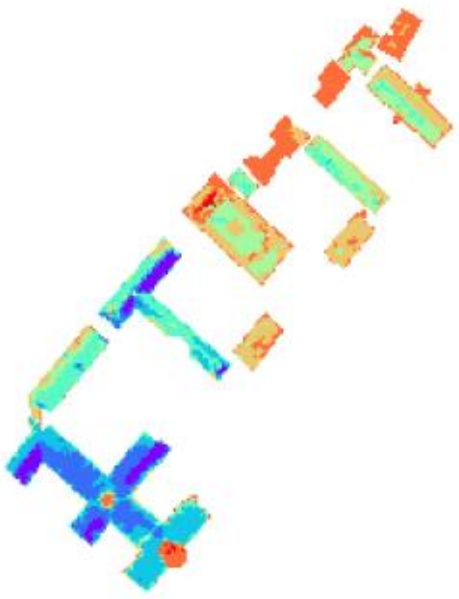 | 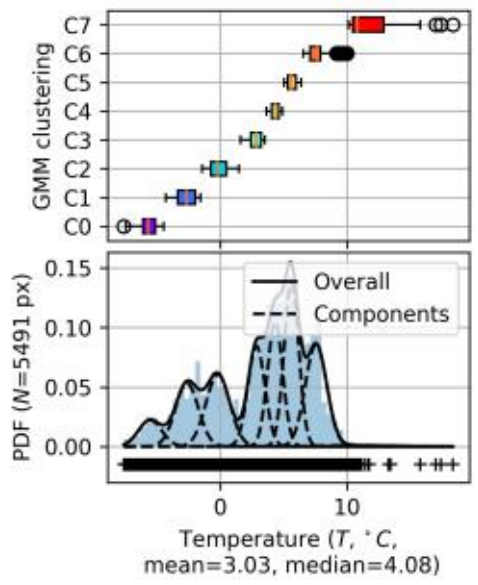  | 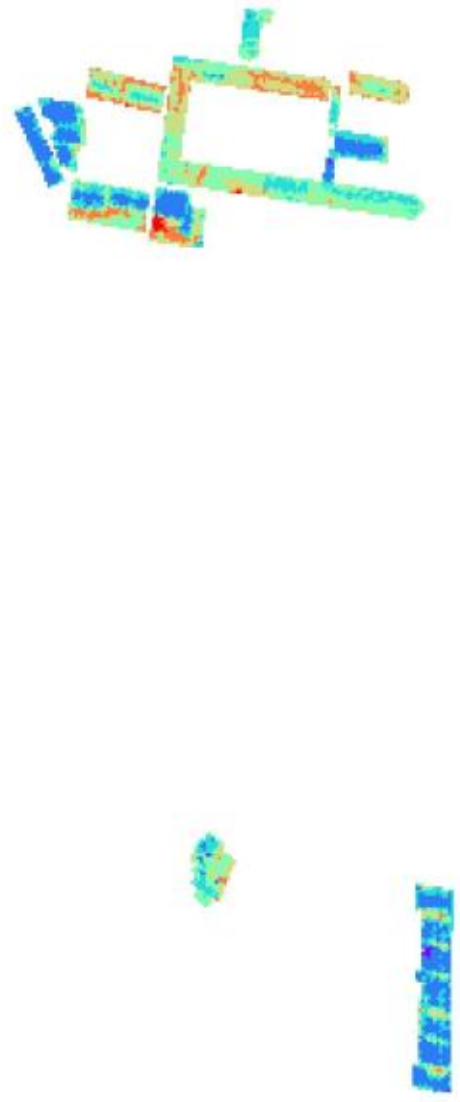 | 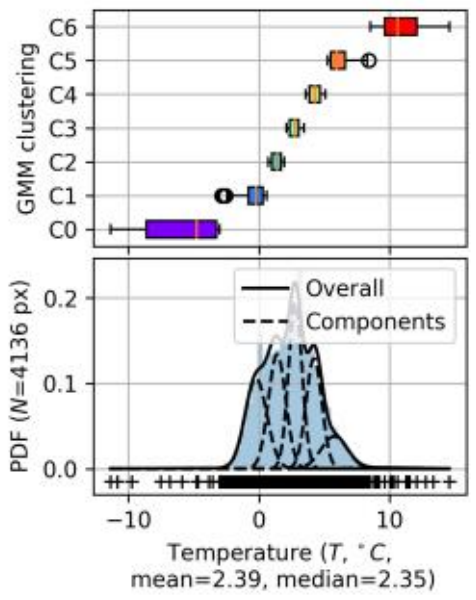  |

<table>
<tr>
<td>3m/pixel</td>
<td>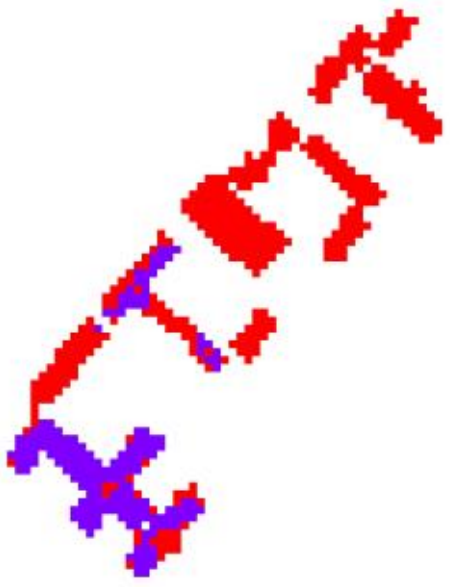</td>
<td>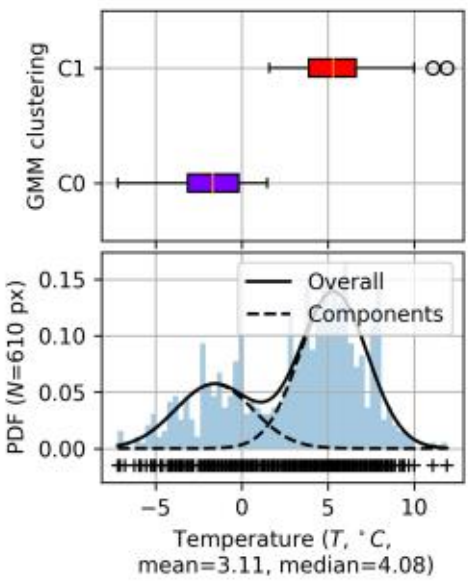<br></td>
<td>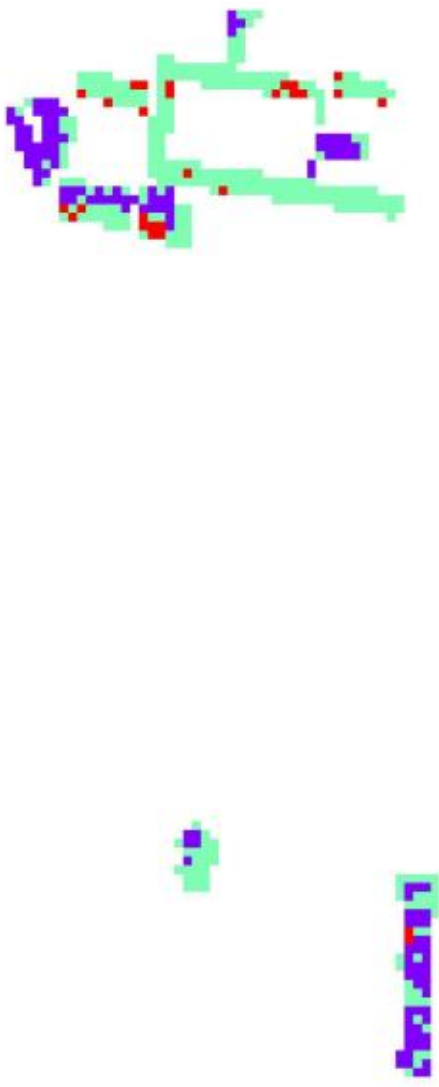</td>
<td>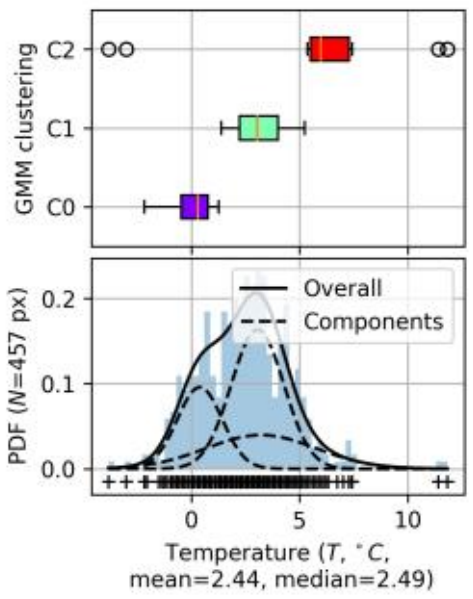<br></td>
</tr>
<tr>
<td>6m/pixel</td>
<td>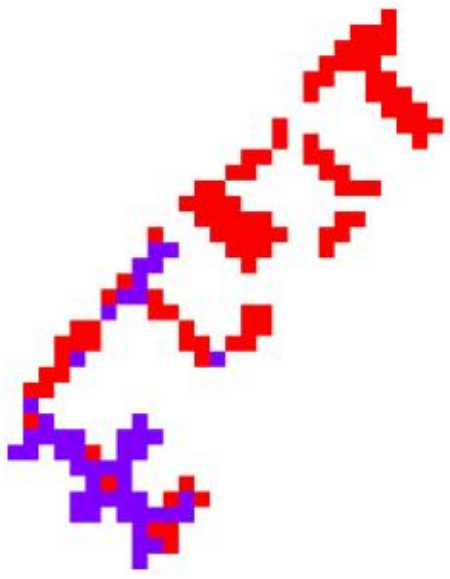</td>
<td>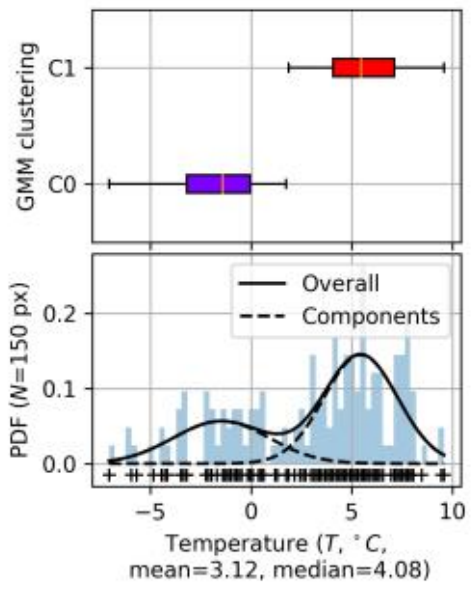<br></td>
<td>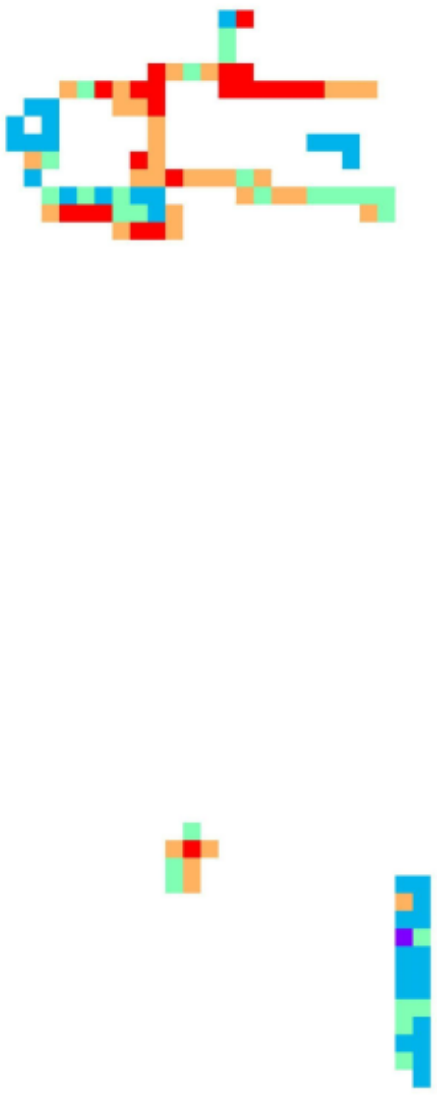</td>
<td>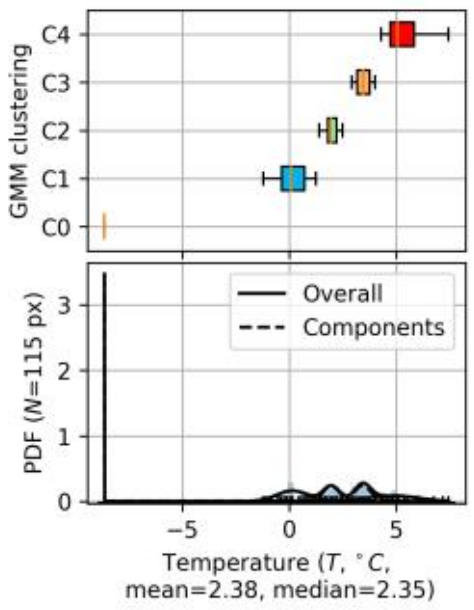<br></td>
</tr>
<tr>
<td colspan="5">* More detailed temperature clustering results regarding individual building rooftops are given in Table A1 in Appendix A.</td>
</tr>
</table>

# 5. Conclusions

Making existing inefficient buildings more efficient and sustainable is crucial to maximising the building sector's potential for energy savings because buildings typically have long lifespans and there is currently low turnover in the sector. To plan and materialise energy retrofit interventions in the existing building stock, we must first understand the energy consumption and the energy saving potential of this stock according to property type, construction period, and specific end-uses, and then have reliable approaches to monitor energy outputs and identify thermal defects of buildings. In this context, energy performance certificate (EPC)-based building archetypes and thermal infrared (TIR)-based building diagnostic techniques are key policy and technology tools to enhance the total energy efficiency of the built environment.

In this study, we first developed the EPC-based building archetypes that aggregated different geometric or non-geometric data sources. Secondly, the TIR images of building rooftops were collected by drones or retrieved from satellite imagery. Finally, we presented an insight into the energy performance of buildings in Cambridge, through the analysis of the 1725 EPCs of existing buildings sampled across the whole city. In addition, we explored machine learning (ML)-assisted thermal anomaly detection based on TIR images of building rooftops in Wolfson College and Peterhouse at the University of Cambridge, and investigated the influencing factors in imaging such as resolution (i.e., ground sample distance, GSD) and angle of view (AOV).

The EPC results suggest that building energy consumption (kWh/m$^2$) is largely affected by the year of construction because modern buildings have to meet higher energy performance requirements stated in the updated thermal building regulations. Moreover, in modern buildings (built after 1980), flats (174.9kWh/m$^2$) are on average more energy-intensive than houses (163.8kWh/m$^2$). Regarding energy end-uses, the energy consumption attributed to heating (over 60%) is much higher than those to hot water (about 20%) and lighting (about 10%). On the other hand, energy saving potential (after specific energy retrofit interventions have been carried out) largely depends on both the construction year and the property type. More specifically, older buildings show greater energy saving potential. Houses are more likely to have an outsized benefit (almost double) than flats from energy retrofits.

For TIR results, we found that thermal anomalies can only be properly identified in thermal images with a GSD of 1m/pixel or less. Images with a GSD of between 1 and 6m/pixel can be used to detect hot areas of building surfaces, instead of more detailed thermal features such as anomalies. TIR images with GSD of more than 6m/pixel (e.g., Landsat), although not accurate enough for characterising individual buildings, can still be used to disclose heat island effects in urbanised areas. Moreover, the AOV has significant impacts on the data. In particular, building temperature results are more sensitive to AOV than to GSD.

In summary, this research provides a full picture of the energy consumption of existing buildings in Cambridge and explores thermography techniques for thermal anomaly detection in building envelopes. The results can help to set priorities for energy retrofitting initiatives and enhance building diagnostic techniques. They could also support construction professionals in modernising their business plans and technological tools, as well as public authorities in developing future energy policies and regulations.

# Acknowledgements

The authors are grateful to colleagues at the University of Cambridge, in particular Department of Architecture (DoA), Institute of Astronomy (IoA), and Cambridge Zero, for their help, discussion and contribution to this work. RD acknowledges funding from the Quadrature Climate Foundation and Keynes Fund [JHVH].

# Competing Interests

The authors declare no competing interests exist.

# Data Availability Statement

Data are synthetic, experiments are reproducible through the GitHub repository: https://github.com/yinglonghe/gee-maps-cam.

# Ethics Statement

The research meets all ethical guidelines, including adherence to the legal requirements of the study country.

# Funding Statement

This research was supported by UK Space Agency (LGAG/360 G113185).

# Appendix A

**Table A1.** GMM clustering of TIR data of individual building rooftops in Wolfson College (WC) and Peterhouse (PH)

<table>
<tr><th rowspan="2">Optical image</th><th colspan="4">Resolutions (GSD, m/pixel)</th></tr>
<tr><th>0.26</th><th>1</th><th>3</th><th>6</th></tr>
<tr>
<td>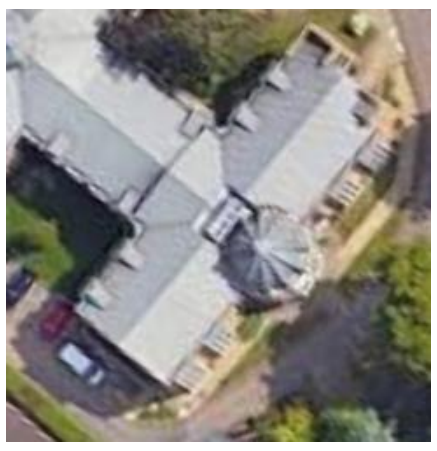<br>WC-1</td>
<td>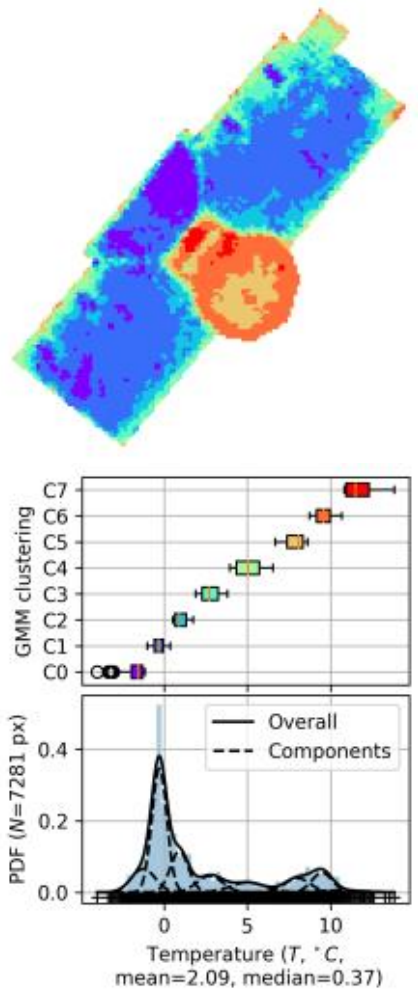
</td>
<td>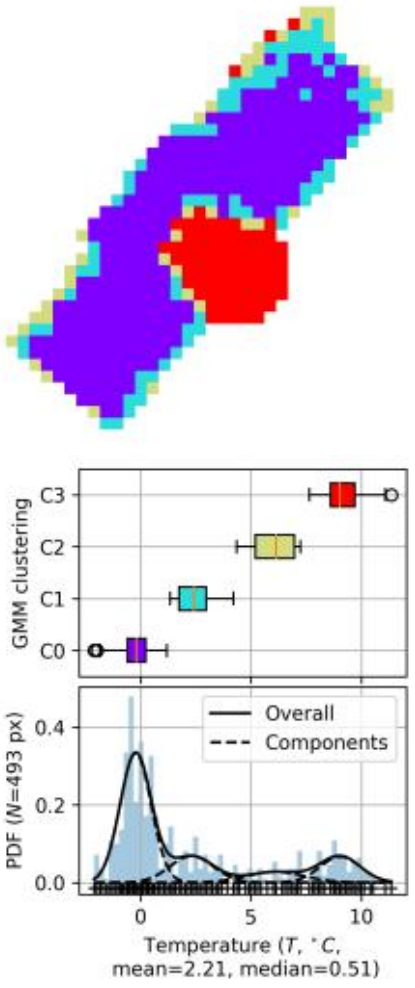
</td>
<td>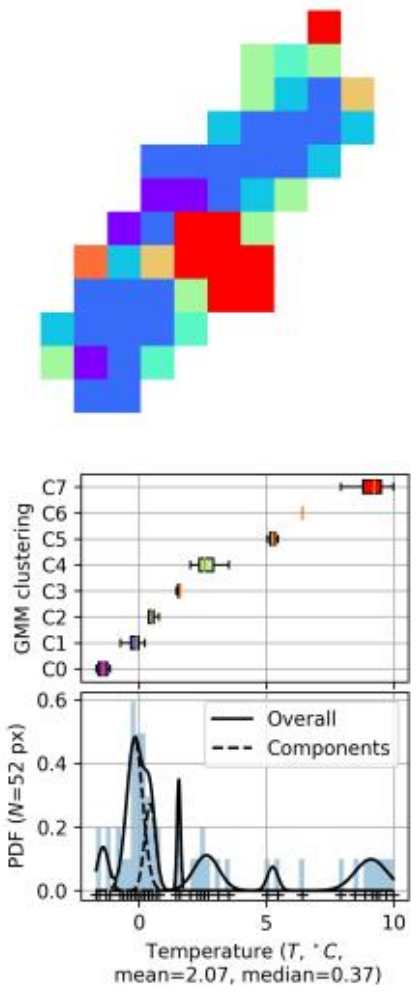
</td>
<td>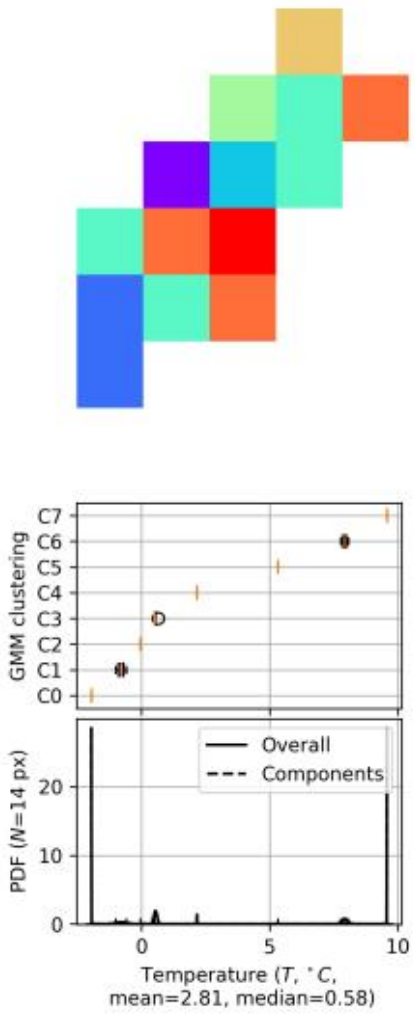
</td>
</tr>
<tr>
<td>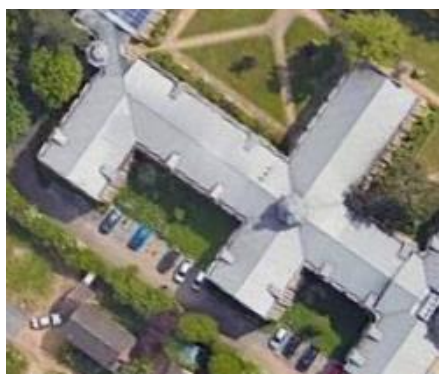<br>WC-2</td>
<td>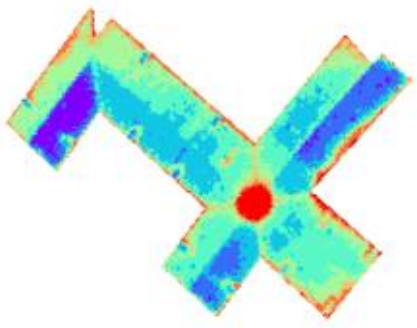</td>
<td>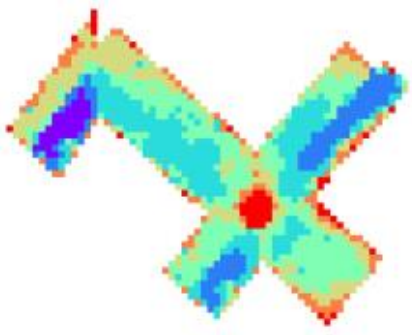</td>
<td>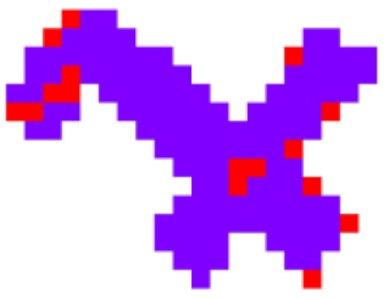</td>
<td>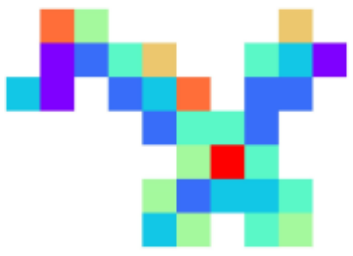</td>
</tr>
</table>

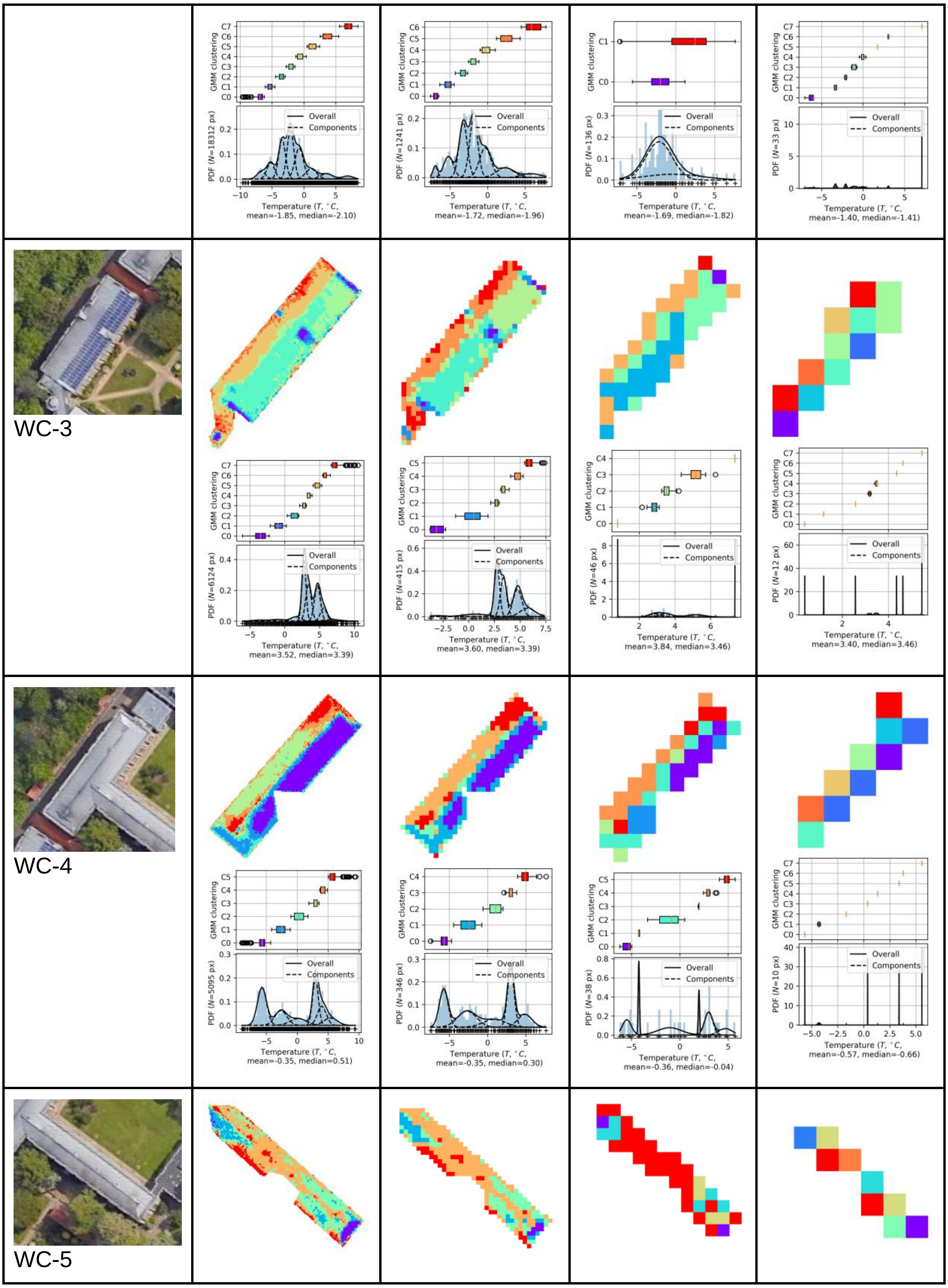
WC-3
WC-4
WC-5
GMM clustering
Overall
Components
PDF (N=18312 px)
Temperature (T, °C, mean=-1.85, median=-2.10)
PDF (N=1241 px)
Temperature (T, °C, mean=-1.72, median=-1.96)
PDF (N=136 px)
Temperature (T, °C, mean=-1.69, median=-1.82)
PDF (N=33 px)
Temperature (T, °C, mean=-1.40, median=-1.41)
PDF (N=6124 px)
Temperature (T, °C, mean=3.52, median=3.39)
PDF (N=415 px)
Temperature (T, °C, mean=3.60, median=3.39)
PDF (N=46 px)
Temperature (T, °C, mean=3.84, median=3.46)
PDF (N=12 px)
Temperature (T, °C, mean=3.40, median=3.46)
PDF (N=5095 px)
Temperature (T, °C, mean=-0.35, median=0.51)
PDF (N=346 px)
Temperature (T, °C, mean=-0.35, median=0.30)
PDF (N=38 px)
Temperature (T, °C, mean=-0.36, median=-0.04)
PDF (N=10 px)
Temperature (T, °C, mean=-0.57, median=-0.66)

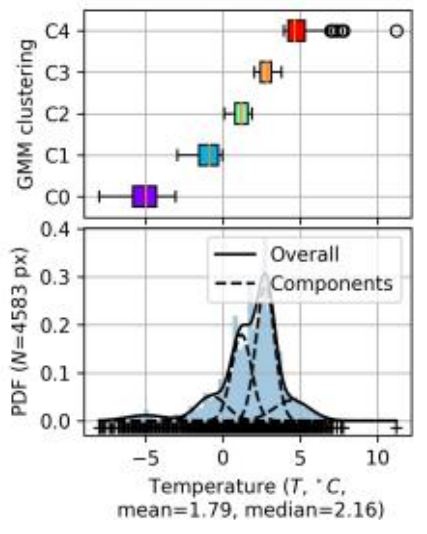

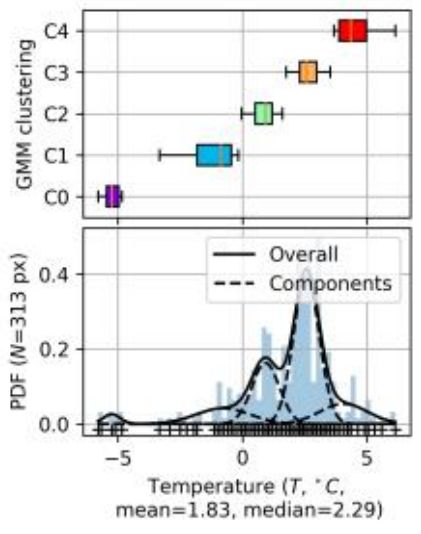

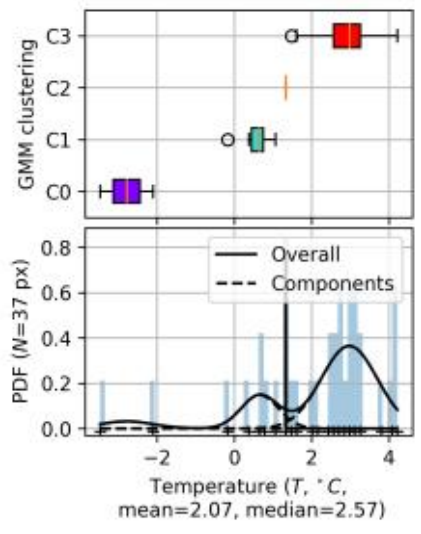

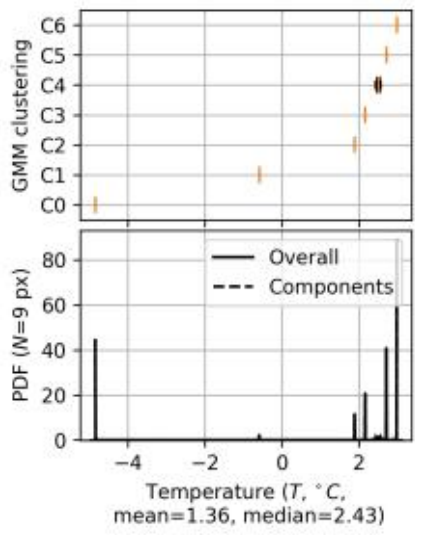

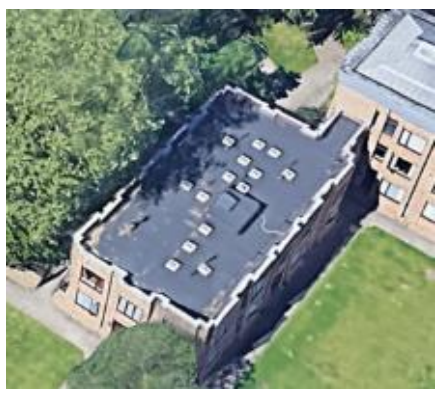

WC-6

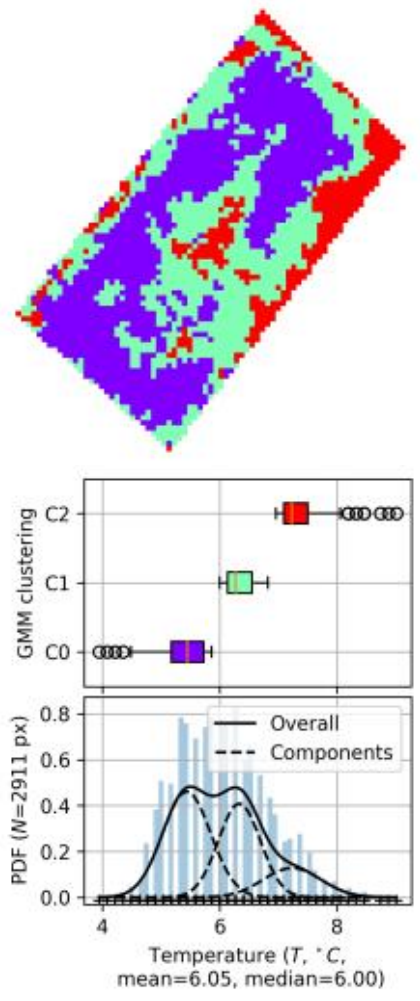

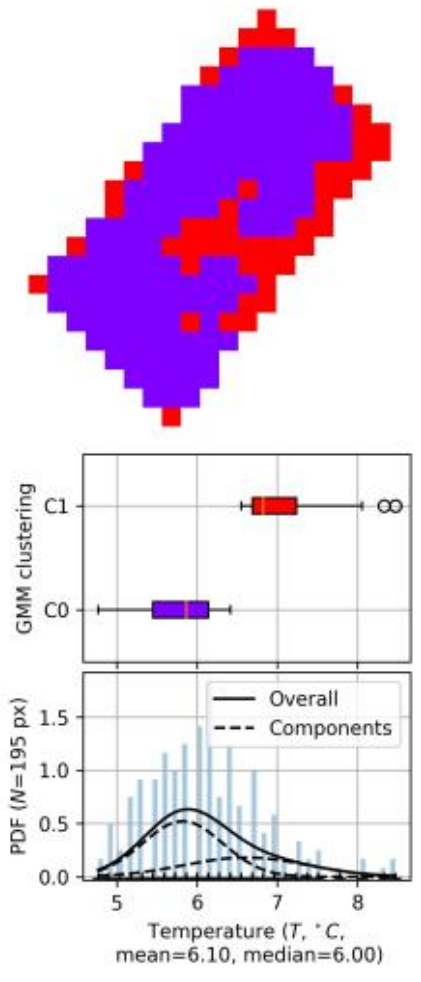

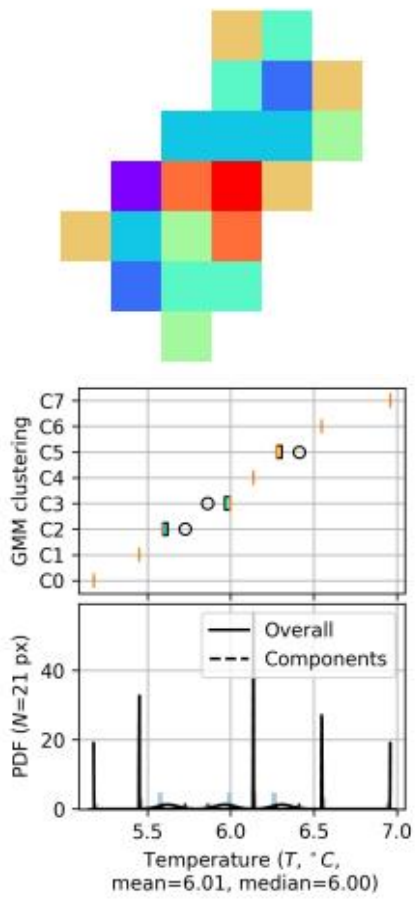

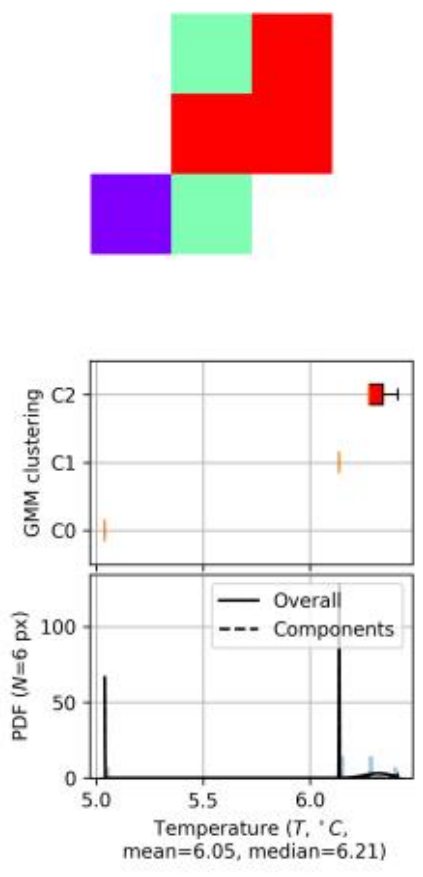

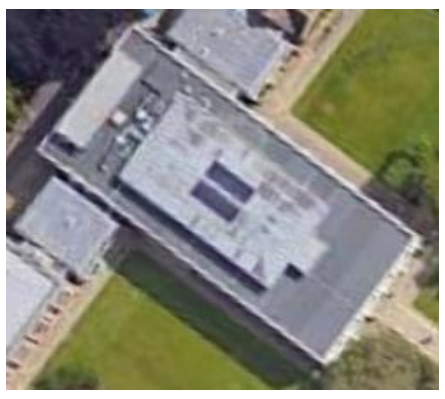

WC-7

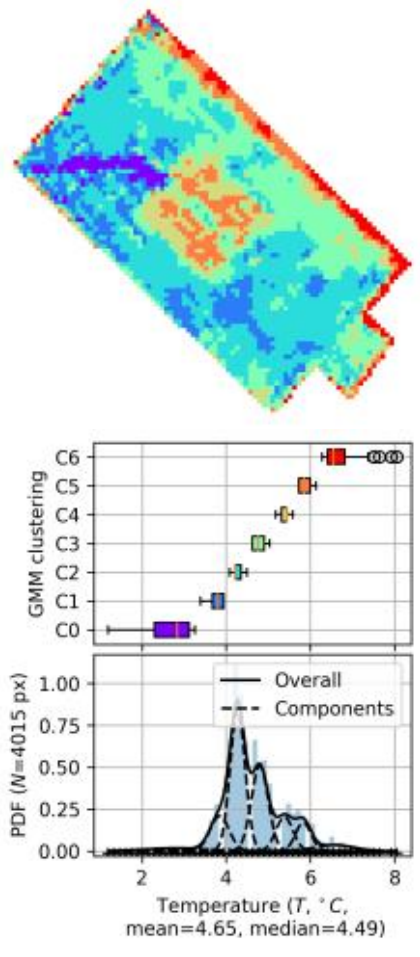

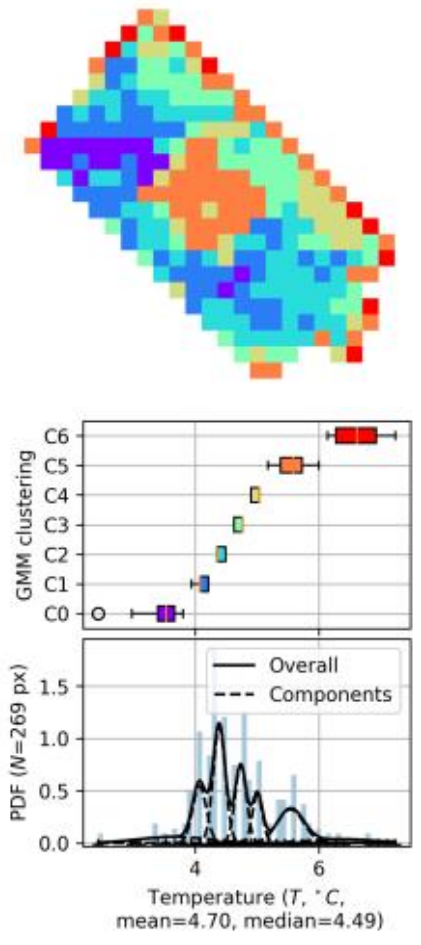

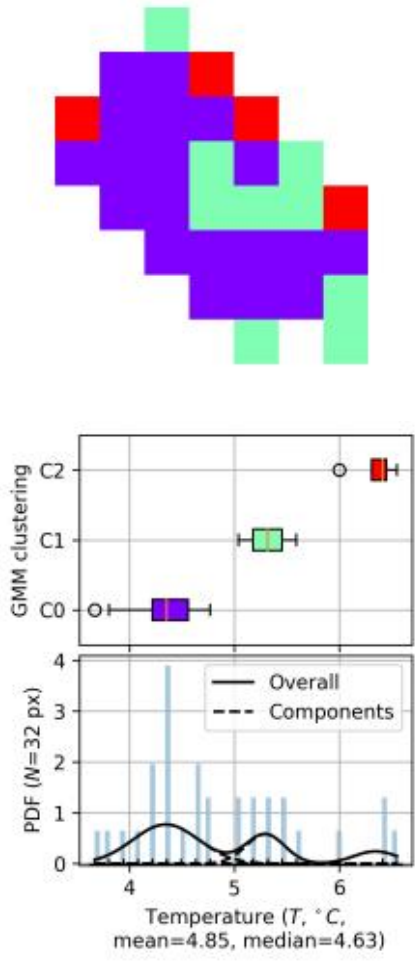

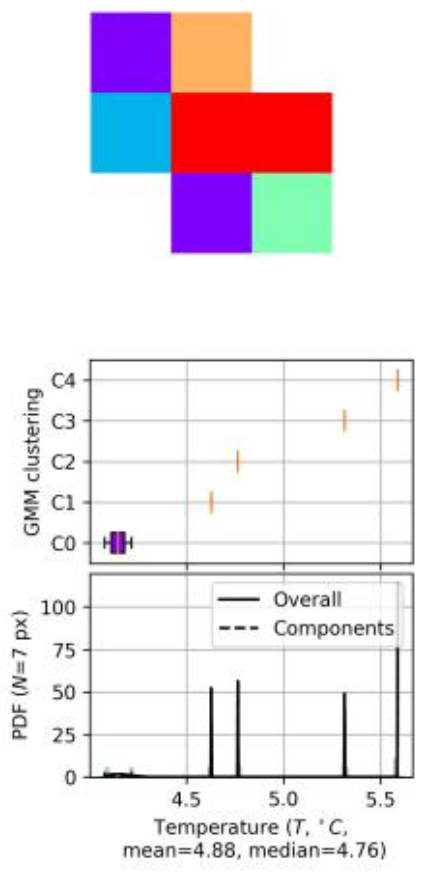

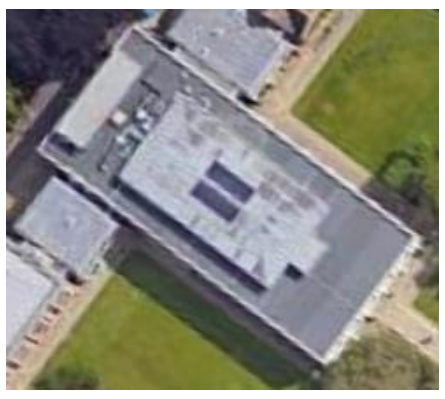

WC-8

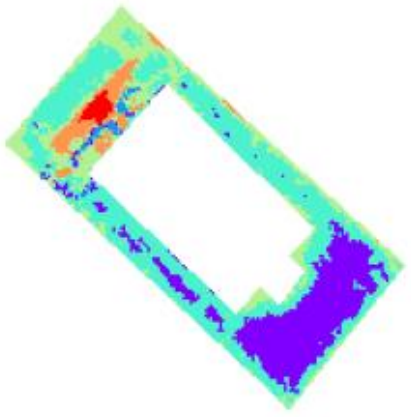

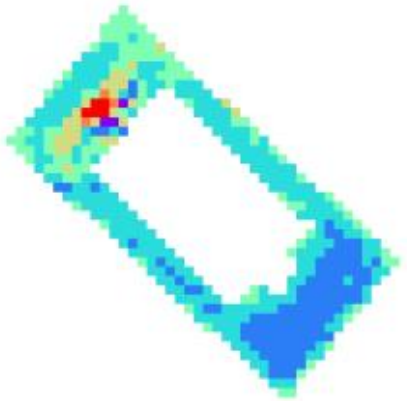

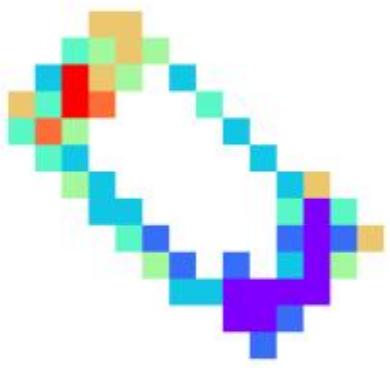

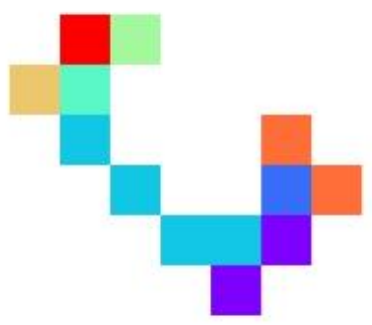

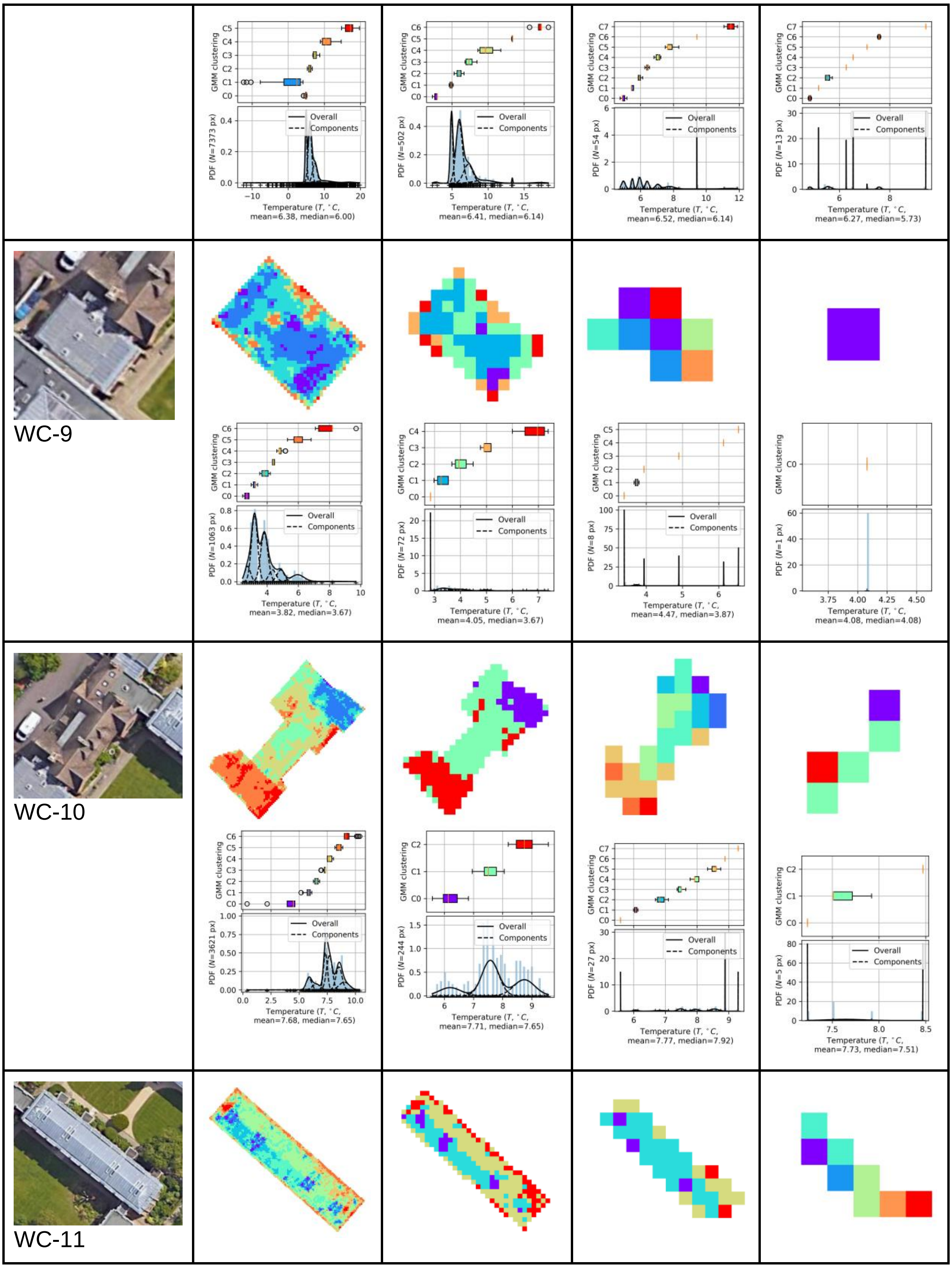

Temperature (T, °C, mean=6.38, median=6.00)
Temperature (T, °C, mean=6.41, median=6.14)
Temperature (T, °C, mean=6.52, median=6.14)
Temperature (T, °C, mean=6.27, median=5.73)
WC-9
Temperature (T, °C, mean=3.82, median=3.67)
Temperature (T, °C, mean=4.05, median=3.67)
Temperature (T, °C, mean=4.47, median=3.87)
Temperature (T, °C, mean=4.08, median=4.08)
WC-10
Temperature (T, °C, mean=7.68, median=7.65)
Temperature (T, °C, mean=7.71, median=7.65)
Temperature (T, °C, mean=7.77, median=7.92)
Temperature (T, °C, mean=7.73, median=7.51)
WC-11

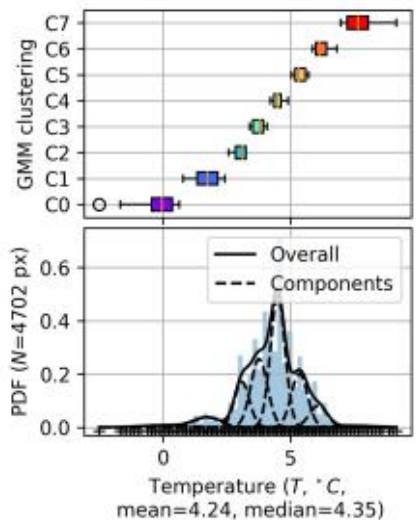

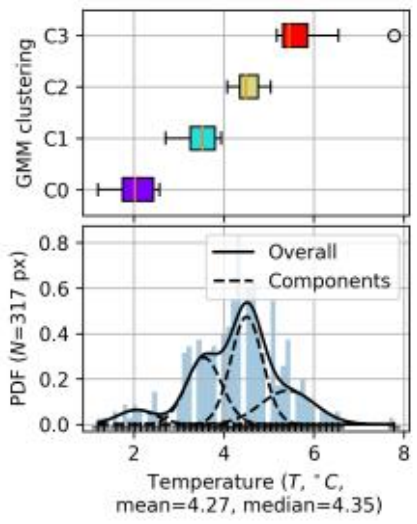

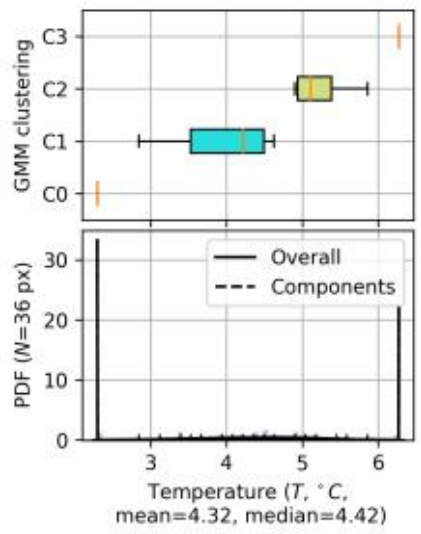

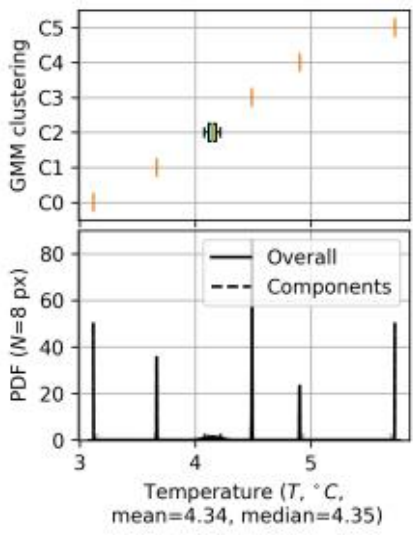

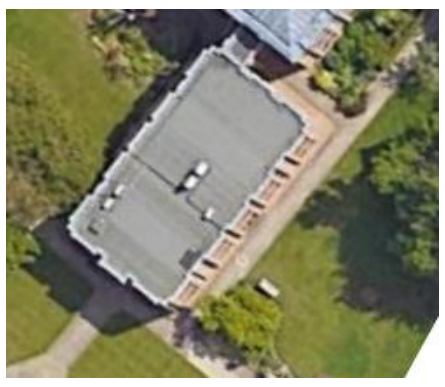

WC-12

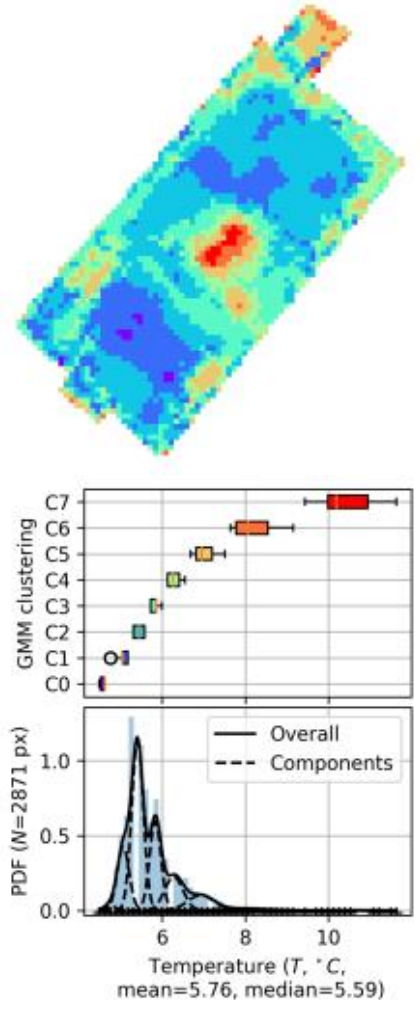

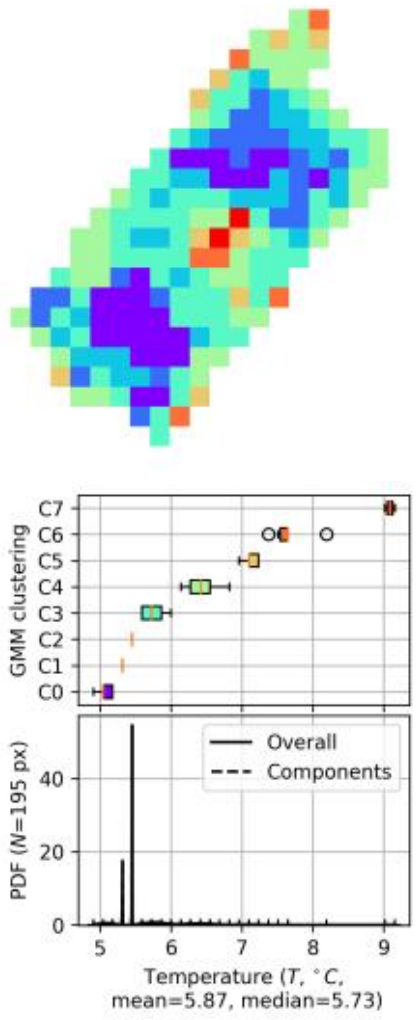

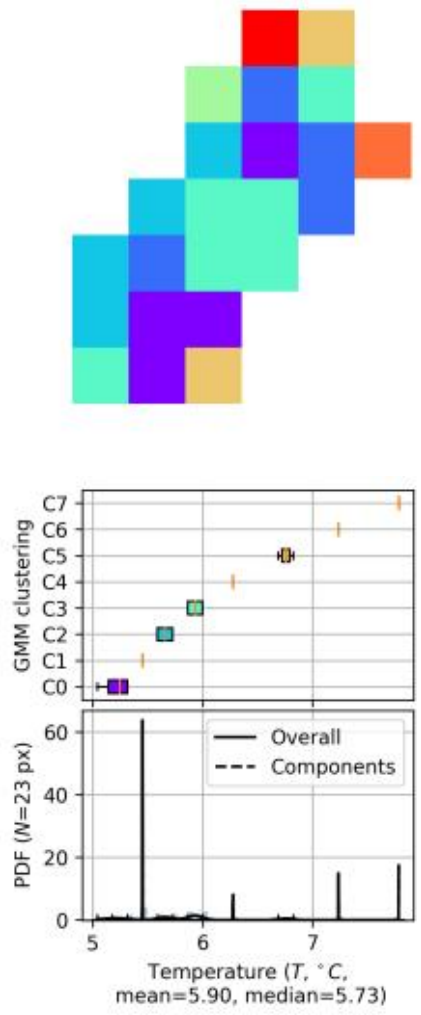

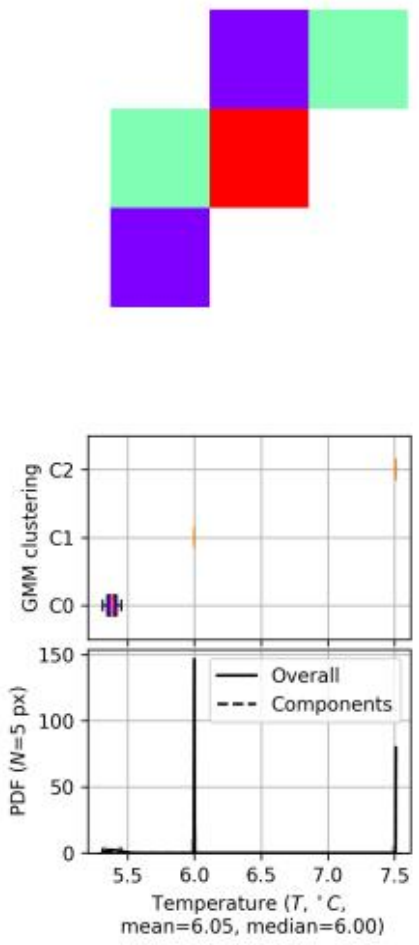

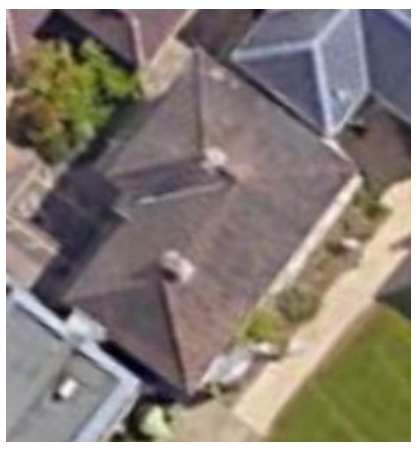

WC-13

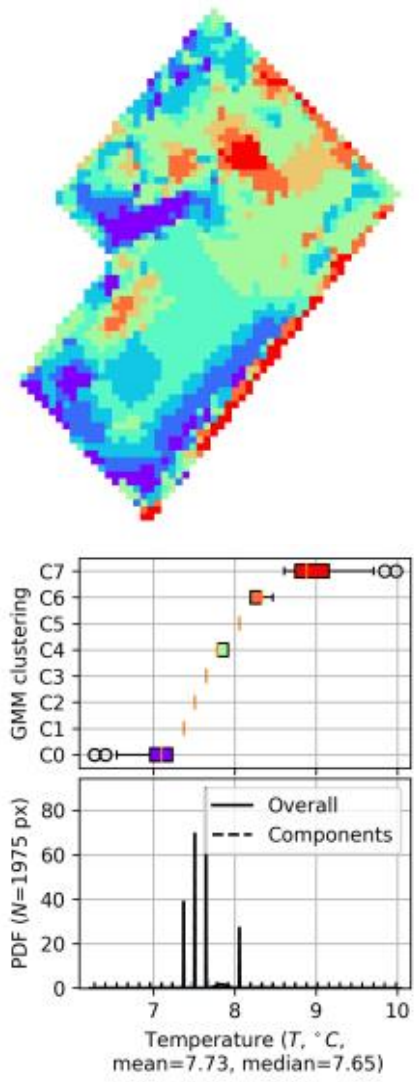

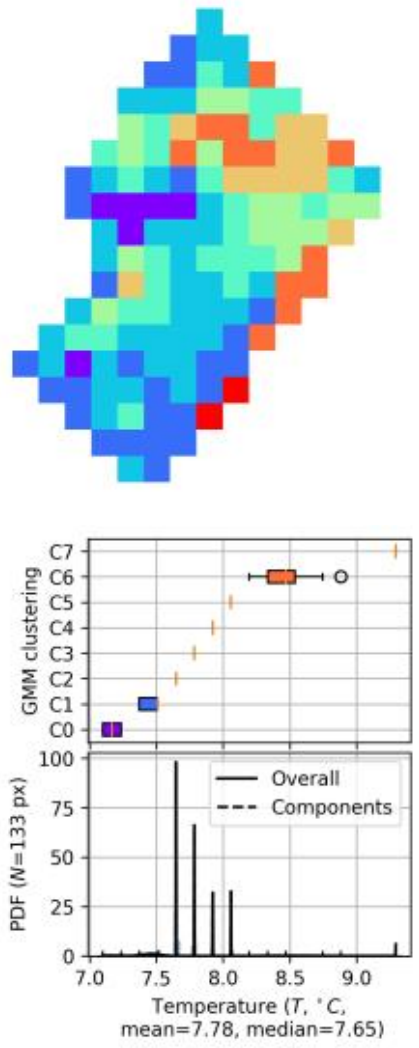

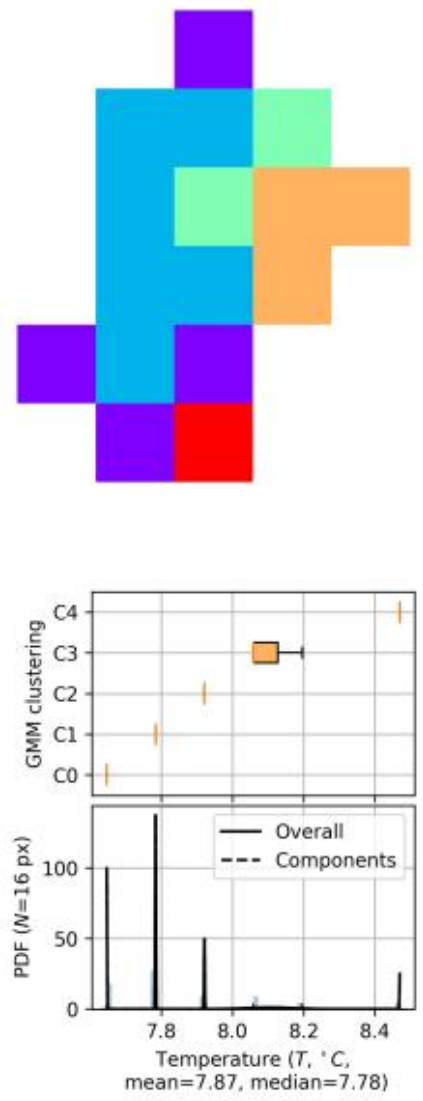

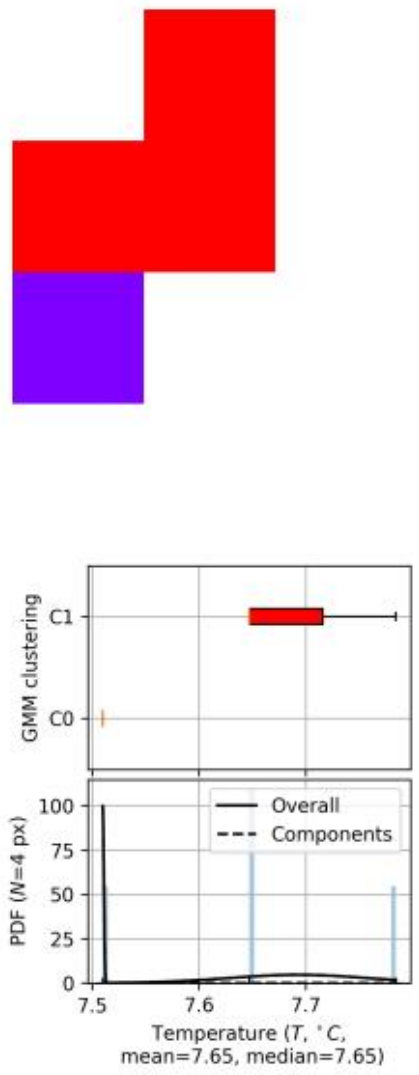

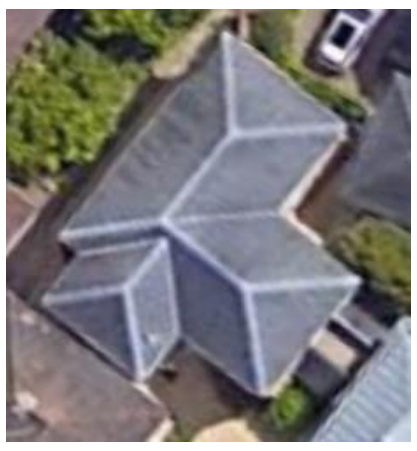

WC-14

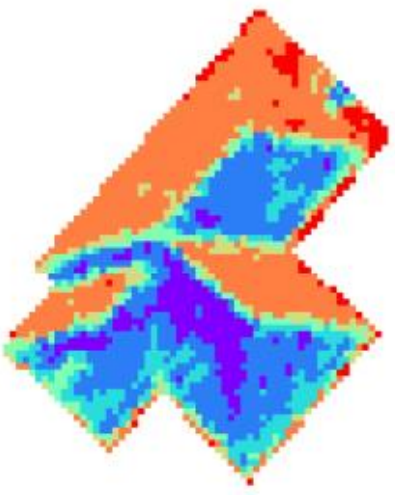

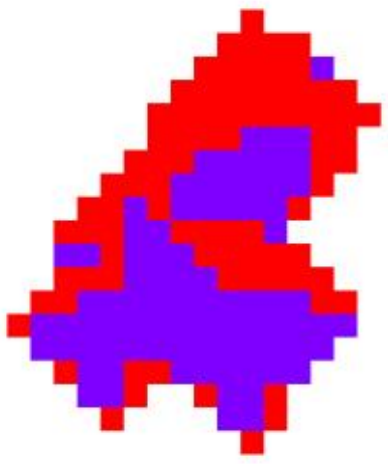

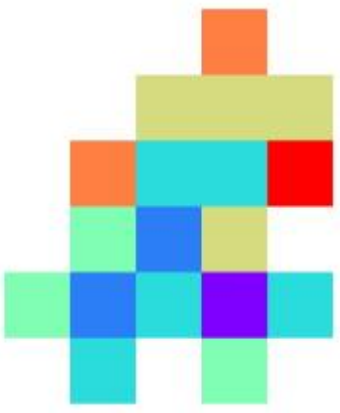

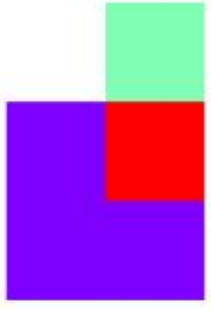

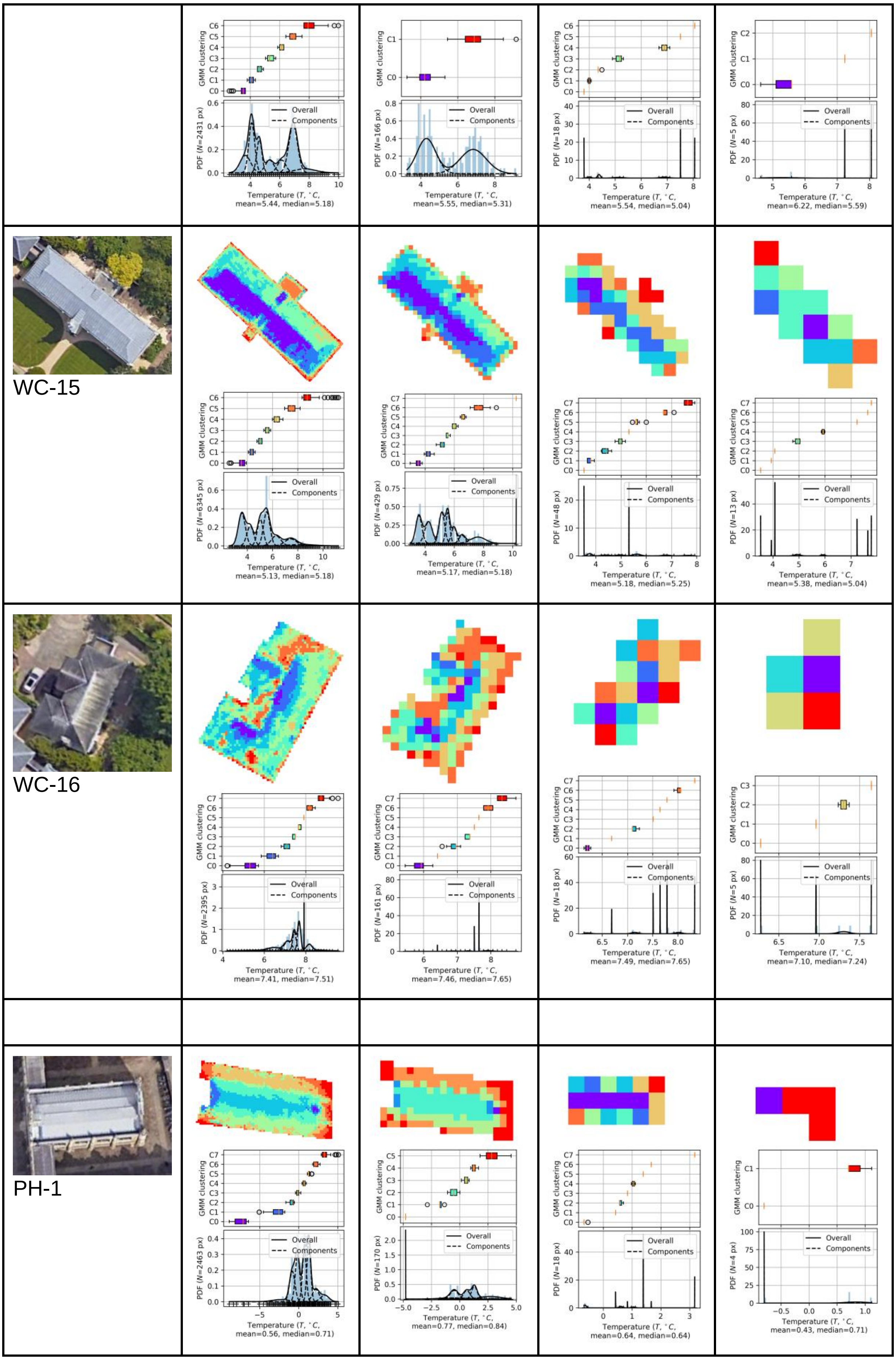

Temperature (T, °C, mean=5.44, median=5.18)
Temperature (T, °C, mean=5.55, median=5.31)
Temperature (T, °C, mean=5.54, median=5.04)
Temperature (T, °C, mean=6.22, median=5.59)
WC-15
Temperature (T, °C, mean=5.13, median=5.18)
Temperature (T, °C, mean=5.17, median=5.18)
Temperature (T, °C, mean=5.18, median=5.25)
Temperature (T, °C, mean=5.38, median=5.04)
WC-16
Temperature (T, °C, mean=7.41, median=7.51)
Temperature (T, °C, mean=7.46, median=7.65)
Temperature (T, °C, mean=7.49, median=7.65)
Temperature (T, °C, mean=7.10, median=7.24)
PH-1
Temperature (T, °C, mean=0.56, median=0.71)
Temperature (T, °C, mean=0.77, median=0.84)
Temperature (T, °C, mean=0.64, median=0.64)
Temperature (T, °C, mean=0.43, median=0.71)

| | | | | |
|---|---|---|---|---|
| 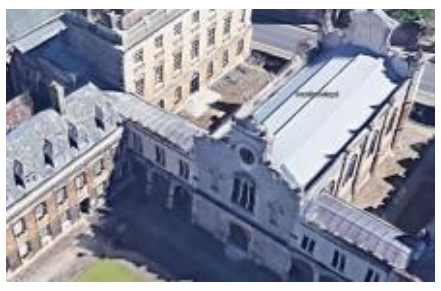 PH-2 | 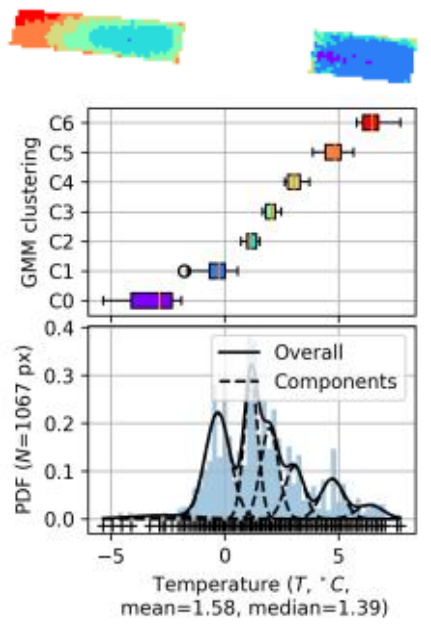 | 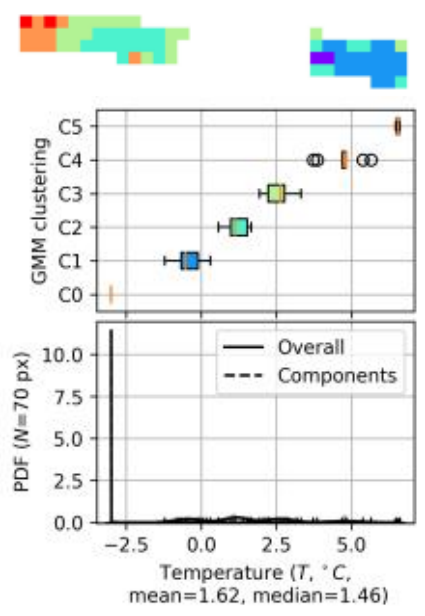 | 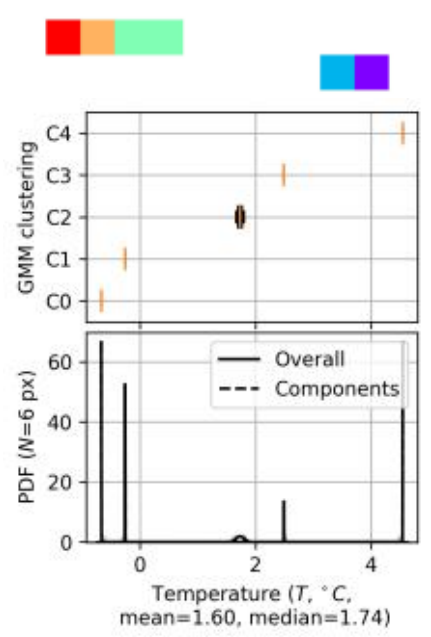 | 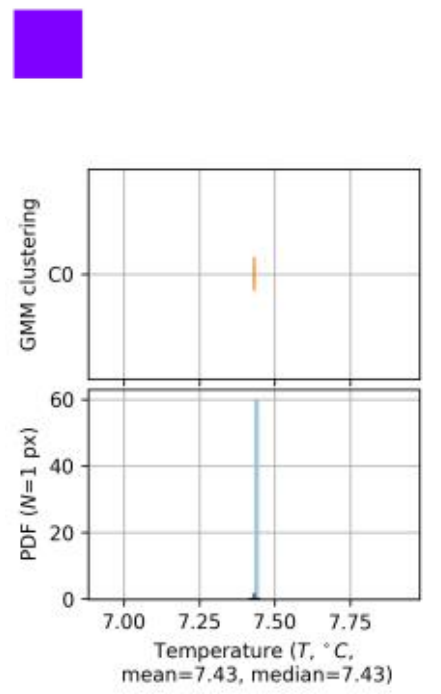 |
| 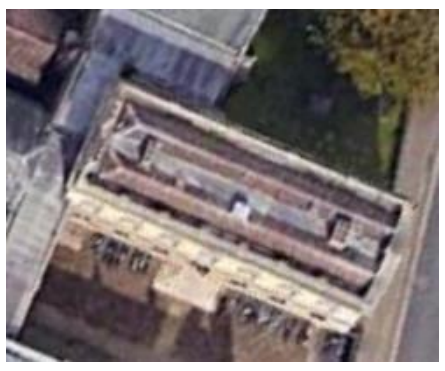 PH-3 | 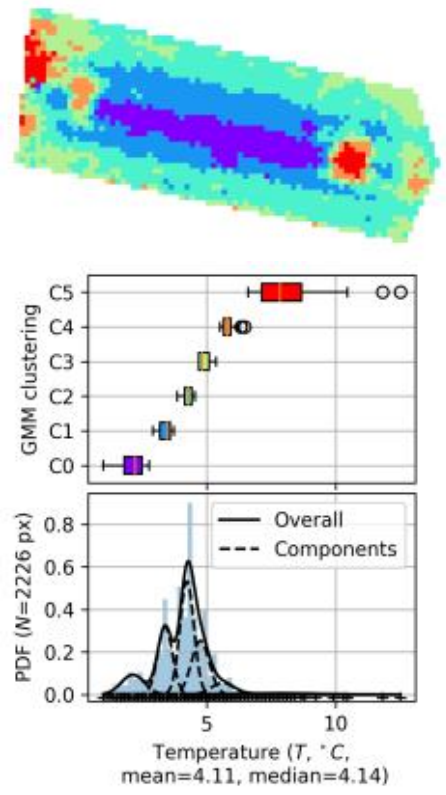 | 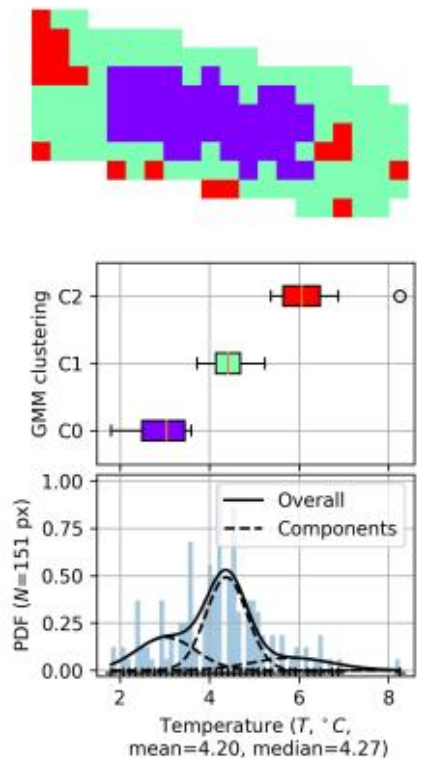 | 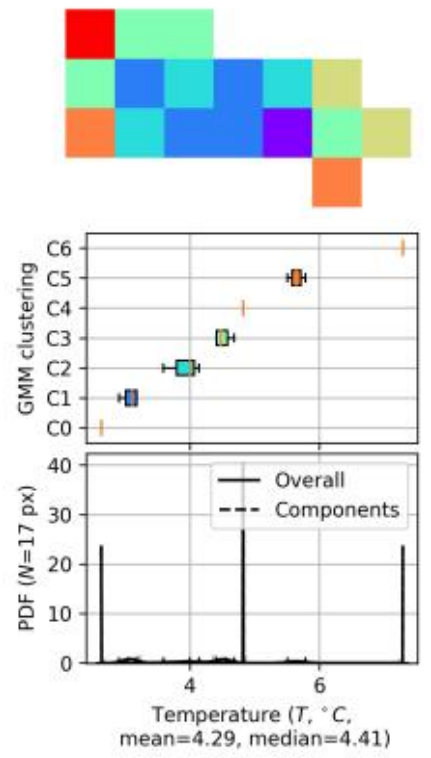 | 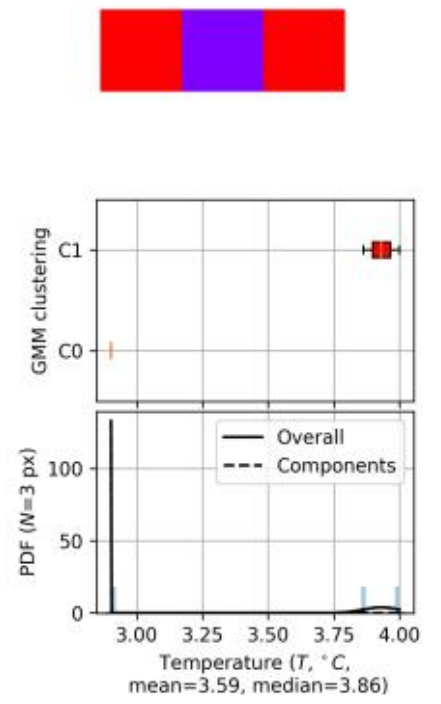 |
| 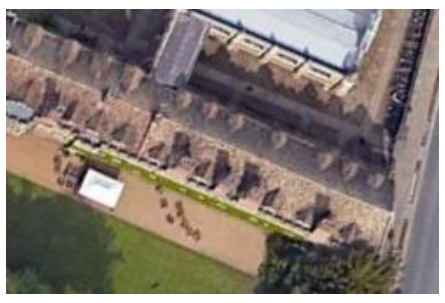 PH-4 | 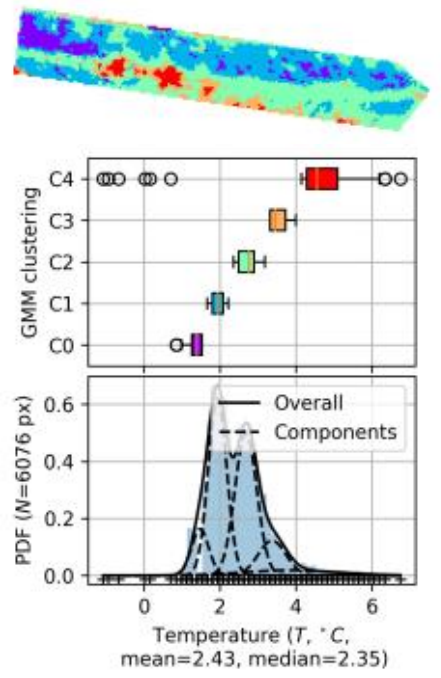 | 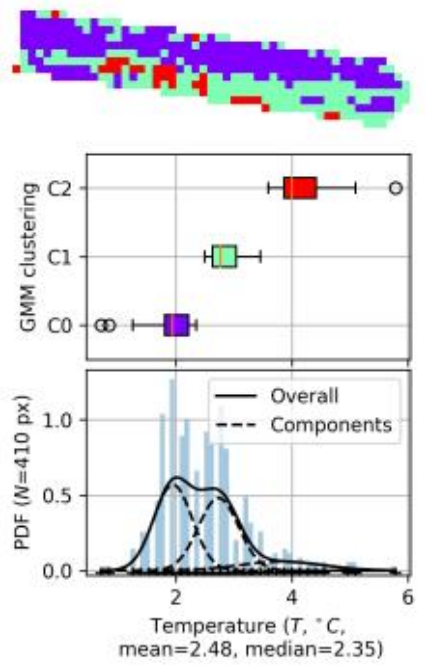 | 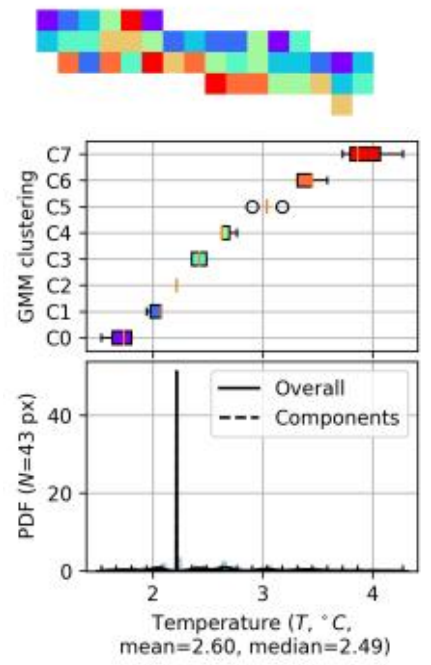 | 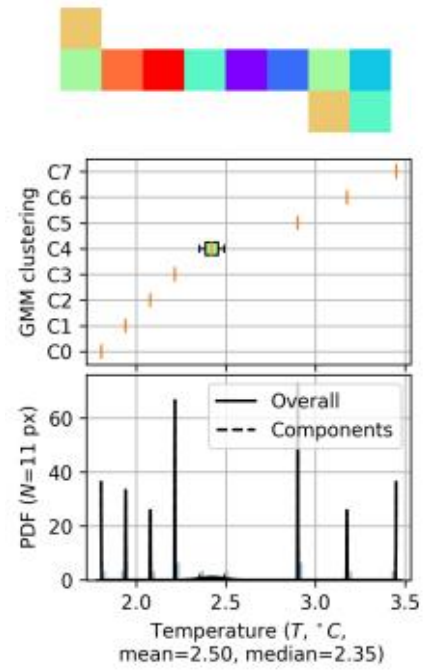 |
| 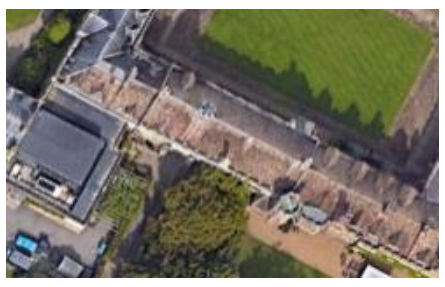 PH-5 | 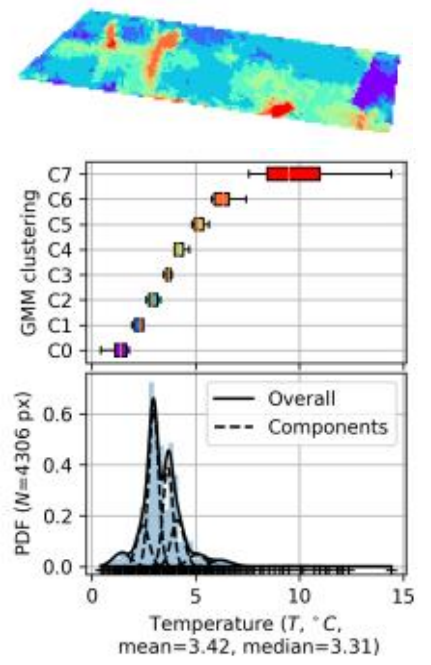 | 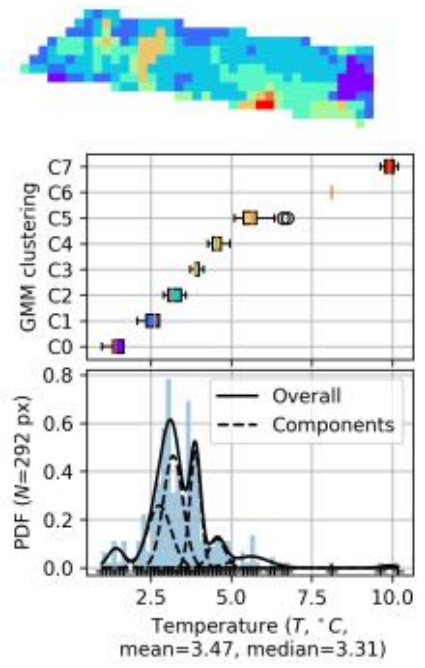 | 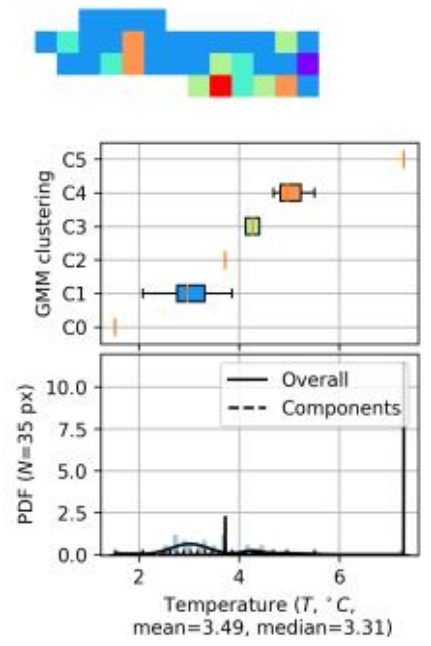 | 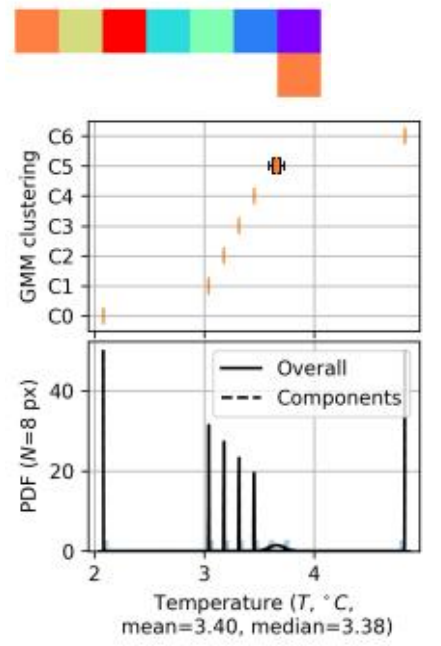 |
| 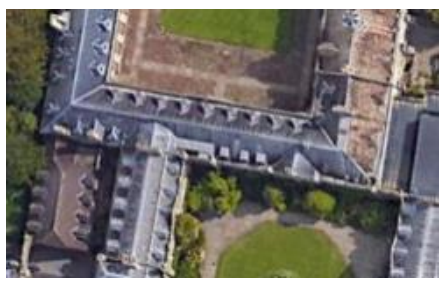 | 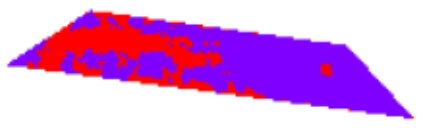 | 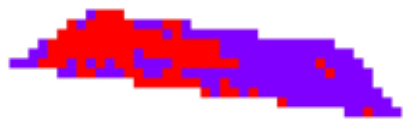 | 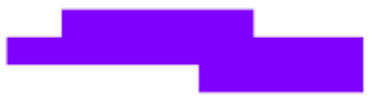 | 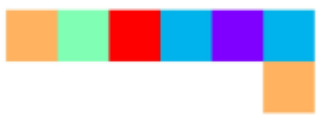 |
| PH-6 | 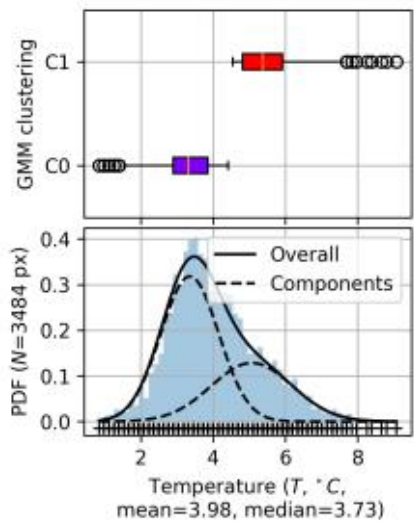 | 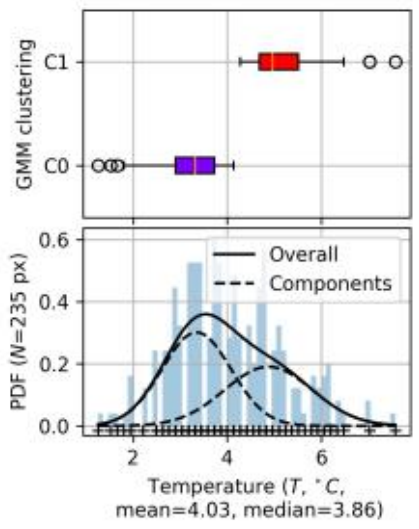 | 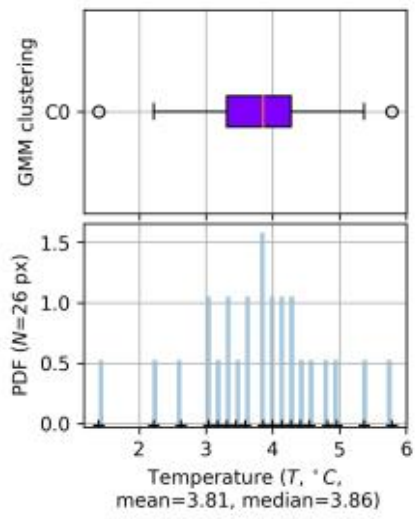 | 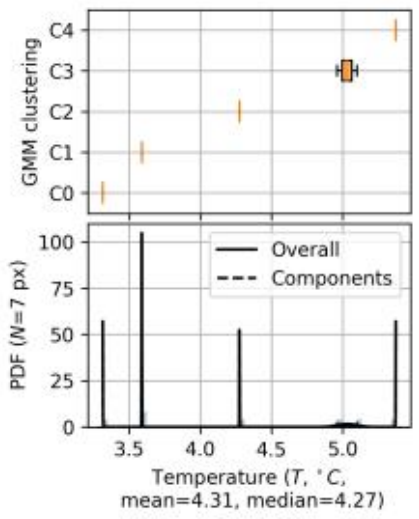 |
| 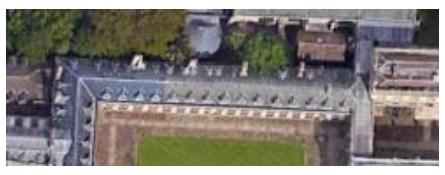 PH-7 | 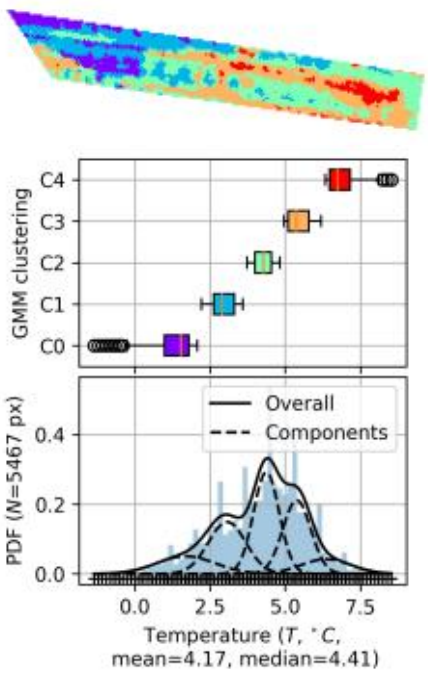 | 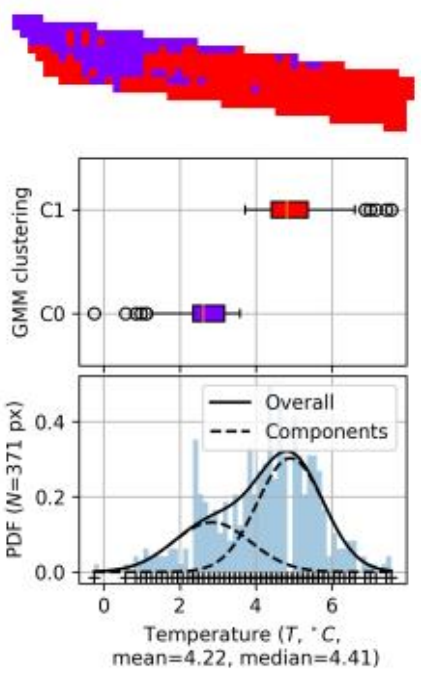 | 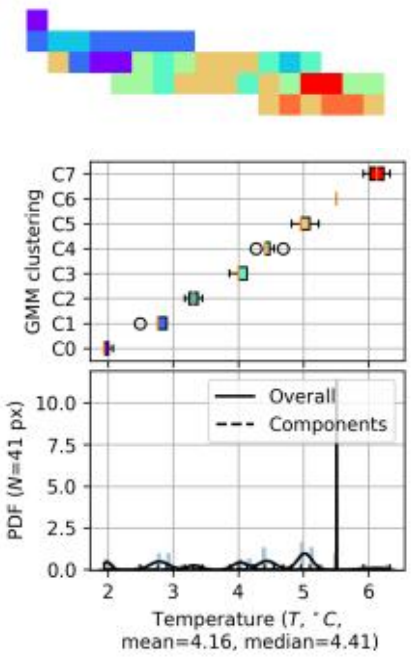 | 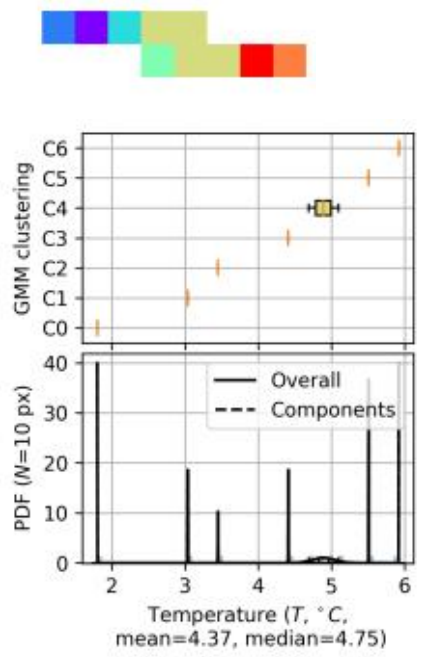 |
| 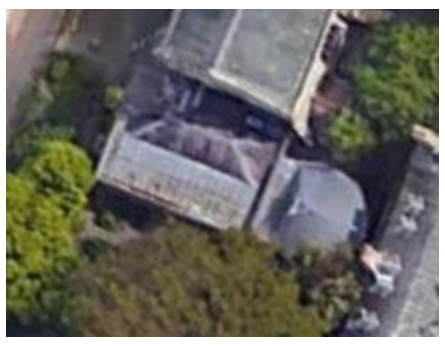 PH-8 | 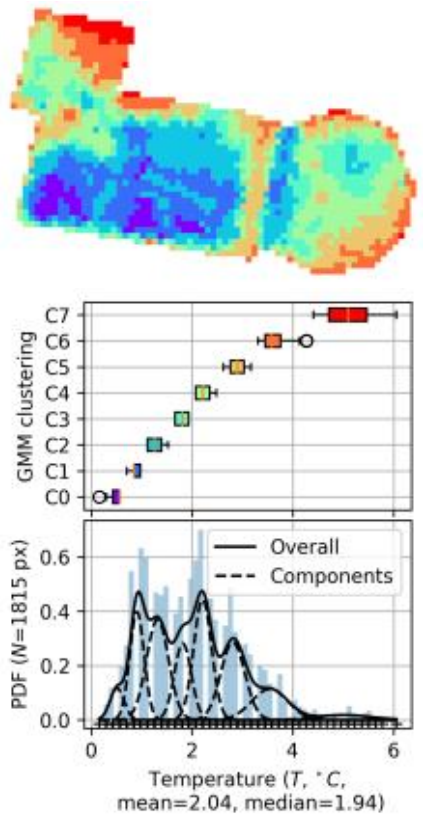 | 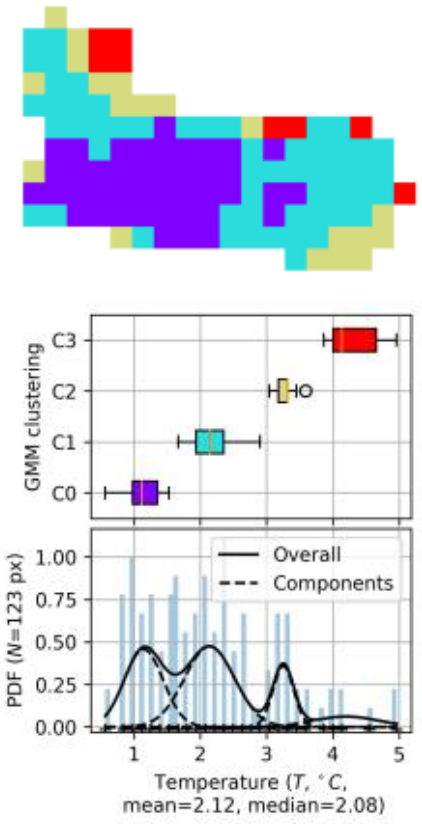 | 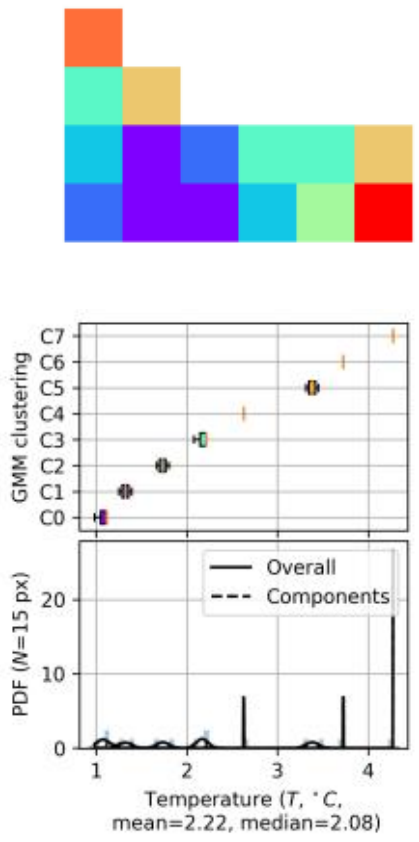 | 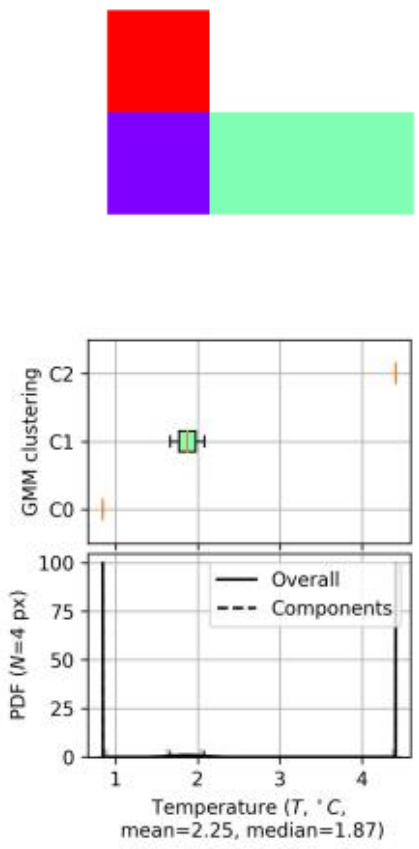 |
| 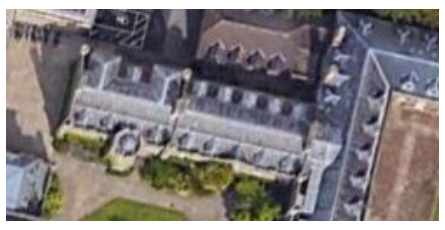 PH-9 | 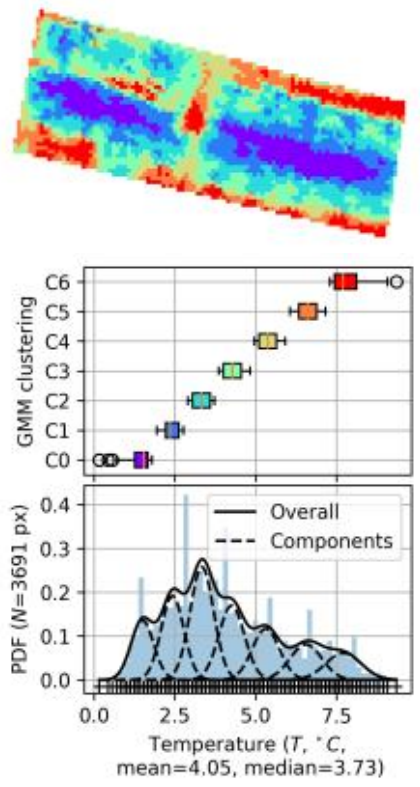 | 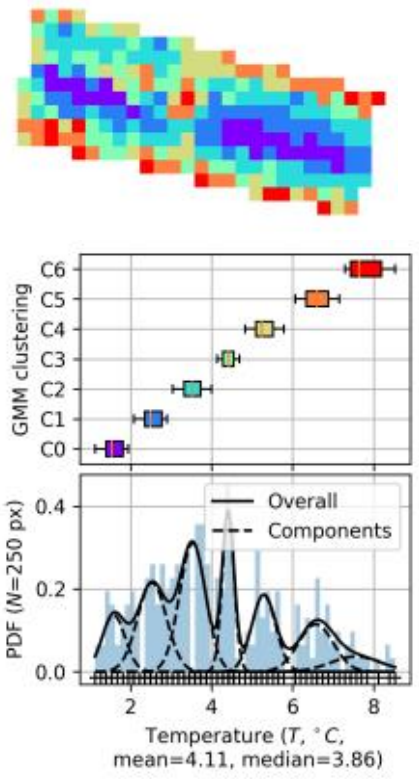 | 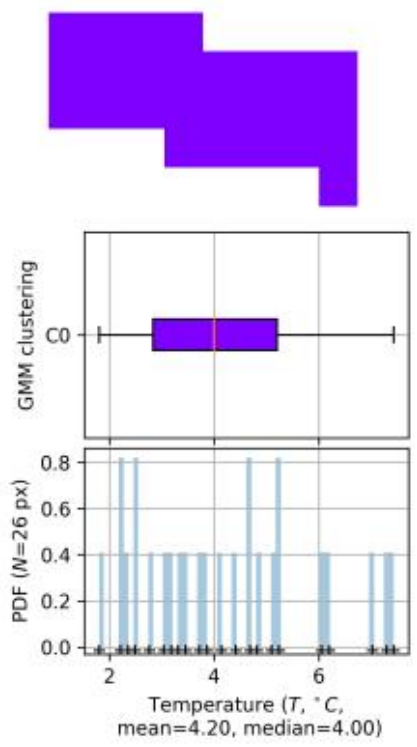 | 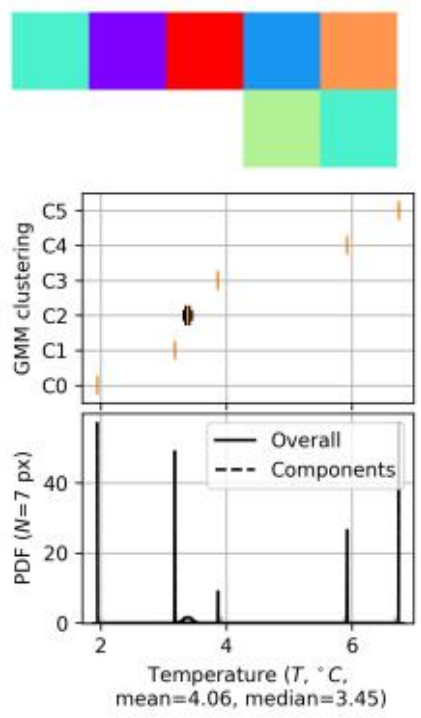 |
| 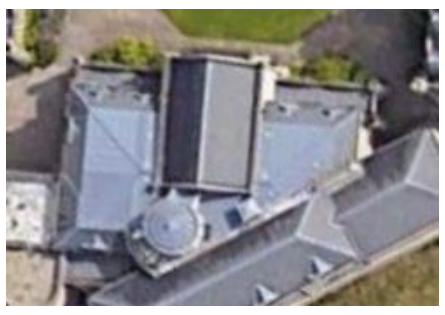 PH-10 | 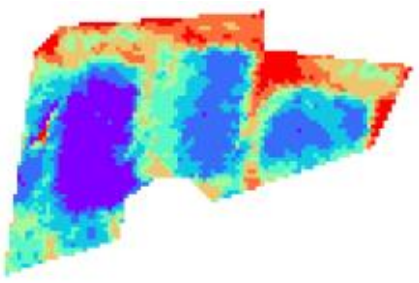 | 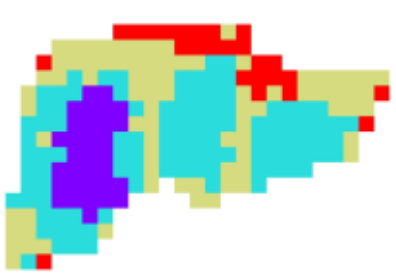 | 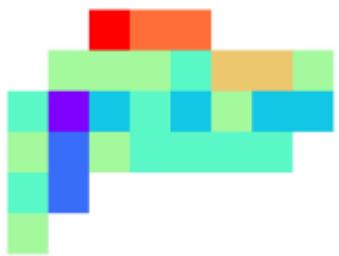 | 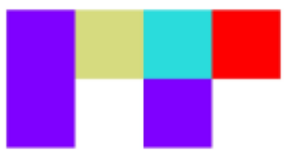 |

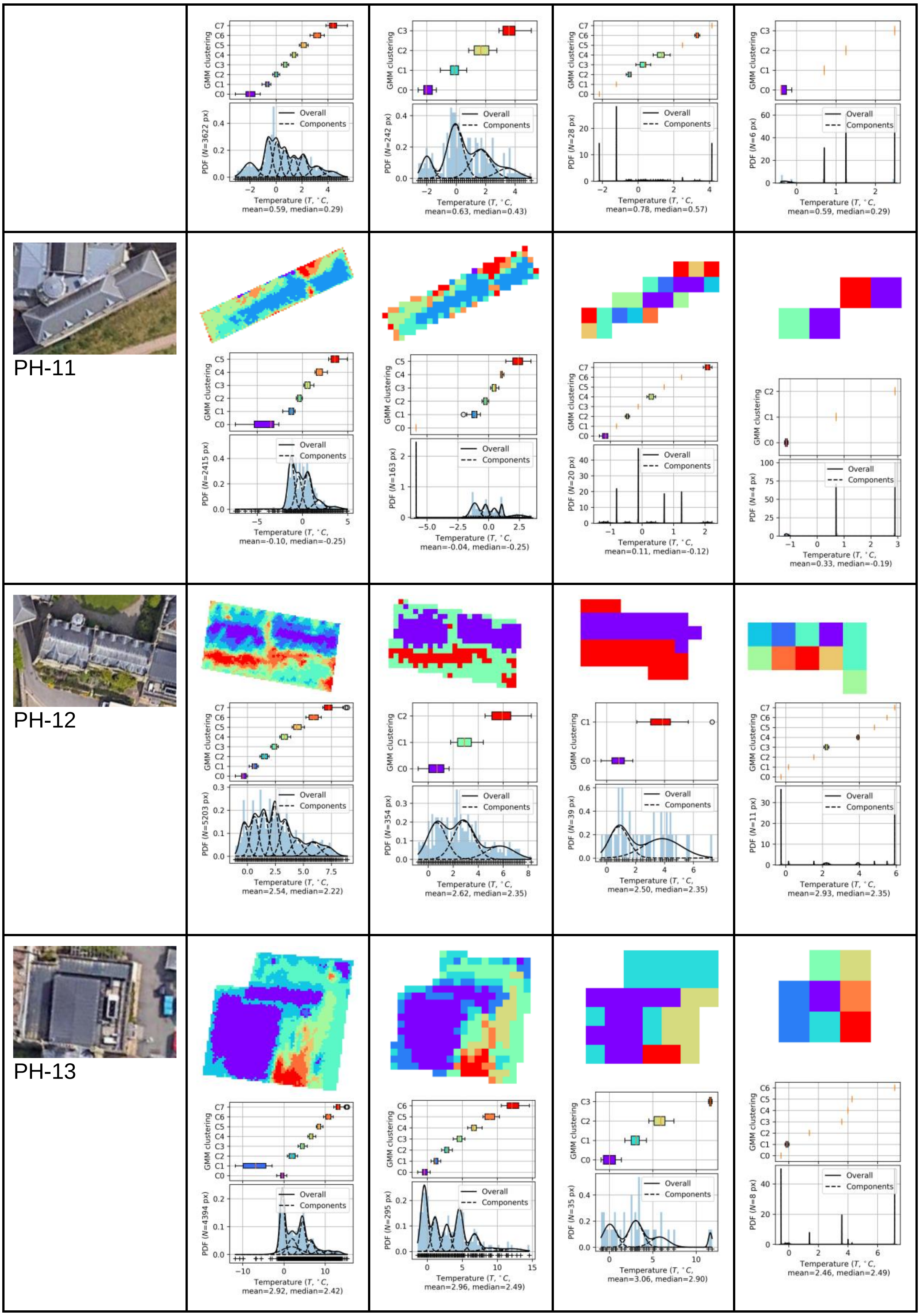

Overall
Components
GMM clustering
Temperature (T, °C, mean=0.59, median=0.29)
Temperature (T, °C, mean=0.63, median=0.43)
Temperature (T, °C, mean=0.78, median=0.57)
Temperature (T, °C, mean=0.59, median=0.29)
PH-11
Temperature (T, °C, mean=-0.10, median=-0.25)
Temperature (T, °C, mean=-0.04, median=-0.25)
Temperature (T, °C, mean=0.11, median=-0.12)
Temperature (T, °C, mean=0.33, median=-0.19)
PH-12
Temperature (T, °C, mean=2.54, median=2.22)
Temperature (T, °C, mean=2.62, median=2.35)
Temperature (T, °C, mean=2.50, median=2.35)
Temperature (T, °C, mean=2.93, median=2.35)
PH-13
Temperature (T, °C, mean=2.92, median=2.42)
Temperature (T, °C, mean=2.96, median=2.49)
Temperature (T, °C, mean=3.06, median=2.90)
Temperature (T, °C, mean=2.46, median=2.49)

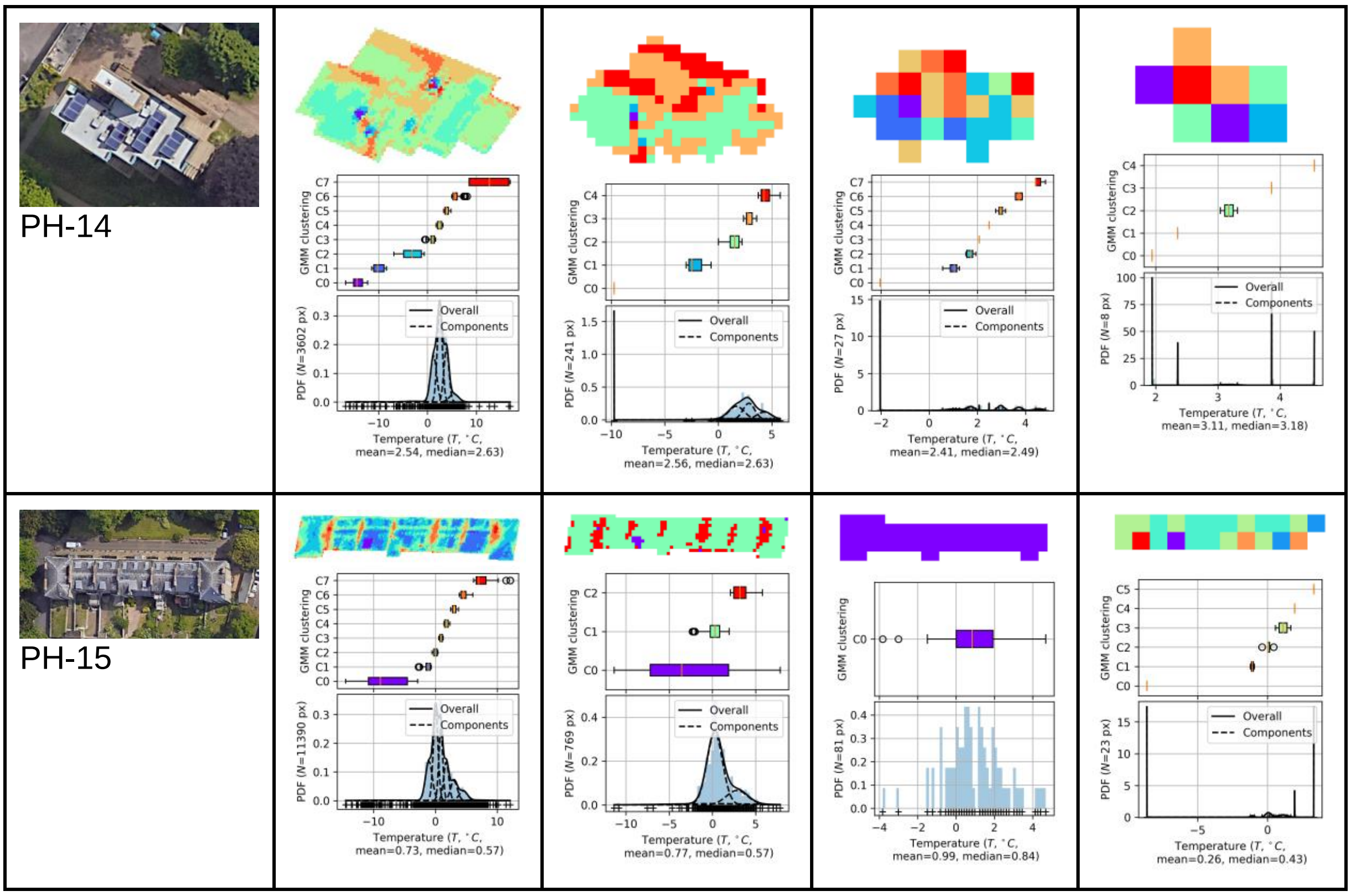